\documentclass[printer]{aa}  
\usepackage{graphicx}
\usepackage{txfonts}
\usepackage{natbib,twoopt}
\usepackage[breaklinks=false]{hyperref} 
\bibpunct{(}{)}{;}{a}{}{,}             

\usepackage{lscape}

\newcommand{\tiunit}[0]{$\mathrm{J}\,\mathrm{m}^{-2}\,\mathrm{s}^{-1/2}\mathrm{K}^{-1}$}
\newcommand{\cho}[0]{$\bar{\chi}^2_{\mathrm{min}}$ }

\begin{document} 

\title{Thermal properties of large main-belt asteroids observed by Herschel PACS}

\author{V. Al\'i-Lagoa\inst{1}
  \and
  T.~G. M{\"u}ller\inst{1}
  \and
  C. Kiss\inst{2,3}
  \and
  R. Szak\'{a}ts\inst{2}
  \and
  G. Marton\inst{2,3}
  \and
  A. Farkas-Tak\'acs\inst{2,4}
  \and
  P. Bartczak\inst{5}
  \and
  M. Butkiewicz-B\k{a}k \inst{5}
  \and
  G. Dudzi{\'n}ski\inst{5}
  \and
  A. Marciniak\inst{5}
  \and
  E. Podlewska-Gaca\inst{5}
  \and
  R. Duffard\inst{6}
  \and
  P. Santos-Sanz\inst{6}
  \and
  J. L. Ortiz\inst{6}
}

\institute{
  Max-Planck-Institut f{\"u}r extraterrestrische Physik, Giessenbachstrasse 1,
  85748 Garching, Germany\\
  \email{vali@mpe.mpg.de}
  \and
  Konkoly Observatory, Research Centre for Astronomy and Earth Sciences, Konkoly Thege Mikl{\'o}s {\'u}t 15-17, H-1121 Budapest, Hungary
  \and
  ELTE E\"otv\"os Lor\'and University, Institute of Physics, P\'azmány P\'eter s\'et\'any 1/A, H-1117 Budapest, Hungary
  \and
  E\"otv\"os Lor\'and University, Faculty of Science, P\'azmány P\'eter s\'et\'any 1/A, H-1117 Budapest, Hungary
  \and
  Astronomical Observatory Institute, Faculty of Physics, A. Mickiewicz University,
  S{\l}oneczna 36, 60-286 Pozna{\'n}, Poland.
  \and  Instituto de Astrof{\'i}sica de Andaluc{\'i}a (CSIC),
  Glorieta de la Astronom{\'i}a s/n, 18008 Granada, Spain
}

\date{Received 12 February 2020; accepted 10 April 2020}

\abstract{
Non-resolved thermal infrared observations enable studies of thermal and physical properties of asteroid surfaces provided the shape and rotational properties of the target are well determined via thermo-physical models.  
We used calibration-programme Herschel PACS data (70, 100, 160 $\mu$m) and state-of-the-art shape models derived from adaptive-optics observations and/or optical light curves to constrain for the first time the thermal inertia of twelve large main-belt asteroids. We also modelled previously well-characterised targets such as (1) Ceres or (4) Vesta as they constitute important benchmarks. 
Using the scale as a free parameter, most targets required a re-scaling $\sim$5\% consistent with what would be expected given the absolute calibration error bars. This constitutes a good cross-validation of the scaled shape models, although some targets required larger re-scaling to reproduce the IR data. 
We obtained low thermal inertias typical of large main belt asteroids studied before, which continues to give support to the notion that these surfaces are covered by fine-grained insulating regolith. Although the wavelengths at which PACS observed are longwards of the emission peak for main-belt asteroids, they proved to be extremely valuable to constrain size and thermal inertia and not too sensitive to surface roughness. 
Finally, we also propose a graphical approach to help examine how different values of the exponent used for scaling the thermal inertia as a function of heliocentric distance (i.e. temperature) affect our interpretation of the results.
}

\keywords{minor planets, asteroids: general --
  surveys --
  infrared: planetary systems
}
\titlerunning{Thermal properties of large asteroids from Herschel PACS data}

\maketitle

%
\section{Introduction}\label{sec:intro}

Non-resolved observations of asteroid thermal infrared (IR) emission provide information about the asteroid sizes and the thermal properties of their surfaces. 
By taking into account asteroid shape, rotation, and the geometry of the observations, thermo-physical models (TPMs) can be used to compute surface temperatures and fit thermal properties, such as thermal inertia, to the IR data. 
The number of asteroids with a thermo-physical characterisation has increased greatly over the last two decades thanks to the ever growing number of available shape and rotational models -- which are necessary input for the TPM -- and great observational efforts, both in the visible and the thermal IR \citep[for recent reviews see e.g.][and references therein]{Delbo2015,Durech2015,Mainzer2015}. 

The number of available shape models, by which we refer to both shape and rotational properties, is dominated by the several hundred models derived from the inversion of non-resolved optical light curves (e.g. \citealt{Durech2010,Durech2018,Hanus2011,Hanus2013}) following the method by \citet{Kaasalainen2001a} (see also \citealt{Kaasalainen2001,Kaasalainen2002}). 
Increasingly more sophisticated algorithms have begun to combine various data types in the inversion \citep{Carry2010,Durech2011,Viikinkoski2015,Bartczak2018}, even thermal IR data \citep{Durech2017, Mueller2017}, and there are now dozens of models based also on stellar occultations, radar, and adaptive optics observations \citep[e.g.][and references therein]{Durech2015,Benner2015}.

As the availability of asteroid shape models continues to increase, we can readily make use of a large collection of thermal IR data from several catalogues like those provided by the Infrared Astronomical Satellite (IRAS), AKARI, or the Wide-field Infrared Survey Explorer (WISE). 
Up to 2018, targeted observations from ground- and space-based facilities \citep[e.g.][]{Mueller2002,Mueller2004,Emery2006,Mueller2017,Landsman2018} and/or from all-sky thermal IR surveys (e.g. \citealt{Delbo2009,Ali-Lagoa2014,Rozitis2014,Hanus2015,Hanus2016,Bach2017,Marciniak2018}) had been used to model $\sim$100 asteroids with TPMs.  
With the recent addition of another 100 \citep{Hanus2018}, WISE data are now the single largest source of asteroid thermal inertias. 

Our aim in this work is to provide a thermo-physical characterisation of main-belt asteroids (MBAs) with Herschel Photodetector Array Camera and Spectrometer (PACS) data taken during the Asteroid Preparatory Programme for Herschel, ASTRO-F and the Atacama Large Millimiter/submillimiter array \citep{Mueller2005a}. 
Only a fraction of this data set has been exploited so far, for example in the context of absolute infrared flux calibration using well-characterised asteroids \citep{Mueller2014}, or in studies of specific targets \citep[e.g.][]{ORourke2012,Marsset2017}. 
The PACS data are important not only in terms of quality and additional wavelength coverage, but because their combination with AKARI and IRAS data allows us to bring the number of large MBAs analysed via a TPM closer to completeness\footnote{Large MBAs were too bright for WISE and their partially saturated fluxes might not be optimal for thermo-physical modelling. On the other hand, partial saturation can be corrected for and reasonably suited for thermal modelling purposes \citep[e.g.][]{Masiero2011,Mainzer2015}.}. 

This work was possible thanks to (i) the availability of a set of new shape models  derived using the All-Data Asteroid Modelling (ADAM) \citep{Viikinkoski2015} and the Shaping Asteroids with Genetic Evolution (SAGE) \citep{Bartczak2014,Bartczak2018} algorithms and (ii) the expert-reduced data products submitted to the Herschel Science Archive, which include some targets for which a non-standard data reduction approach was required\footnote{\url{http://archives.esac.esa.int/hsa/legacy/UPDP/SBNAF_MBA/SBNAF_MBA_Release_Note.pdf}}. 
Such data reduction and the creation of a large database featuring IRAS, AKARI, and WISE data \citep{Szakats2020} have been carried out in the framework of the ``Small Bodies: Near and Far'' (SBNAF) project \citep{Mueller2018}, funded by the European Commission. 
SBNAF aimed at exploiting synergies between different small-body modelling techniques that use a wide range of data types from ground and space observatories (stellar occultations, visible and thermal infrared photometry, etc.) to produce physical models of the selected targets, which range from near-Earth asteroids to trans-Neptunian objects. 

In Sect.~\ref{sec:data} we briefly describe the IR data set and in Sect.~\ref{sec:TPM} the TPM approach and simplifying assumptions. In Sect.~\ref{sec:results} we present our results and in Sect.~\ref{sec:discussion} we provide further discussion and comments. Appendix \ref{app:tpm} contains plots relevant to the TPM analysis and Appendix \ref{app:fluxes} a table with all the Herschel PACS data used in this work.

\section{Data}\label{sec:data}

In this section we give a brief summary of the IR data.
Further details of the Herschel PACS \citep{Pilbratt2010,Poglitsch2010} catalogue of asteroid observations \citep{Mueller2005} can be found in \citet{Mueller2014}, who already published the PACS data of (1) Ceres, (2) Pallas, (4) Vesta, and (21) Lutetia, which were used as IR calibrators in that work. 
In addition, information about the more specialised data reduction approach required for some PACS measurements is already available from the User Provided Data Products release note of the main-belt asteroid expert-reduced data products submitted to the Herschel Science Archive by Kiss, M\"{u}ller and Farkas-Tak{\'a}cs$^2$.  

We also used whatever data was available from the IRAS and Midcourse Space Experiment (MSX) \citep{Tedesco2002,Tedesco2002b}, AKARI Infrared Camera \citep{Murakami2007,Takita2012,Usui2011,Hasegawa2013}, and WISE catalogues (W3 and W4 data; \citealt{Mainzer2011,Masiero2011,Wright2010})
\footnote{For (21) Lutetia (see Sect.~\ref{sec:lutetia}), we also included most of the data analysed in \citet{ORourke2012}. }. 
We used all colour-corrected flux densities and absolute calibration error bars compiled in the SBNAF thermal infrared database (IRDB). 
\citet{Szakats2020} provide all relevant details about the production of the IRDB\footnote{https://ird.konkoly.hu/}, including the colour correction approach, sources and references of each catalogue, and useful auxiliary quantities such as observation geometry, and light-travel time. For completeness, Table~\ref{tab:fluxes} provides all previously unpublished PACS observations used in this work. 

\section{Thermo-physical model implementation}\label{sec:TPM}

Our TPM implementation is the one used by \citep{Ali-Lagoa2015} based on that of \citet{Delbo2007} and \citet{Delbo2009}. Said version was upgraded to account for the effects of shadowing, but global self-heating was neglected for this work \citep[see e.g. the discussion about average view factors by][]{Rozitis2013}. 
Below we provide only a brief summary of the technique that we use and the approximations that we make. 
Our methodology was described in more length in \citet{Marciniak2018} and \citet{Marciniak2019}. There, details about the thermo-physical modelling of each target were provided in separate sections. Here, we only present some relevant plots in the main text and include all TPM-related plots and additional comments in Appendix~\ref{app:tpm}. 

We take a shape model as input for our TPM (see
Table~\ref{tab:shapes}) with the main goal of modelling the surface
temperature distribution at epochs at which we have thermal IR observations
and constrain the target's diameter, thermal inertia and, whenever possible,
surface roughness. 
From the known geometry of observation we identify which surface elements
(usually triangular facets) were visible to the observer at each epoch
and compute the model fluxes from the surface temperature distribution.

To account for heat conduction towards the subsurface, we solve the 1D heat
diffusion equation for each facet and we use the Lagerros approximation when
modelling the surface roughness via hemispherical craters of different depth
coveraging 0.6 of the area of each facet  
(\citealt{Lagerros1996I}, \citealt{Lagerros1998}, \citealt{Mueller1998},
\citealt{Mueller2002}).
We also consider the spectral emissivity to be 0.9 regardless of the wavelength
(see e.g. \citealt{Delbo2015}). 

For each target, we estimated the Bond albedo (necessary input) as the
average value obtained from the different radiometric diameters available from
AKARI and/or WISE \citep{Usui2011,Ali-Lagoa2018,Mainzer2016}, and all
available $H$-$G$, $H$-$G_{12}$, and $H$-$G_1$-$G_2$ values from the
Minor Planet Center, \citet{Oszkiewicz2011}, or \citet{Veres2015}. 

This approach leaves us with two free parameters, the scale of the shape
(interchangeably called the diameter, $D$) and the thermal inertia ($\Gamma$).
The diameters and other relevant information related to the  TPM analyses of
our targets are provided  in Table \ref{tab:tpm}.
Whenever the data are too few to provide realistic error bar estimates, we
report the best fitting diameter so that the models can be scaled and
compared to the scaling given by the occultations. 
On the other hand, if we have multiple good-quality thermal data (with absolute
calibration errors below 10\%) then this typically translates to a size accuracy
of around 5\% as long as the shape is not too extreme and the spin vector
is reasonably well established. 
This rule of thumb certainly works for large MBAs like the Gaia mass targets. 
We do not consider the errors introduced by the pole orientation uncertainties
or the shapes (see \citealt{Hanus2015} and \citealt{Bartczak2019}), so our
TPM error bars estimates are lower values. 

\begin{table}
  {\footnotesize 
    \centering
    \caption{Origin and references for the shape models.}\label{tab:shapes}
    \begin{tabular}{l l l}
      \hline
      Asteroid & Model & Reference\\
      \hline
      (1) Ceres       & Dawn         & \citet{Park2019}\tablefootmark{a}\\
      (2) Pallas      & ADAM         & \citet{Hanus2017AO}\tablefootmark{b}\\
      (3) Juno        & ADAM         & \citet{Viikinkoski2015}\tablefootmark{b}\\
      (3) Juno        & SAGE         & \citet{Podlewska-Gaca2020}\tablefootmark{c}\\
      (4) Vesta       & Dawn         & Gaskell\tablefootmark{d}\\
      (8) Flora       & ADAM         & \citet{Hanus2017AO}\tablefootmark{b}\\
      (10) Hygiea     & ADAM         & \citet{Vernazza2019} \\
      (18) Melpomene  & ADAM         & \citet{Hanus2017AO}\tablefootmark{b}\\
      (19) Fortuna    & ADAM         & \citet{Hanus2017AO}\tablefootmark{b}\\
      (20) Massalia   & SAGE 1, 2   & \citet{Podlewska-Gaca2020}\tablefootmark{c}\\
      (21) Lutetia    & Rosetta      & Jorda \citep{Farnham2013} \\
      (29) Amphitrite & ADAM         & \citet{Hanus2017AO}\tablefootmark{b}\\
      (52) Europa     & ADAM         & \citet{Hanus2017AO}\tablefootmark{b}\\
      (54) Alexandra  & ADAM         & \citet{Hanus2017AO}\tablefootmark{b}\\
      (65) Cybele     & ADAM         & \citet{Viikinkoski2017}\tablefootmark{b}\\
      (88) Thisbe     & ADAM         & \citet{Hanus2017AO}\tablefootmark{b}\\
      (93) Minerva    & ADAM         & \citet{Hanus2017AO}\tablefootmark{b}\\
      (423) Diotima   & ADAM         & \citet{Hanus2018eos}\tablefootmark{b}\\
      (511) Davida    & ADAM         & \citet{Viikinkoski2017}\tablefootmark{b}\\    
      \hline
    \end{tabular}
    \tablefoot{
      \tablefoottext{a}{Downloaded from the PDS. \url{https://pds.nasa.gov/}}
      \tablefoottext{b}{
        Downloaded from the DAMIT database \url{https://astro.troja.mff.cuni.cz/projects/damit/}
        \citep{Durech2010}}        
      \tablefoottext{c}{
        Available from the ISAM service \url{http://isam.astro.amu.edu.pl}
        \citep[see][for details]{Marciniak2012}.
      }
      \tablefoottext{d}{
        Downloaded from the Dawn Public Data site
        \url{http://dawndata.igpp.ucla.edu/tw.jsp?section=geometry/ShapeModels}
      }
    }
  }
\end{table}

\section{Results}\label{sec:results}

\begin{figure}
  \centering
  \includegraphics[width=0.95\linewidth]{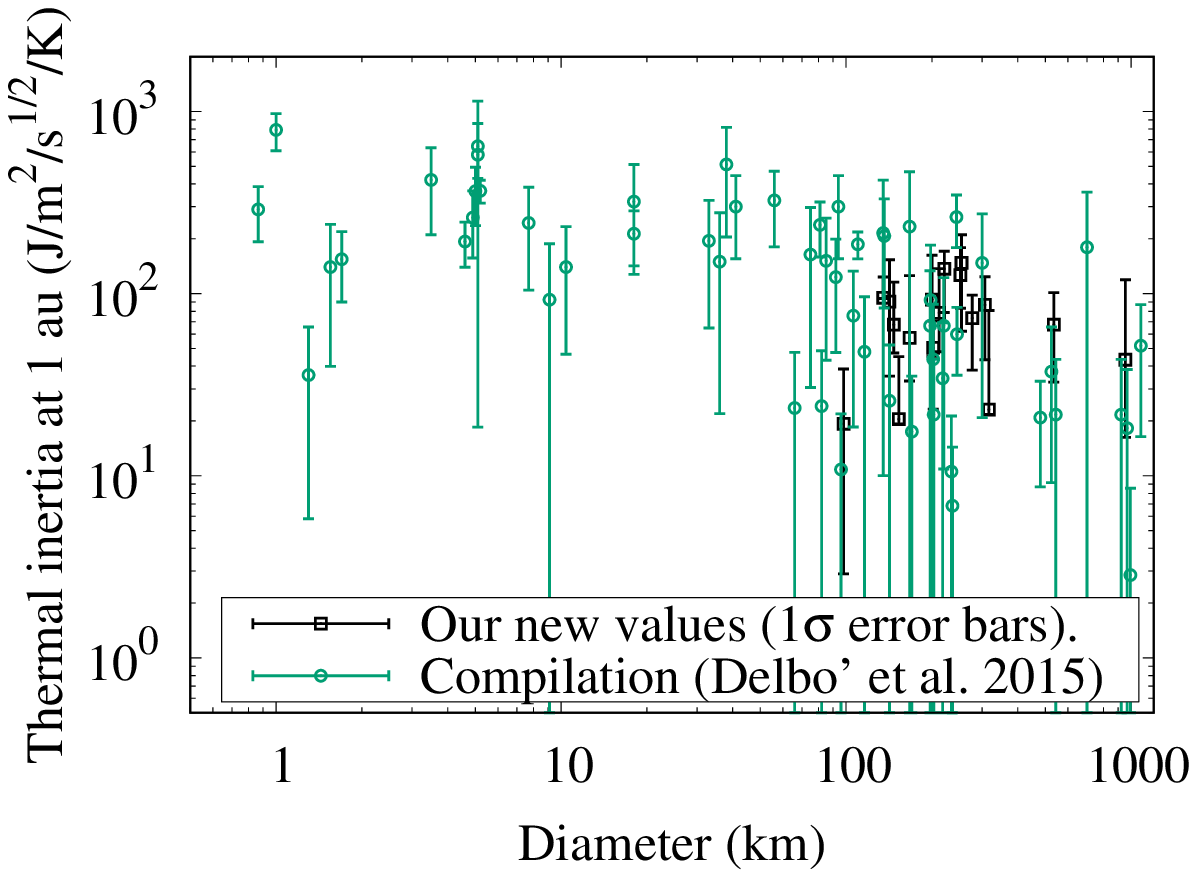}

  \caption{Thermal inertia normalised at 1 au versus size.
    The PACS targets (black symbols) follow the trend in the compilation by Delbo' et al.
    and \citet{Hanus2018}.
  }\label{fig:gamma_size}
  
  \includegraphics[width=0.95\linewidth]{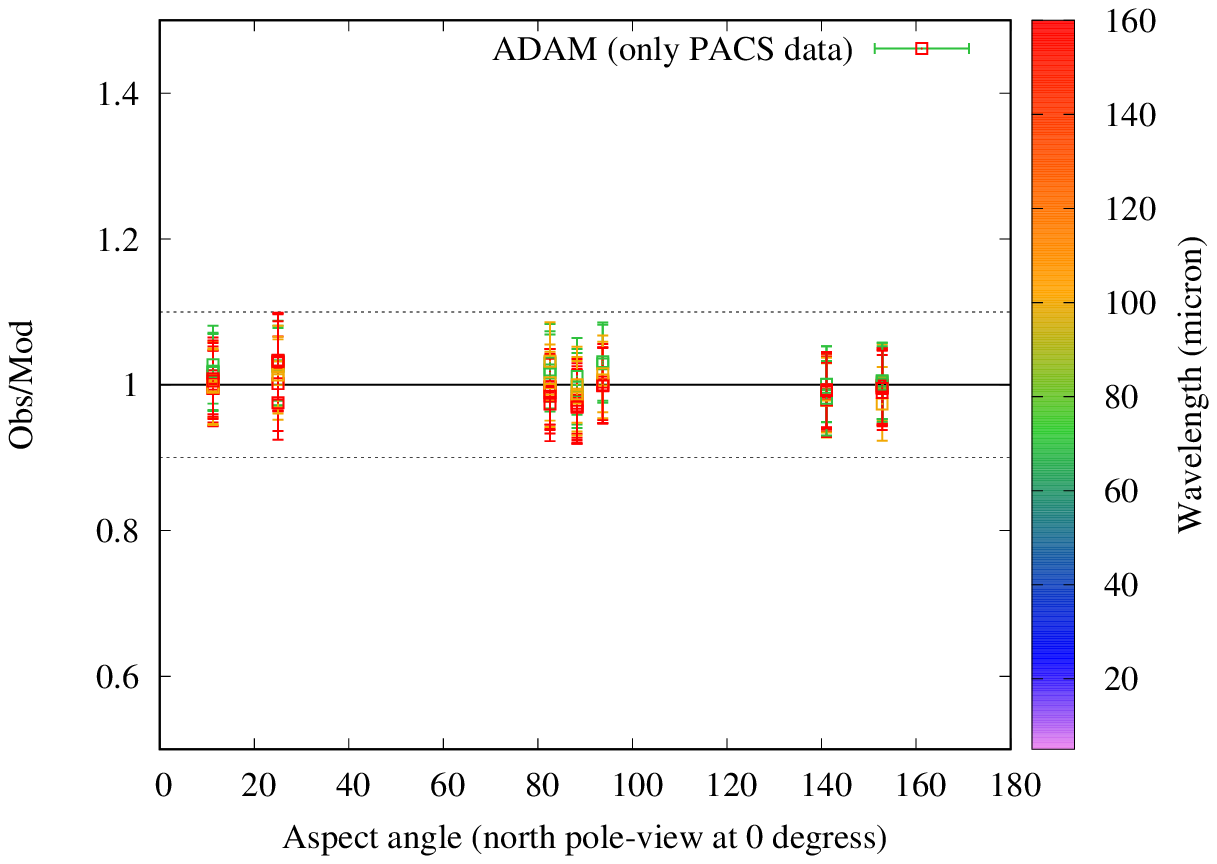}

  \caption{
    Observation to model ratios vs. aspect angle for the best fitting TPM model to the PACS data of
    (2) Pallas. The best fit was obtained by 
    $\Gamma=$\,30 \tiunit and a re-scaling of 1.03 
    (see Table~\ref{tab:tpm}). 
  }\label{fig:OMR_alpha_fitted}
\end{figure}

Table~\ref{tab:tpm} summarises the thermophysical properties obtained for each target in our sample.
For comparison, we also include the results for the spheres with the same rotational properties as the shape models. 
All plots produced during the modelling and further details about some selected targets are provided in Appendix \ref{app:tpm}.
In this section we provide a summary of the results and focus on several aspects relevant to the sample and/or a larger catalogue of thermo-physical properties retrieved from the literature. 
For example, the thermal inertia versus size plot (Fig.~\ref{fig:gamma_size}) shows that our new thermal inertias fall within the same range of those of other large MBAs found in previous works. 

\begin{figure}
  \centering
  \includegraphics[width=0.95\linewidth]{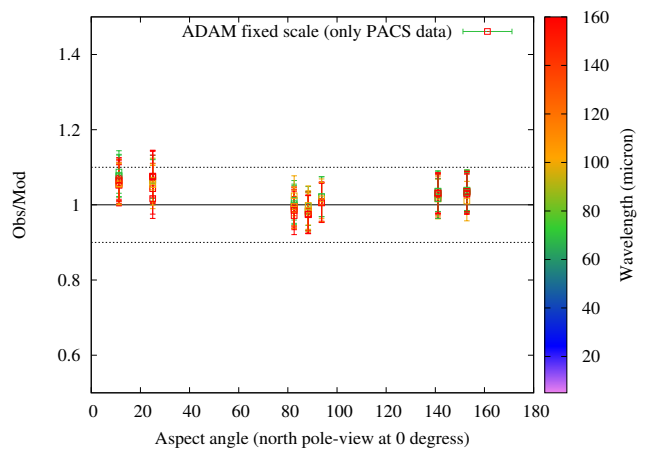}

  \caption{
    Same as Fig.~\ref{fig:OMR_alpha_fitted} but for the case when we
    kept the scale of the ADAM model fixed. 
    Here, the best-fitting thermal inertia was 8 \tiunit (cf. 30 \tiunit), but
    the data taken with a sub-observer point far from the equator could not be
    reproduced equally well. 
  }\label{fig:OMR_alpha_fixed}

\end{figure}

\begin{figure}
  \centering
  \includegraphics[width=0.95\linewidth]{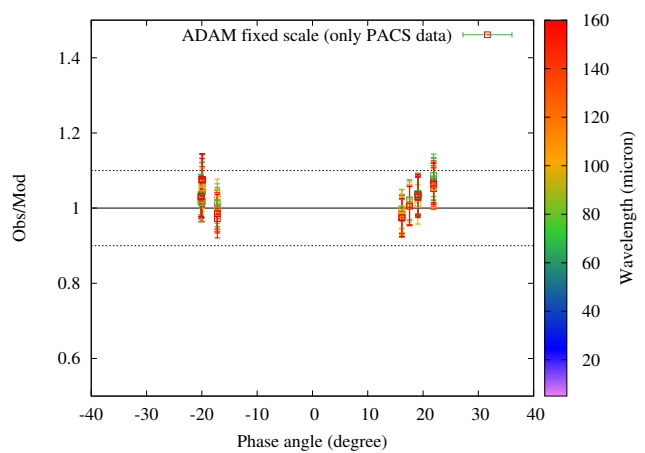}
  \caption{Observation to model ratios vs. phase angle for the best fit to PACS data of (2) Pallas
    keeping the scale of the shape model fixed. A lower $\Gamma$ fitted 
    the data better than when the scale was optimised
    (cf. Fig.~\ref{fig:00002_OMR}, lower panel). 
  }\label{fig:OMR_phase_pallas}  
\end{figure}

Most ADAM shape models only required a small re-scaling of the size to fit the data with low \cho, which we take as an additional confirmation of their high quality.
However, in the cases of (65) Cybele, (18) Melpomene, and (54) Alexandra, we required re-scalings of the order of 10\% to fit the data, which are larger than expected for the available high-quality thermal data and shape models \citep[e.g.][]{Delbo2015}. 
Thus, regardless of their quality, it is still worthwhile keeping the scale as a free parameter when performing thermo-physical modelling using already-scaled shape models in order to check potential issues. 
For example, the ADAM shape of Pallas with a fixed scale could not reproduce well PACS data taken closer to pole-on, as illustrated in Figs.~\ref{fig:OMR_alpha_fitted} and \ref{fig:OMR_alpha_fixed}. 
The first one shows the observation-to-model ratios (OMR) versus aspect angle for our best fit, that is $\Gamma = 30$\,\tiunit and a re-scaling of +3\%.
The second one shows the best fit obtained with a fixed scale, which led to $\Gamma = 8$ \tiunit. Although this fit is also formally acceptable, the OMRs are only close to one at equatorial views (aspect angles around 90 degrees).
No other combination of thermal inertia and roughness could bring the other ratios 
closer to unity, which suggests that the shape model could be less accurate at
pole-on views. 
In this particular case, this could also explain the lower thermal inertia
needed to fit the data when we kept the scale fixed, because the data taken at
higher phase angles were coincidentally the ones taken at non-equatorial
sub-observer latitudes, which makes it more difficult for the shape model to
reproduce the thermal phase curve (see Fig.~\ref{fig:OMR_phase_pallas}). 

We found formally good fits and thermal properties within the expected
range for the two targets with SAGE models available for our study as well, albeit with slightly higher error bars.
Additional SAGE models scaled using stellar occultation chords and further discussion are provided by \citet{Podlewska-Gaca2020}. 
It is worth mentioning that the TPM did not help us to favour any of the two mirror shape solutions of (20) Massalia. 

For each shape model, we also fitted the data using spheres with the corresponding rotational parameters with the aim of comparing the resulting thermo-physical parameters (see Table \ref{tab:tpm}). 
On the one hand, this approximation seems to provide reasonable estimates for the diameters that would be expected for large objects with relatively low-amplitude light curves. 
On the other hand, spheres lead to scales $\sim$5\% larger on average than the ADAM shapes, and several of the thermal inertia values were up to an order of magnitude higher than those of the corresponding shape models.  
With the current sample we could not identify any single cause that should lead to such systematic effects, but perhaps a future larger sample could help explore this further.
Also, this could be used to create a reliable benchmark to estimate thermal inertias of objects with more limited shape models.

\subsection{Thermal inertia, pole-on views, and rotational variability}\label{sec:poleonviews}

Only those asteroids with highly oblique rotational axes (or alternatively, unusually high orbital inclinations) can be seen with aspect angles close to pole-on, that is, 0 or 180 degrees for the north and the south pole. 
In these cases, seasonal effects on the surface temperature distribution are expected, since the depth at which the heat wave penetrates is higher at close to pole-on illuminations.
In principle, data taken in such circumstances could require a model accounting for thermal inertia variation with depth (such as those used for the Moon; \citealt{Hayne2017}), but our OMR versus aspect angle plots do not suggest problems in fitting data taken at close to pole-on views (e.g. see Fig.~\ref{fig:OMR_alpha_fitted}). We did not find any correlation between $\Gamma$ and the pole ecliptic latitude either. 
The vast majority of OMR versus rotational phase plots suggest that the shape models and the assumption of homogeneous thermal properties throughout the surface can still reproduce the IR data well (see Figs.~\ref{fig:00001_OMR}--\ref{fig:00511_OMR}, third panel from the top).

\subsection{Constant thermal inertia with temperature}\label{sec:temperature}

\begin{figure}
  \centering
  \includegraphics[width=0.95\linewidth]{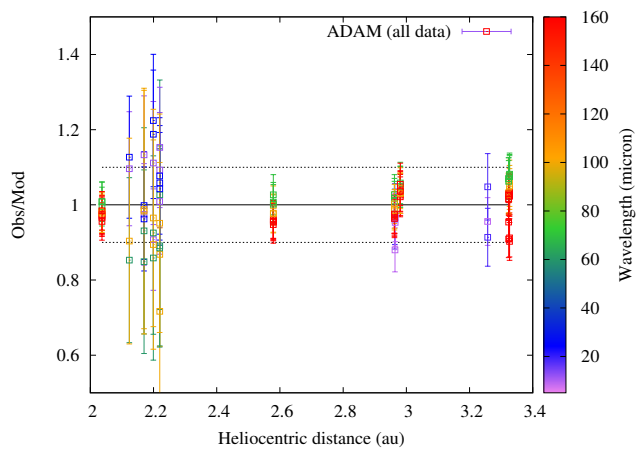}
  \caption{Observation-to-model ratios vs. heliocentric distance
    for (3) Juno using the ADAM model. 
    Although there is a slight curvature in the plot, it shows that
    with reasonable accuracy we can fit  data taken at different
    temperatures with a constant thermal inertia. The more scattered
    points at about 2.2 au are IRAS data, which did not have a strong
    weight on the fit given their significantly larger error bars.
    Section \ref{sec:juno} provides further discussion. }
  \label{fig:OMR_helioc_juno}
\end{figure}

Given the dependence of conductivity on temperature, thermal inertia obtained from IR data taken at an average heliocentric distance $r$ needs to be normalised to some reference heliocentric distance (i.e. temperature), usually 1 au, following
\begin{equation}
  \Gamma_{1au} = \Gamma(r) r^{\alpha}, \label{eq:gamma}
\end{equation}
where $\alpha=-0.75$ if we consider a radiative conduction term in the thermal conductivity $\kappa$. 
Marsset et al. (2017) analysed thermal IR data of asteroid (6) Hebe taken at a wide range of heliocentric distances and found indications that higher thermal inertias fitted low-$r$ data better. 
However, even though our sample includes some highly eccentric asteroids, our OMR versus heliocentric distance plots show no systematics or biases due to the assumption that $\Gamma=$ constant (see e.g. Fig.~\ref{fig:OMR_helioc_juno}), so it seems that the temperature range covered by most MBAs might be such that this effect is not critical.

\citet{Rozitis2018} studied three highly eccentric near-Earth objects observed by WISE at widely different $r$ and fitted the exponent $\alpha$ separately for each case. 
This work clearly showed that the physical interpretation of thermal inertia and the comparison between different objects and populations without good constraints for $\alpha$ is quite complicated. 
Here we propose a graphic approach to analyse a large catalogue of thermal inertias with the necessary assumption that all asteroids can be modelled using a single value of  $\alpha$, which is arguable.
The aim, nonetheless is to further explore the impact of the value of this parameter on our interpretation of the thermal inertias with a larger catalogue of values (see also \citealt{Szakats2020}). 

Instead of normalising the $\Gamma$-values to 1 au, we can plot  the thermal inertias compiled in \citet{Delbo2015} plus those in \citet{Hanus2015}, \citet{Hanus2018eos}, \citet{Hanus2018}, \citet{Marciniak2018}, and \citet{Marciniak2019} as a function of heliocentric distance and compare them with different curves with $\Gamma_{1au} =10,50,\ldots,2000$ and a given value of $\alpha$ using Eq. \ref{eq:gamma}. 
With such a plot, we can visually estimate $\Gamma_{1au}$ for each asteroid as the curve going through its corresponding point. 

\begin{figure*}
  \centering
  \includegraphics[width=0.80\linewidth]{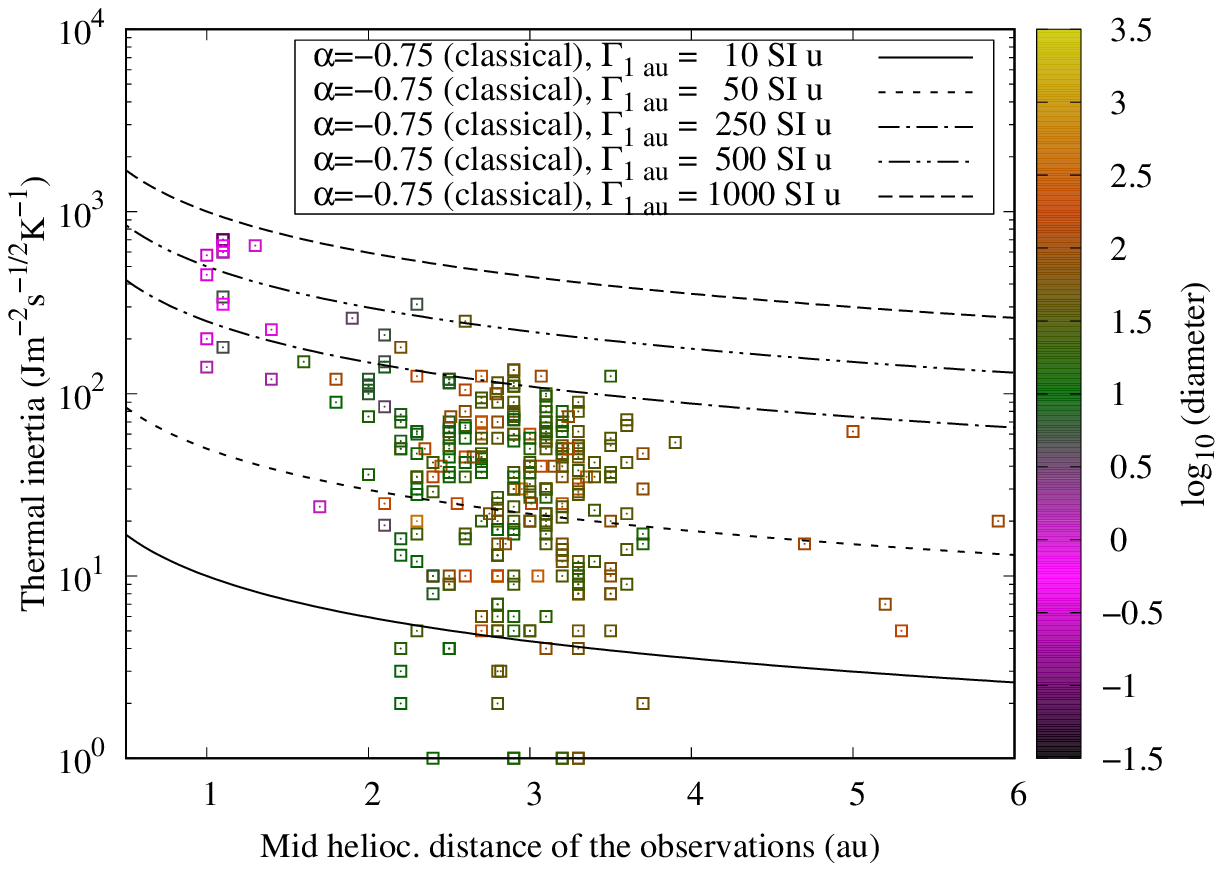}
  \caption{Thermal inertia vs. average heliocentric distance.
    We include values compiled from the literature (references given in the
    text).
    The lines correspond to Eq. \ref{eq:gamma} using different values of $\Gamma_{1\,au}$ and the classically assumed $\alpha=-0.75$.
  }\label{fig:gamma_helioc_alpha_class}
  
  \includegraphics[width=0.80\linewidth]{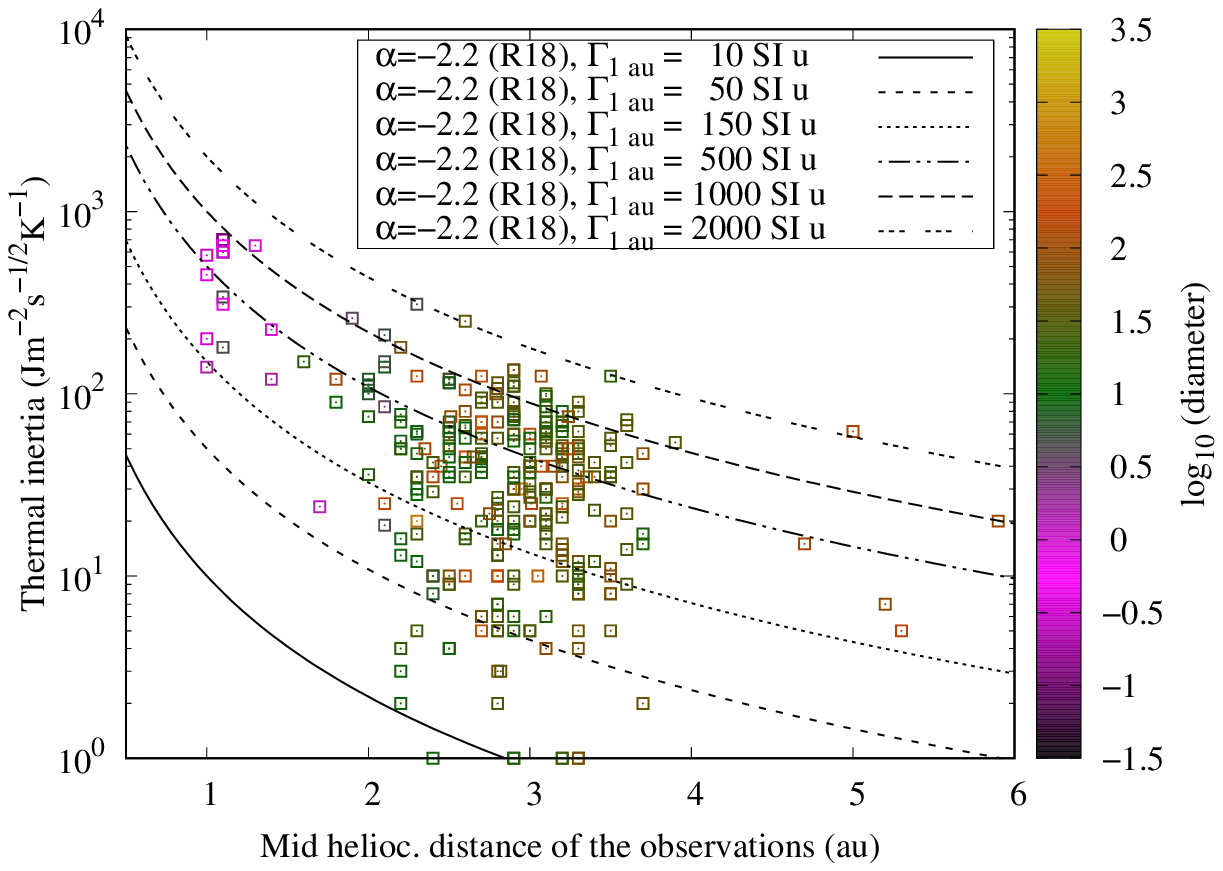}

  \caption{Same as Fig. \ref{fig:gamma_helioc_alpha_class} but $\alpha=-2.2$, which was found for near-Earth asteroid (1036) Ganymed by \citet{Rozitis2018}.
  }\label{fig:gamma_helioc_alpha_R18}
\end{figure*}

On the one hand, $\alpha=-0.75$ (Fig.~\ref{fig:gamma_helioc_alpha_class}) leads to the conclusion that most $D>10$-km asteroids (i.e. the green, brown and yellow symbols) have thermal inertias at 1 au between 10 and 250 \tiunit (dotted line), whereas those of most sub-km objects (pink symbols) lie between $\sim$200 and 1000 \tiunit.  
On the other hand, with $\alpha=-2.2$ (Fig.~\ref{fig:gamma_helioc_alpha_R18}), we would conclude that there is a much higher degree of thermal inertia variability amongst $D>10$ km asteroids, since $\Gamma_{1au}$ spans the range $<50$--2000 \tiunit. 

Finally, we also examined plots like Figs.~\ref{fig:gamma_helioc_alpha_class} and ~\ref{fig:gamma_helioc_alpha_R18} but using the rotational period instead of size in the colour axis: 
they do not show any appreciable excess of slow rotators in the regions of higher $\Gamma_{1au}$ or, conversely, fast rotators in the regions with lower $\Gamma_{1au}$, regardless of the value of $\alpha$. 
This is consistent with the conclusions of \citet{Marciniak2019}, whose sample shows that slow rotators do not always present higher thermal inertias (cf. \citealt{Marciniak2018}, who had a smaller sample, and \citealt{Harris2016} based on an empirical relation between thermal inertia and the infrared beaming parameter $\eta$ used in the near-Earth asteroid thermal model of \citealt{Harris1998}).

\section{Discussion}\label{sec:discussion}

One of the original motivations for this work within the framework of the SBNAF project was to test how the TPM could be used to evaluate the quality of shape models. While the comparison of the results with those obtained for a sphere is always a useful benchmark for discussion of bad or borderline acceptable fits, we found a relatively small systematic offset of the equivalent diameters obtained for the spherical models. However, after a thorough examination of each individual case, we failed to identify any cause that could explain why the spheres would occasionally provide good approximations of the thermal inertia, while being up to an order of magnitude higher in other cases. Neither the irregularity of the shape nor the particular orientation of the spin axis seem to produce worse results, at least within our sample. 

We fully neglect the uncertainties in the shape models even though they would certainly contribute to the final error bars of our TPM parameters \citep[see ][]{Hanus2015}. In this sense, our error bars are lower limits. One of the reasons for this limitation is that the methods to estimate errors in the shape are either labour intensive, computationally costly, or both, but there are already some works that explore this direction. 
  Hanus et al. (2015) proposed a way to assess the effects of shape uncertainty
  in TPM modelling by bootstrapping the visible data used to obtain a shape model from light-curve inversion and producing a set of shape models, each one of which is in turn modelled with a TPM to provide statistics for the inferred TPM parameters. More recently \citet{Bartczak2019} used an approach featuring millions of slightly perturbed shape clones to examine how the uncertainties and inaccuracies of the inversion models map over the whole surface. 
  As the number of shape models determined from adaptive optics observations and stellar occultations continues to rise, we could reach a point where we have sufficient ground-truth information about the shapes to have a more empirical estimate of the errors of thermal inertia and diameter.
 \citet{Podlewska-Gaca2020} provide a larger sample of SAGE models scaled with stellar occultation chords and, whenever possible, thermo-physical modelling.

 Finally, we note that our thermal inertias for Ceres and Pallas are higher than previous values (M\"{u}ller \& Lagerros 1998; see also Secs.\,\ref{sec:ceres} and \ref{sec:pallas}). 
 Thermo-physical modelling of Uranian satellites by Detre et al. (submitted to Astronomy \& Astrophysics) suggests this could be a small systematic trend: the radiometric diameters are 3-5\% larger than the ones derived from direct methods. 
We have not identified a clear unique cause for such an offset, but it could be related to the absolute flux calibration of the PACS data. On the other hand, we obtained only slightly higher values than \citet{ORourke2012} for Lutetia and \citet{Capria2014} for Vesta. Another source of error might be the assumption of a constant emissivity of 0.9, but the need for much lower thermal emissivities has so far been pointed out at significantly longer wavelengths (M\"{u}ller \& Lagerros 1998, \citealt{Muller2007a}, \citealt{Mueller2014}).

\section{Conclusions}\label{sec:conclusions}

In this work we derive thermo-physical properties of 18 large main-belt asteroids, 12 of which did not have any previous constraints on thermal inertia.
This was enabled by previously unpublished Herschel PACS data (Table~\ref{app:fluxes}) and recently available state-of-the art non-convex shape models (the term also includes the rotational properties). Most of our targets' shape models (see Table~\ref{tab:shapes}) were derived from a combination of adaptive optics, occultation and optical light-curve data using the ADAM algorithm \citep{Viikinkoski2015}; three of them, which we used as benchmarks, from direct images from spacecraft, and two of them from only optical light curves via SAGE \citep{Bartczak2018}. 

We find that the ADAM shapes can reproduce the thermal IR data in spite of the usual simplifying assumptions of the thermo-physical modelling -- constant properties over the surface, and thermal inertia also independent of depth and temperature\footnote{A discussion of the case of (6) Hebe  can be found in \citet{Marsset2017}}-- even in the case of (10) Hygiea, which has been recently shown to have a variable albedo over the surface \citep{Vernazza2019}.  
We optimised the scale of the shape models (i.e. the scale was a free parameter in our thermo-physical model, as usual) and found that most cases required an average re-scaling of 5\%, which serves as a cross-validation. Nonetheless, there were notable exceptions, especially (65) Cybele (see Table \ref{tab:tpm}), that require $\sim$10\% scaling, which means these targets are interesting for follow-up observations and modelling. 

From the example of (2) Pallas, a target extensively observed at a wide range of aspect angles due to the high obliquity of its rotation axis, we also examined how potential inaccuracies in the shape can bias the results of thermo-physical modelling when the scale of the shape model is not allowed to vary, so we suggest that results from both approaches (fixed and fitted scale) should always be compared and a careful examination of the modelling results should be performed on a case-by-case basis. 
Also, more work is needed to quantify and parameterise shape model errors \citep{Hanus2015,Bartczak2019} so that they can be propagated into the thermo-physical properties derived from them in a practical way.

In addition, we found relatively low thermal inertias in the same range as previous results \citep[][and references therein]{Delbo2015} for similarly-sized asteroids, which continues to support the notion that these bodies are covered by fine regolith. This, rather than composition, is the dominant effect governing the thermal emission of the large asteroids. 
Although the peak of the emission of main-belt asteroids is closer to the 10-micron region of the spectrum, high-quality data at longer wavelengths like PACS data are also extremely useful to determine good-quality sizes and thermal inertias; objects with large data sets especially benefit from the fact that surface roughness is not as dominant as thermal inertia at this wavelength range.

\begin{sidewaystable*}{\small
    \centering
    \caption{Summary of TPM results. $D_0$ is the sphere-equivalent diameter of the ADAM or in-situ shape models.
      Spherical model refers to a sphere ($\sim$3000 facets) with the same spin axis. 
      The symbols $D$, $\Gamma$ (SI units = \tiunit) and $\rho$ denote the best-fitting diameter, thermal inertia and surface roughness (rms) of the corresponding model. 
      To normalise $\Gamma$ at 1 au  ($\Gamma_{1au}$), we took the mid-point ($\bar{r}$) between the shortest and longest heliocentric distance at which the data were taken. 
      The surface roughness (rms) were not constrained at the 1$\sigma$ level unless otherwise stated. 
      For more information we refer to Appendix A and Table~\ref{tab:key}. 
    }
    \label{tab:tpm}    
    \begin{tabular}{|l l c | c  c  c  c|  c c | l|}
    \hline
    \hline
    Asteroid        & Model              & $D_0$      &  $D$           &   $\Gamma$         &  $\rho$ (rms)  & $\bar{\chi}^2_{m}$  & $\bar{r}$ (au) &  $\Gamma_{1au}$ &   Comments                   \\
                    &                    & (km)       &  (km)          &  (SI units)             &                &                    &                &  (SI units)        &                               \\
    \hline
    \hline
    (1) Ceres       & Dawn SPC            & 939.5      & $951^{+9}_{-7}$   & $25^{+15}_{-10}$     &  $\sim$1       &  0.2              & 2.80          &  43           & Modelled PACS data only                         \\
                    & Spherical model             &            & $949^{+20}_{-13}$  & $25^{+35}_{-10}$     &                &  0.5              &             &                & Similar conclusion but larger error bars                    \\
    \hline
    (2) Pallas      & ADAM               & 520        & $536^{+5}_{-5}$     & $30^{+15}_{-15}$   & $\sim$0.65     &  0.4              &  2.95       &  67          & AKARI data show a small slope in the OMR vs. $\lambda$ plot   \\
                    & Spherical model             & --         & $545^{+3}_{-4}$     & $50^{+15}_{-15}$   &                &  0.6              &             &              & $D\sim$5\% larger than ADAM $\rightarrow$ higher $\Gamma$    \\
    \hline
    (3) Juno        & ADAM               & 248        & $252^{+2}_{-3}$     & $60^{+25}_{-20}$   & $\sim$1        &  0.5              &  2.70       &  126          & Feature in the OMR-vs-aspect angle plot \\
                    & Spherical model             & --         & $255^{+3}_{-3}$     & $80^{+20}_{-30}$   &                &  1.0               &             &              & Formally acceptable fit, consistent thermal properties \\
                    & SAGE               & --         & $254^{+3}_{-4}$     & $70^{+30}_{-40}$   & $\sim$1        &  1.27              &         &  147          & Formally acceptable fit, consistent thermal properties \\
    \hline
    (4) Vesta       & Gaskell            &  522       & $520^{+12}_{-6}$   & $35^{+55}_{-23}$   & $\sim$0.9      &   0.8              & 2.36            & 66           &  Shallow $\chi^2$ minimum, slope in the OMR-vs-wavelength plot  \\
                    & Spherical model             &  --        & $520^{+21}_{-9}$   & $70^{+70}_{-45}$   &                &   1.1               &                &            &  Formally acceptable fit             \\
    \hline
    (8) Flora        & ADAM             & 143         & $142^{+2}_{-2}$   & $50^{+35}_{-30}$   & $\sim$0.4        &  0.4                  &  2.20        &  90          & IRAS data: slope in the OMR vs. $\lambda$ plot   \\
                     & Spherical model           & --          & $147^{+2}_{-1}$   & $120^{+40}_{-40}$  &                  &  0.5                  &              &            & Low $\chi^2$ too, but significantly higher $\Gamma$           \\
    \hline
    (10) Hygiea      & ADAM             & 433.6       & $441^{+7}_{-4}$   & $50^{+20}_{-25}$   & $\sim$0.9        &  0.6                  &  3.02        &  114         & IRAS data: slope in the OMR vs. $\lambda$ plot. IR insensitive to albedo variegation  \\
                     & Spherical model           & --          & $445^{+5}_{-6}$   & $55^{+25}_{-25}$   &                  &  0.65                  &              &            & Virtually the same results          \\
    \hline
    (18) Melpomene  & ADAM              & 146         & $135^{+4}_{-1}$   & $50^{+15}_{-44}$   &  $\sim$1         &  0.3                  & 2.35        &  95           &  Shallow and asymmetric $\chi^2_m$. Requires 8\% rescaling  \\
                    & Spherical model            & --          & $143^{+2}_{-1}$   & $80^{+30}_{-40}$   &                  &  0.3                  &             &             & Similar fit, but only small rescaling required (2\%) \\
    \hline
    (19) Fortuna    & ADAM              & 212         & $219^{+3}_{-2}$   & $40^{+30}_{-15}$   & $\sim$0.50       &  0.5                  &  2.45        &  78          & rms$>$0.2 at 3$\sigma$ level. Few data but low pole obliquity \\
                    & Spherical model            & --          & $219^{+2}_{-9}$   & $20^{+45}_{-5}$    &                  &  1.1                    &             &            & Lower $\Gamma$, formally acceptable fit                     \\
    \hline
    (20) Massalia   & SAGE1            & --         & $147^{+2}_{-2}$     & $35^{+25}_{-10}$   & $\sim$0.2        &  0.45                  &  2.40       &   67         & No mirror solution can be rejected. Very low roughness favoured \\
                    & Spherical model           & --         & $146$             & $35$            &                  &  1.6                  &             &            & $\chi^2_m$ too high, although in agreement with the SAGE results \\
    \hline
    (21) Lutetia    & Jorda            & 98.15      & $98^{+1}_{-1}$      & $10^{+10}_{-2}$   & $\sim$0.6         & 0.7                   &  2.40       &   20         & $0.4<$rms$<0.9$. Possible rotational phase shift (see Sect.\ref{sec:lutetia}) \\
                    & Spherical model           &            & 104               & 60              &                  & $>3$                  &             &            & Spherical approximation fails             \\
    \hline
    (29) Amphitrite & ADAM             & 205.5       & $202^{+3}_{-2}$   & $25^{+10}_{-13}$  & $\sim$0.4          &  0.4                     &  2.55       &  50          &  IRAS data: strong slope in the OMR vs. $\lambda$ plot   \\
                    & Spherical model           &             & $208^{+3}_{-3}$   & $100^{+40}_{-50}$ &                    &  0.9                     &             &           & Formally acceptable fit but significantly higher $\Gamma$           \\
    \hline
    (52) Europa     & ADAM             & 313.7       & $317^{+4}_{-3}$   & $10^{+25}_{-10}$ & $\sim$0.5           &  0.5                     &  3.05     &  23          & Very shallow $\chi^2_m$, despite large dataset. $\Gamma\propto T$ effect?    \\
                    & Spherical model           & --          & 342             &  200          &                     &  $>2$                    &           &              & Sphere greatly overestimates diameter and $\Gamma$                           \\
    \hline
    (54) Alexandra   & ADAM            & 143.        & $153^{+2}_{-2}$   & $10^{+22}_{-10}$  & $\sim$0.3          &  0.2                     &  2.60       &  20          & Few data, Southern hemisphere not well sampled in the IR data        \\
                     & Spherical model          & --          & $161^{+4}_{-3}$   & $75^{+30}_{-45}$  &                    &  0.8                     &             &            & Formally acceptable fit, but inconsistent $\Gamma$  \\
    \hline
    (65) Cybele   & ADAM               & 313.3      & $277^{+4}_{-2}$    & $30^{+10}_{-15}$   & $\sim$0.45        &  0.7                    & 3.30         &  73          & Required 12\% rescaling. Southern hemisphere not well sampled in the IR            \\
                  & Spherical model             &            & $292^{+5}_{-6}$    & $50^{+20}_{-15}$   &                   &  1.1                    &              &             & Formally acceptable fit, size in better agreement with ADAM               \\
    \hline
    (88) Thisbe   & ADAM               & 220        & $221^{+2}_{-2}$    & $60^{+15}_{-25}$  & $\sim$0.9          &  0.3                    &  3.00        &  137          &  IRAS data: slight slope in the OMR vs. $\lambda$ plot                          \\
                  & Spherical model             &            & $219^{+3}_{-2}$    & $60^{+15}_{-35}$  &                    &  0.5                    &              &               &  Very low $\chi^2_m$ and similar $D$-$\Gamma$ despite irregular shape                  \\
    \hline
    (93) Minerva   & ADAM              & 160        & $167^{+3}_{-3}$    & $25^{+30}_{-10}$  & $\sim$0.2          &  0.5                    &  3.01        &   57         &  Removed MSX and IRAS 12-$\mu$m data from analysis                              \\
                   & Spherical model            &            & $162^{+2}_{-2}$    & $100^{+10}_{-40}$ &                    &  0.2                    &              &              &  Sphere gets lower $\chi^2_m$, but too high $\Gamma$                             \\
    \hline
    (423) Diotima   & ADAM             & 209        & $200^{+3}_{-4}$    & $40^{+30}_{-20}$  & $\sim$0.45         &  0.6                    &  3.07        &   93         &  IRAS data: slope in the OMR vs. $\lambda$ plot. Southern hemisph. not sampled   \\
                    & Spherical model           &            & 205              &  150           &                    &  $>2$                   &              &               &  Bad fit                              \\
    \hline
    (511) Davida   & ADAM              & 313        & $307^{+7}_{-4}$    & $35^{+15}_{-17}$  & $\sim$0.5          &  0.4                   &  3.35         &  87           &  IRAS data: slight slope in the OMR vs. $\lambda$ plot. North. Hemisph. not sampled   \\
                   & Spherical model            &            & $328^{+10}_{-7}$   & $120^{+80}_{-50}$ &                    &  0.6                    &              &               &  Low $\chi^2_m$ but unrealistically high $\Gamma$, perhaps because shape is elongated   \\
    \hline
    \hline
  \end{tabular}
  }
\end{sidewaystable*}

\begin{acknowledgements}
  The research leading to these results has received funding from the European Union's Horizon 2020 Research and Innovation Programme, under Grant Agreement number 687378 (SBNAF). 
C.K., R.S., A.F-T., and G.M. have been supported by the K-125015 and 
GINOP-2.3.2-15-2016-0000 grants of the National Research, Development and Innovation Office (NKFIH), Hungary.

  P. Santos-Sanz acknowledges financial support by the Spanish grant AYA-RTI2018-098657-J-I00 (MCIU/AEI/FEDER, UE). R. Duffard and P. Santos-Sanz acknowledge financial support from the State Agency for Research of the Spanish MCIU through the ``Center of Excellence Severo Ochoa'' award for the Instituto de Astrof{\'i}ica de Andaluc{\'i}a (SEV-2017-0709); they also acknowledge financial support by the Spanish grant AYA-2017-84637-R and the Proyecto de Excelencia de la Junta de Andaluc{\'i}a J.A. 2012-FQM1776. 
\end{acknowledgements}

\bibliographystyle{aa}
\bibliography{/Users/valilagoa/Dropbox/AsteroidsGeneral}

\begin{appendix}

\section{Thermo-physical model analysis}\label{app:tpm}

In this section we provide all observation-to-model ratio (OMR) plots to help further examine the fits. Table~\ref{tab:key} links each target to its corresponding plots. Some targets warrant a short subsection with relevant comments in addition to those of Table\,\ref{tab:tpm}. 
\begin{table}[!h]
  \centering
  \caption{List of targets and references to the corresponding figures.}
  \label{tab:key}
  \begin{tabular}{l l }
    \hline
    \hline
    Target  & OMR plots \\
    \hline
    \hline
    &  \\
    (1) Ceres           & Fig.~\ref{fig:00001_OMR} \\
    (2) Pallas          & Fig.~\ref{fig:00002_OMR} \\
    (3) Juno            & Figs.~\ref{fig:00003_OMR} and \ref{fig:00003_OMRSAGE} \\
    (4) Vesta           & Fig.~\ref{fig:00004_OMR} \\
    (8) Flora           & Fig.~\ref{fig:00008_OMR} \\
    (10) Hygiea         & Fig.~\ref{fig:00010_OMR} \\
    (18) Melpomene      & Fig.~\ref{fig:00018_OMR} \\
    (19) Fortuna        & Fig.~\ref{fig:00019_OMR} \\
    (20) Massalia       & Fig.~\ref{fig:00020_OMR} \\
    (21) Lutetia        & Fig.~\ref{fig:00021_OMR} \\
    (29) Amphitrite     & Fig.~\ref{fig:00029_OMR} \\
    (52) Europa         & Fig.~\ref{fig:00052_OMR} \\
    (54) Alexandra      & Fig.~\ref{fig:00054_OMR} \\
    (65) Cybele         & Fig.~\ref{fig:00065_OMR} \\    
    (88) Thisbe         & Fig.~\ref{fig:00088_OMR} \\
    (93) Minerva        & Fig.~\ref{fig:00093_OMR} \\
    (423) Diotima       & Fig.~\ref{fig:00423_OMR} \\
    (511) Davida        & Fig.~\ref{fig:00511_OMR} \\
    &  \\
    \hline
    \hline
  \end{tabular}
\end{table}
  
\subsection{(1) Ceres}\label{sec:ceres}

The 180 PACS data points are fitted with a very low \cho of 0.2 with $\Gamma = 25$\,\tiunit, an extremely high roughness of rms$\sim$1, and a rescaling of about +2\% of the original ADAM shape. 
This thermal inertia is higher than but compatible within the error bars with classical TPM results ($10 \pm 10$\,\tiunit; \citealt{Mueller1998}) and Dawn-based analyses ($<20$\,\tiunit; \citealt{Rognini2018}). 
If we fit the data while keeping the Dawn stereo-photoclinometry (SPC) shape fixed, we obtain a lower thermal inertia of 6$^{+9}_{-6}$\,\tiunit and roughness values lower than rms$\sim$0.35 are rejected at the 3$\sigma$ level. 

We find systematically higher thermal inertias when we model only PACS data and optimise the diameter, although the results are not statistically significantly different (1-$\sigma$ regions in the $\chi^2$ vs. $\Gamma$ plots overlap). 
Actually, for targets with such a rich PACS data set as Ceres, we can fit data from the three PACS filters separately. We found $\Gamma=$\,35, 40, and 50 \tiunit with the 70, 100, and 160 $\mu$m data. Although the trend is not statistically significant -- our $\chi^2$ minima still overlap within the ranges of formally acceptable fits -- it could be caused by the fact that the longer wavelengths probe deeper and perhaps more compacted layers of the subsurface. However, it could also be an artefact related to our assumption of a constant emissivity of 0.9 if the (spectral) emissivity at the PACS wavelengths was lower, but previous work by \citet{Mueller2002} suggests that the effect should only be significant at wavelengths much longer than that of PACS. Indeed, \citet{Mueller2014} were able to reproduce even Herschel Spectral and Photometric Imaging Receiver (SPIRE) data at 250, 350 and 500 $\mu$m with the same modelling approach we use. Nonetheless, we leave it to future work to revisit the data with additional observations and better constraints on spectral emissivity.

\begin{figure}
  \centering
  \includegraphics[width=0.7\linewidth]{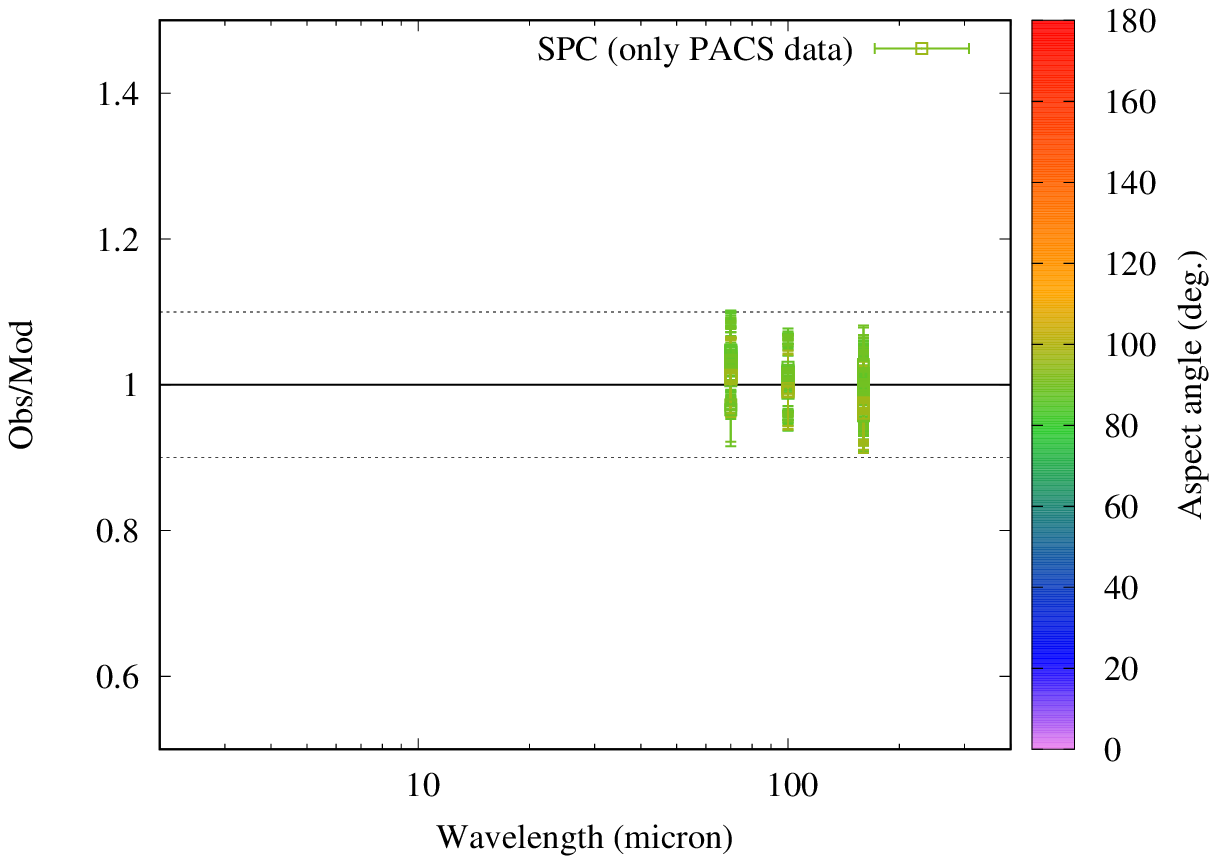}

  \includegraphics[width=0.7\linewidth]{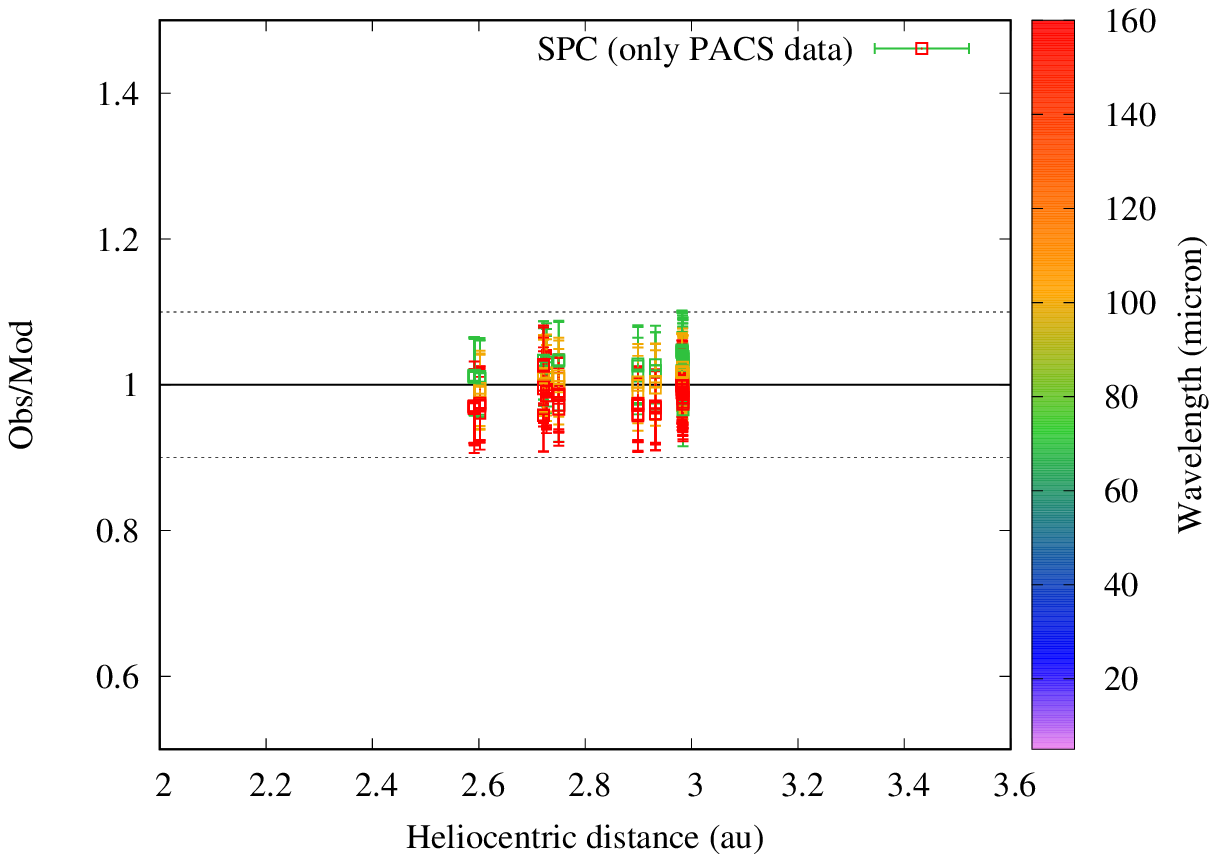}

  \includegraphics[width=0.7\linewidth]{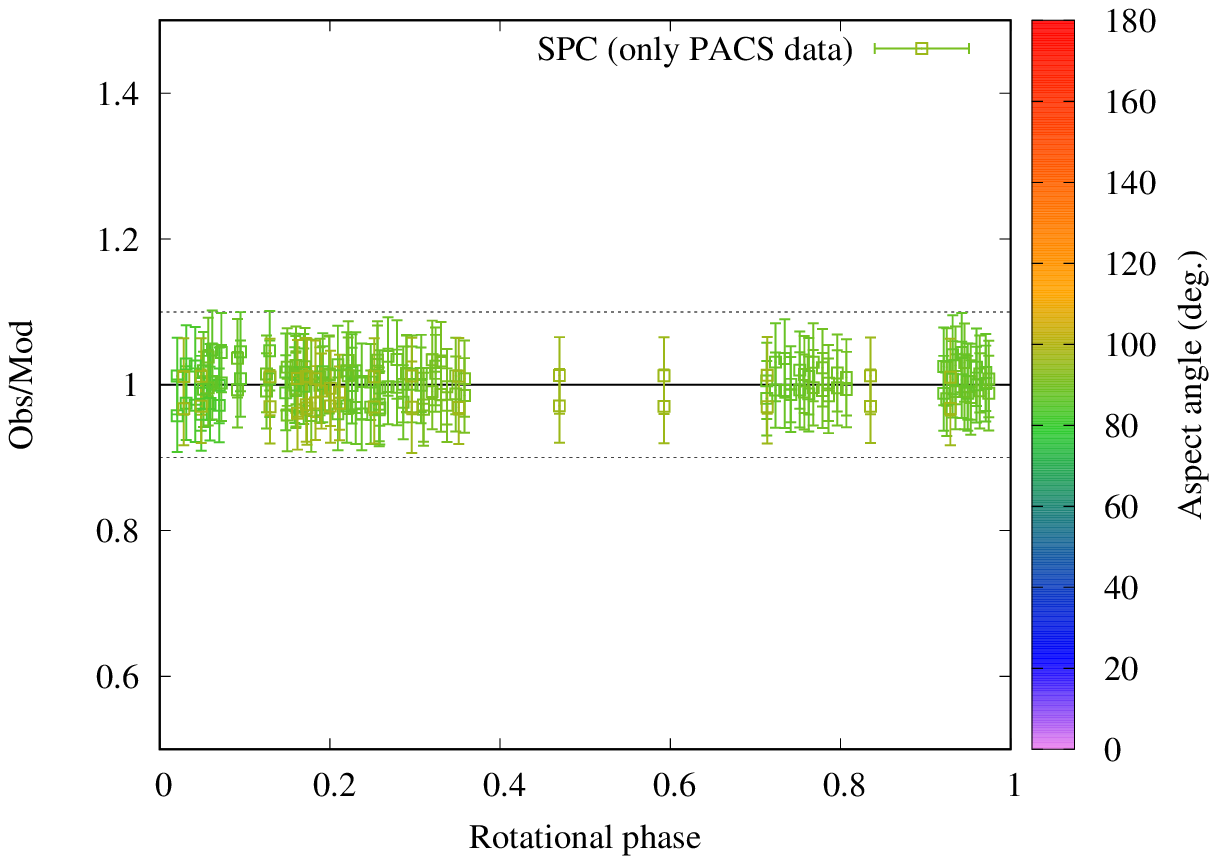}

  \includegraphics[width=0.7\linewidth]{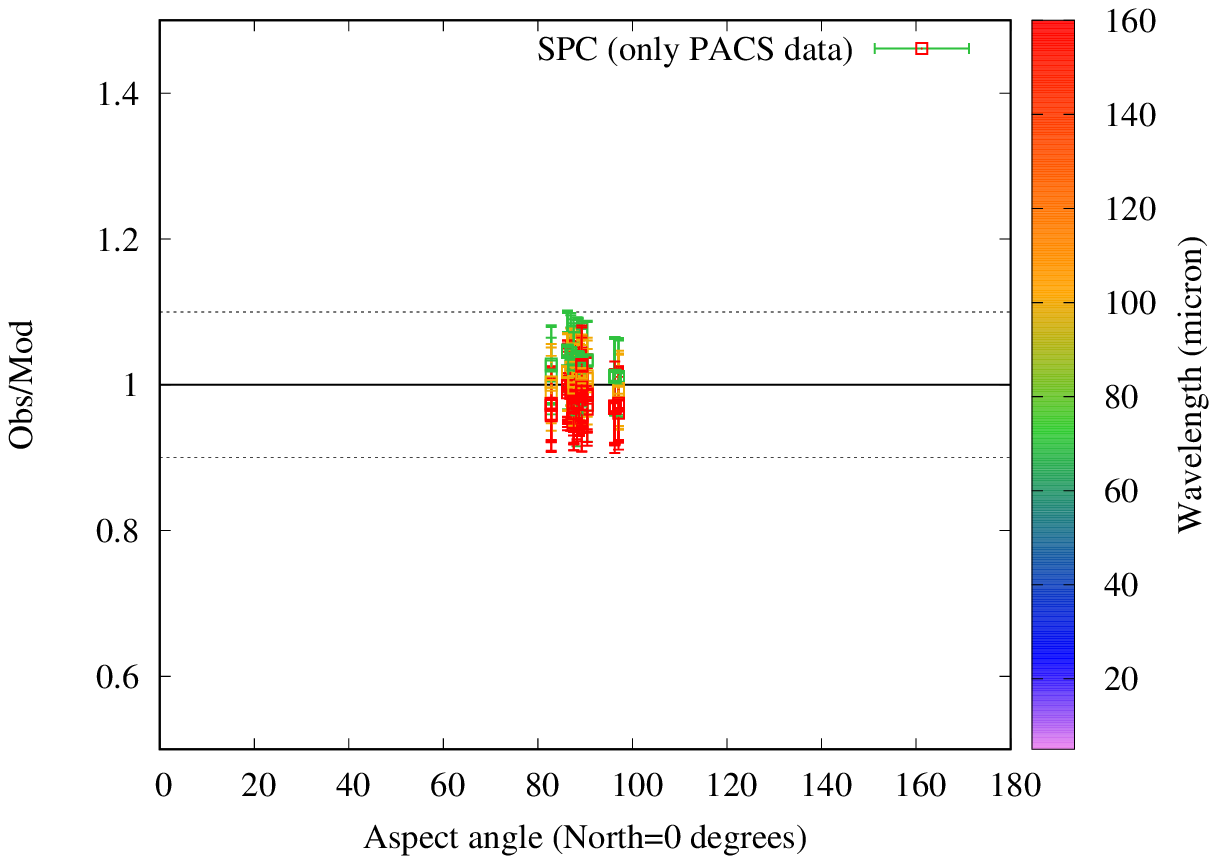}

  \includegraphics[width=0.7\linewidth]{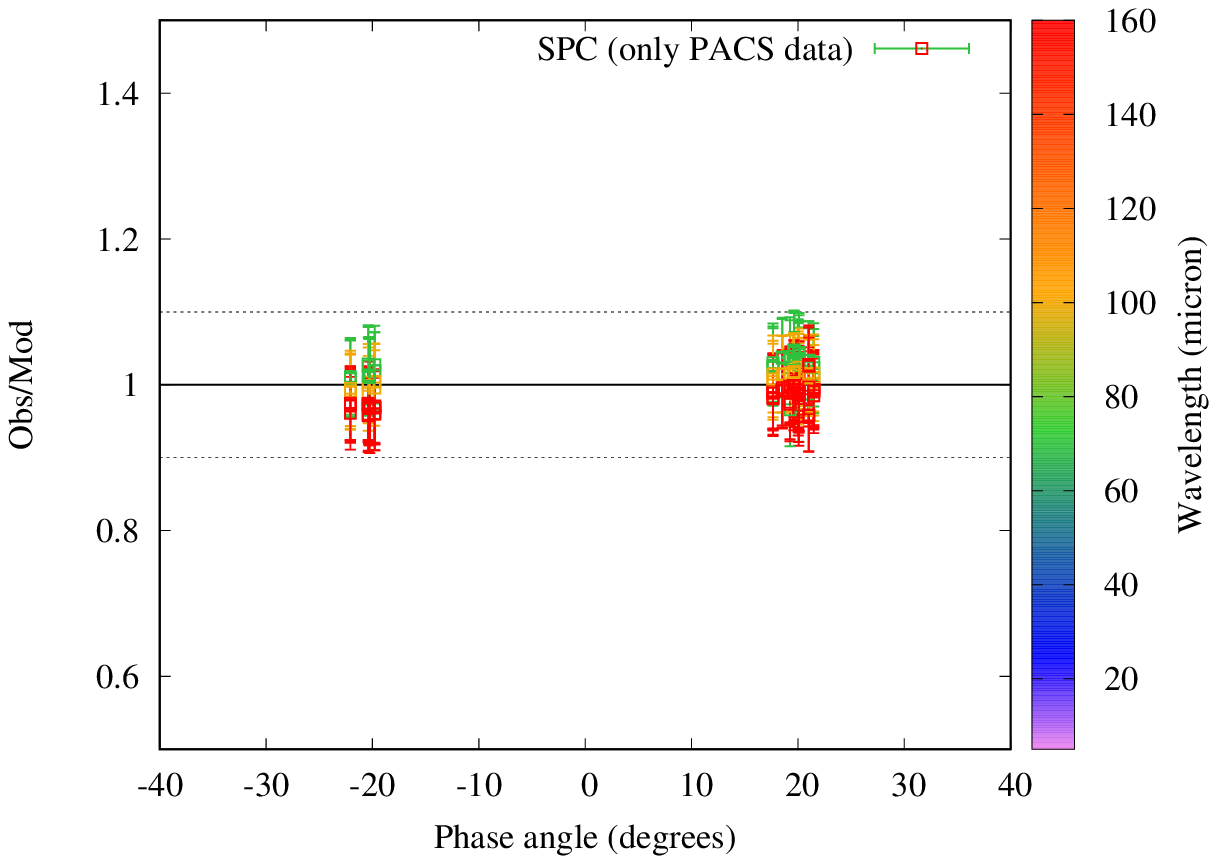}

  \caption{(1) Ceres: From top to bottom,
    observation-to-model ratios vs. wavelength,
    heliocentric distance, rotational phase, and phase angle. 
    The colour bar corresponds either to the aspect angle or to the
    wavelength at which each observation was taken.
  }\label{fig:00001_OMR}
\end{figure}

\subsection{(2) Pallas}\label{sec:pallas}

The ADAM shape fits the 80 PACS data with a very low \cho (lower than 0.1) with extremely high roughness and a thermal inertia of 30 \tiunit. 
Nevertheless, we find indications of small shape model inaccuracies given it cannot reproduce all data equally well when we fix the scale (see the main text). 

\begin{figure}
  \centering
  \includegraphics[width=0.7\linewidth]{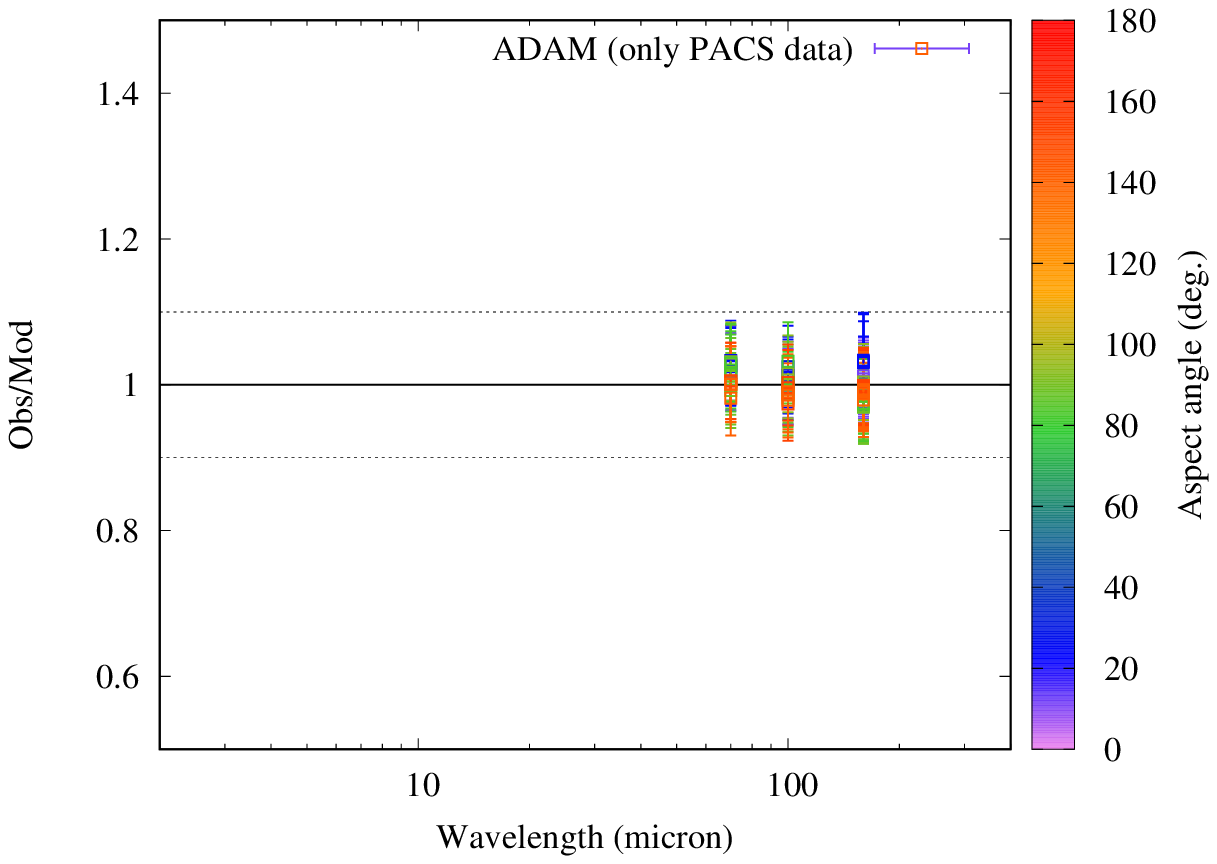}

  \includegraphics[width=0.7\linewidth]{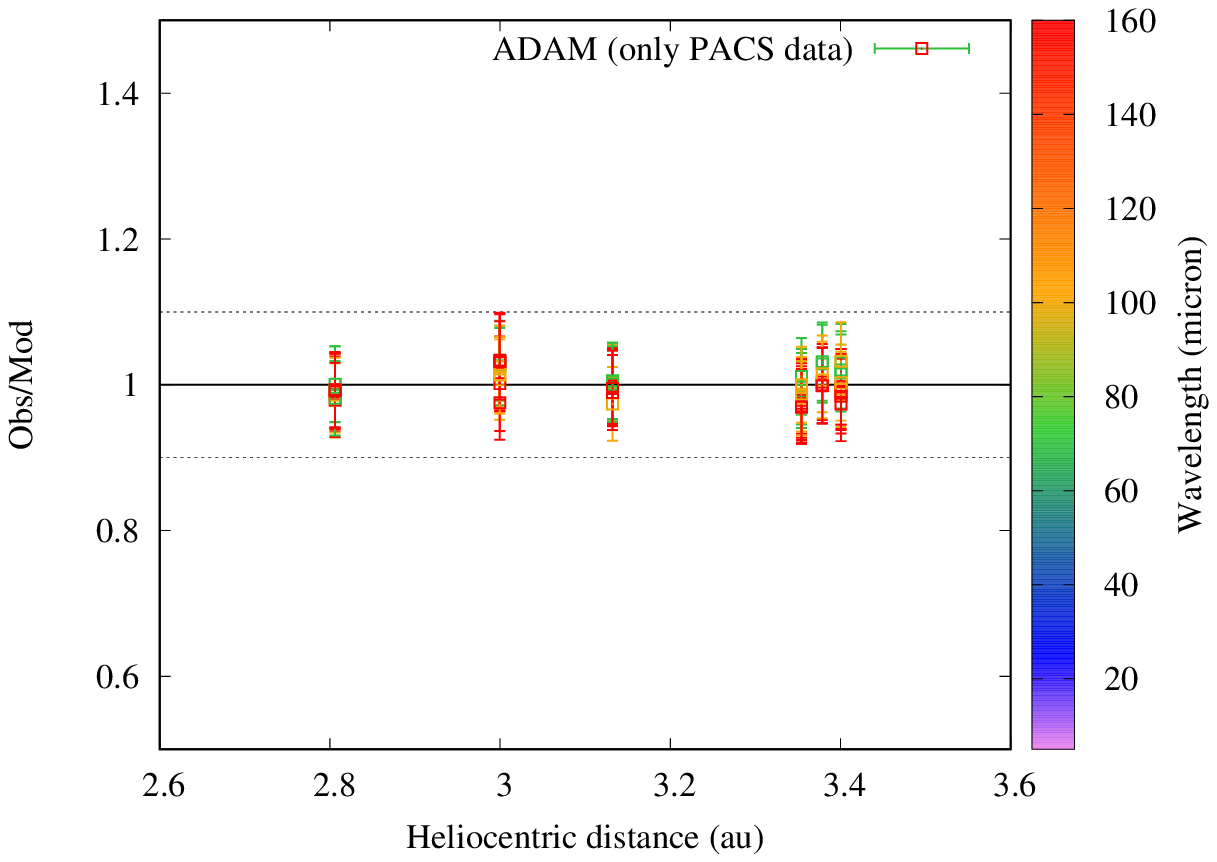}

  \includegraphics[width=0.7\linewidth]{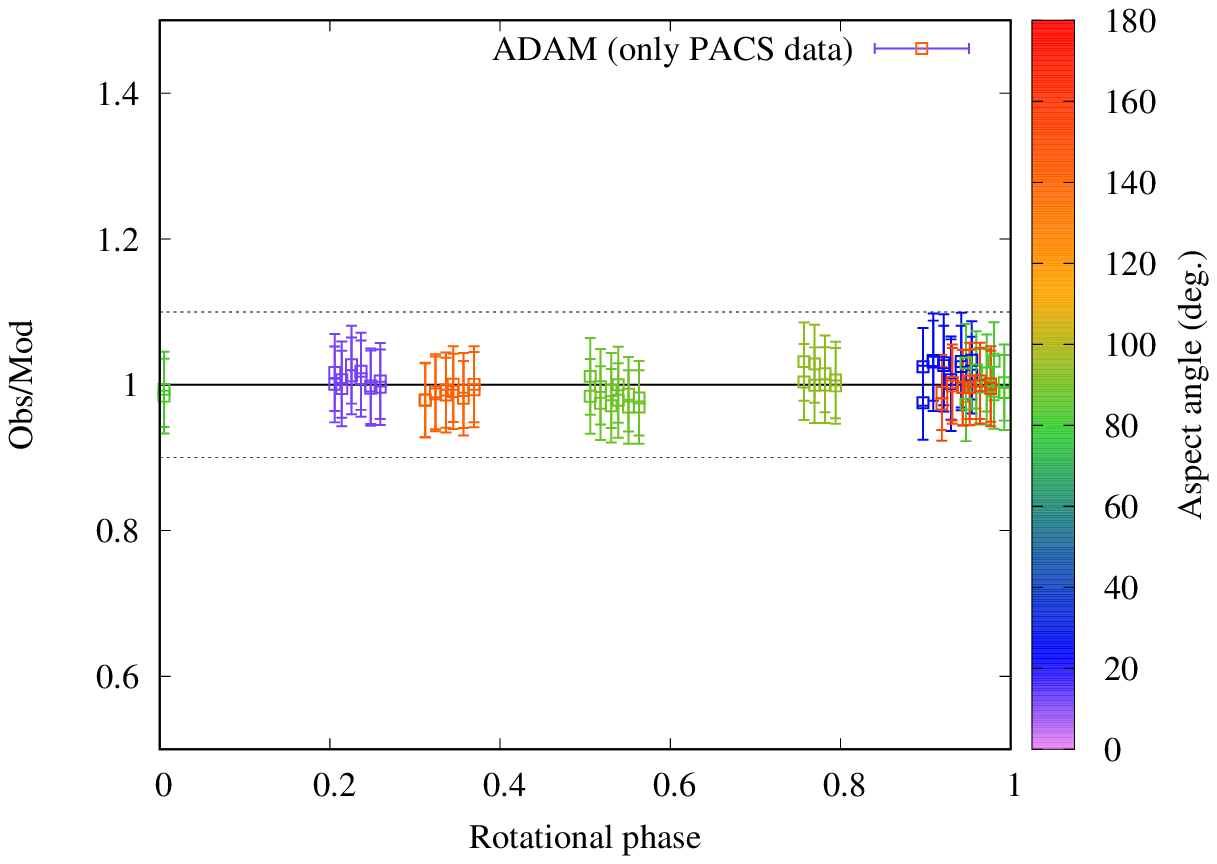}

  \includegraphics[width=0.7\linewidth]{OMR_aspect_00002_ADAM_O02.eps}

  \includegraphics[width=0.7\linewidth]{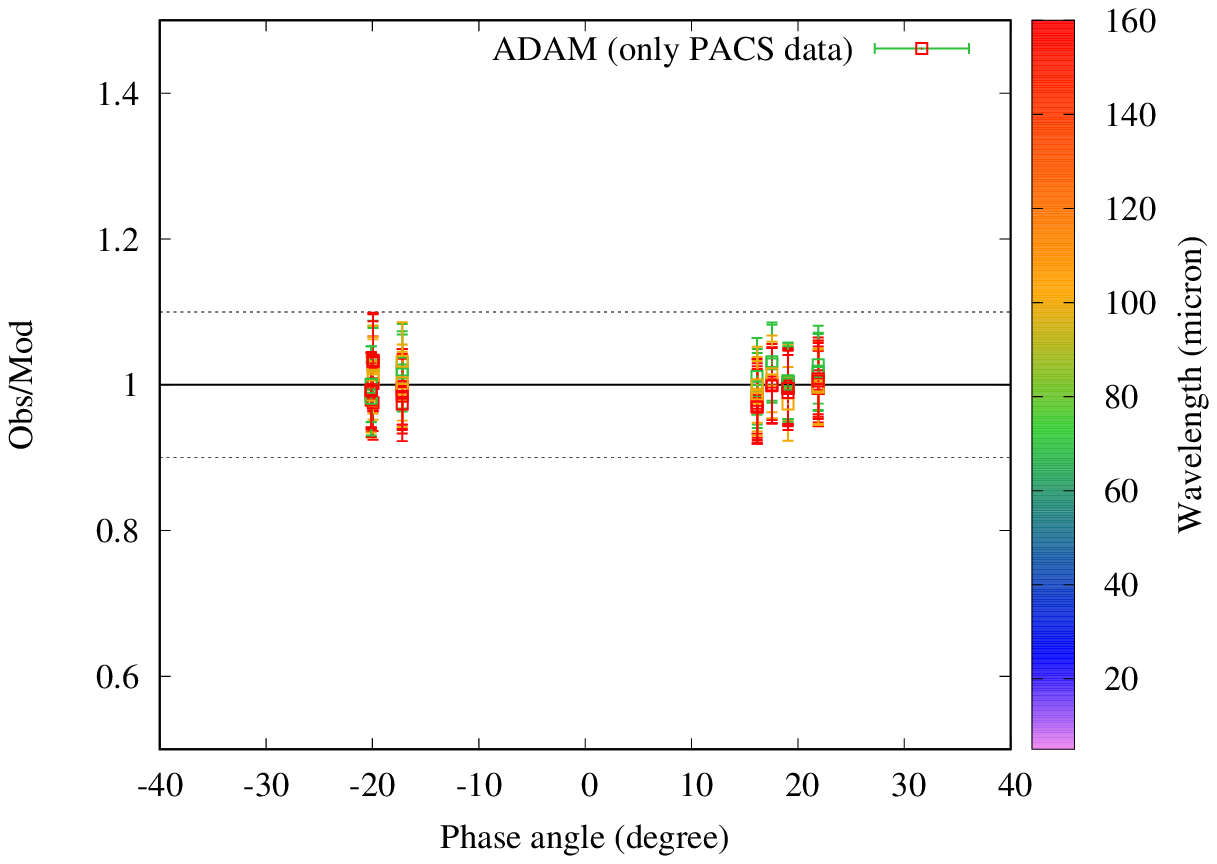}

  \caption{(2) Pallas. See the caption in Fig.~\ref{fig:00001_OMR}. 
  }\label{fig:00002_OMR}
\end{figure}

\subsection{(3) Juno}\label{sec:juno}
The OMR plots present systematics within the 10\% error margins. 
From the aspect angle plot, we infer that the northern hemisphere
part of the shape model could have an artefact, and the wavelength plot shows a
slight "convex up" curvature. 
The SAGE model provides a borderline formally acceptable fit and shows more
scatter in the OMR plots and larger parameter error bars, which means the shape
is not optimal (the rotational parameters are virtually the same). The trends
in the OMR plots are slightly blurred to the eye due to the higher scatter.
Nonetheless, the best-fitting values of size and gamma are fully compatible
within the error bars, but the ADAM ones are still better constrained. 
The same can be said about the scatter in the plots corresponding to the sphere
(which does slightly better than SAGE in terms of $\chi^2$, but not
statistically significantly). 

\begin{figure}
  \centering
  \includegraphics[width=0.7\linewidth]{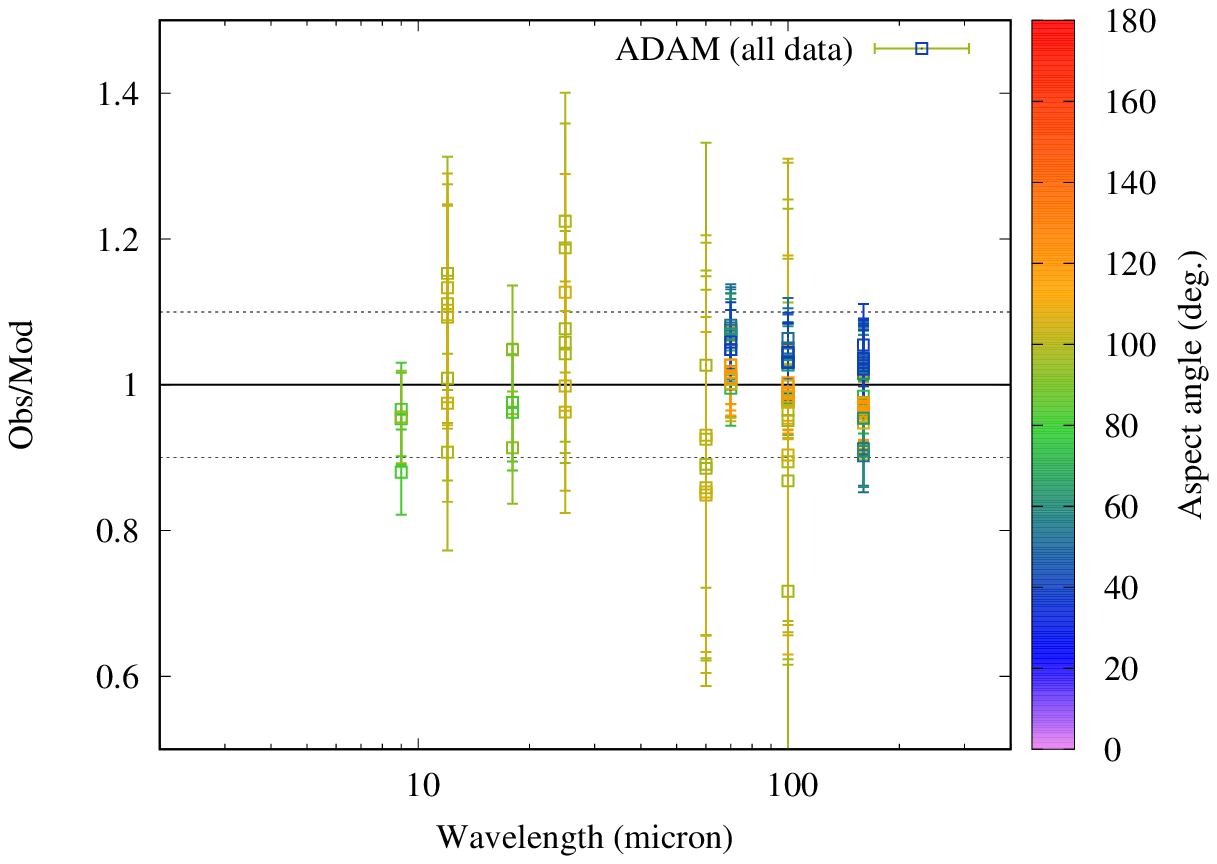}

  \includegraphics[width=0.7\linewidth]{OMR_helioc_00003_ADAM_O00.eps}

  \includegraphics[width=0.7\linewidth]{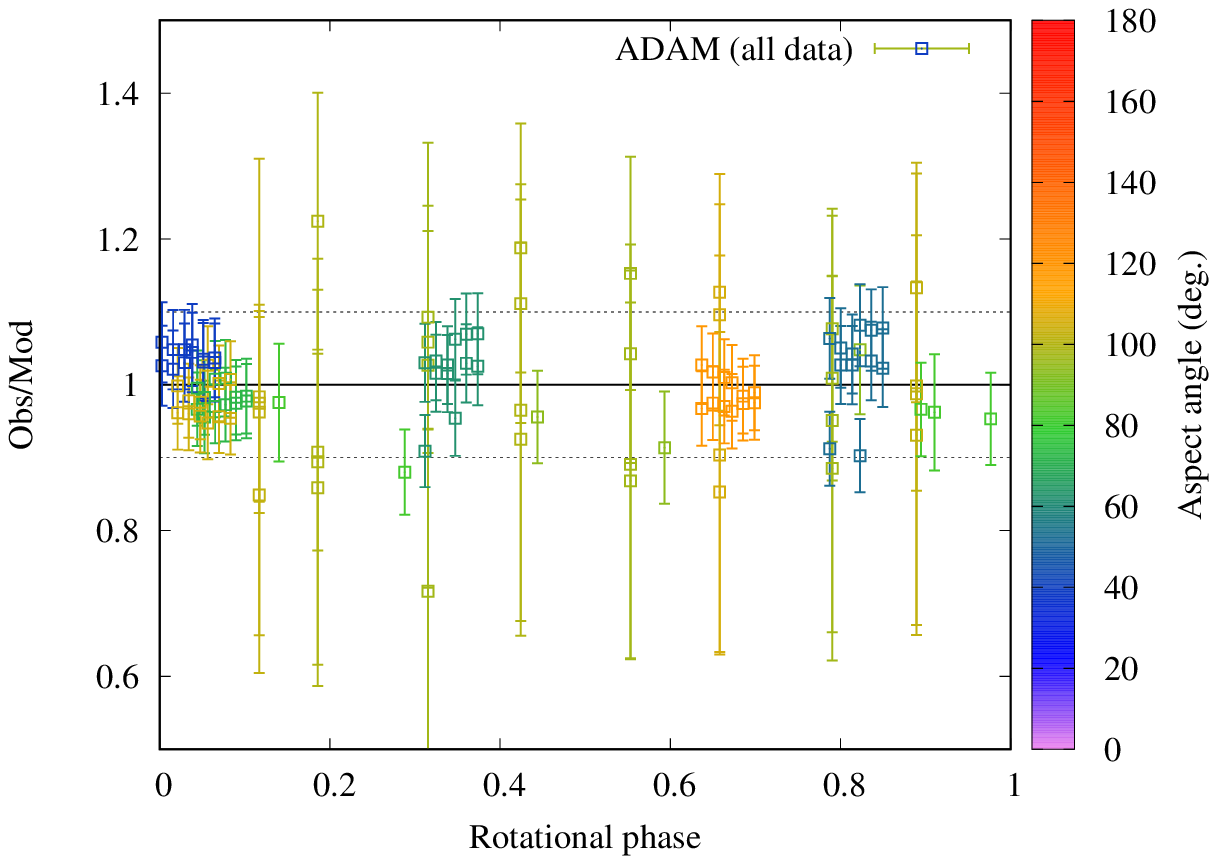}

  \includegraphics[width=0.7\linewidth]{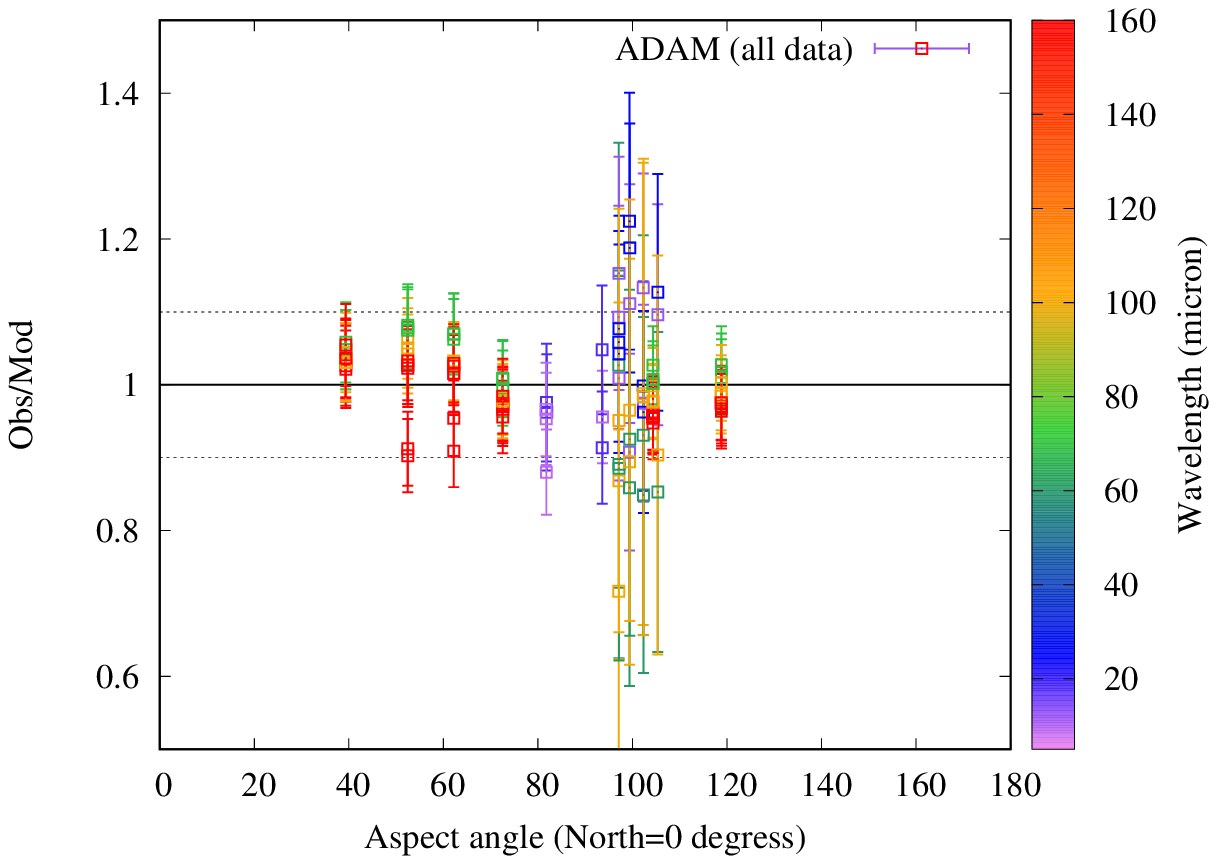}

  \includegraphics[width=0.7\linewidth]{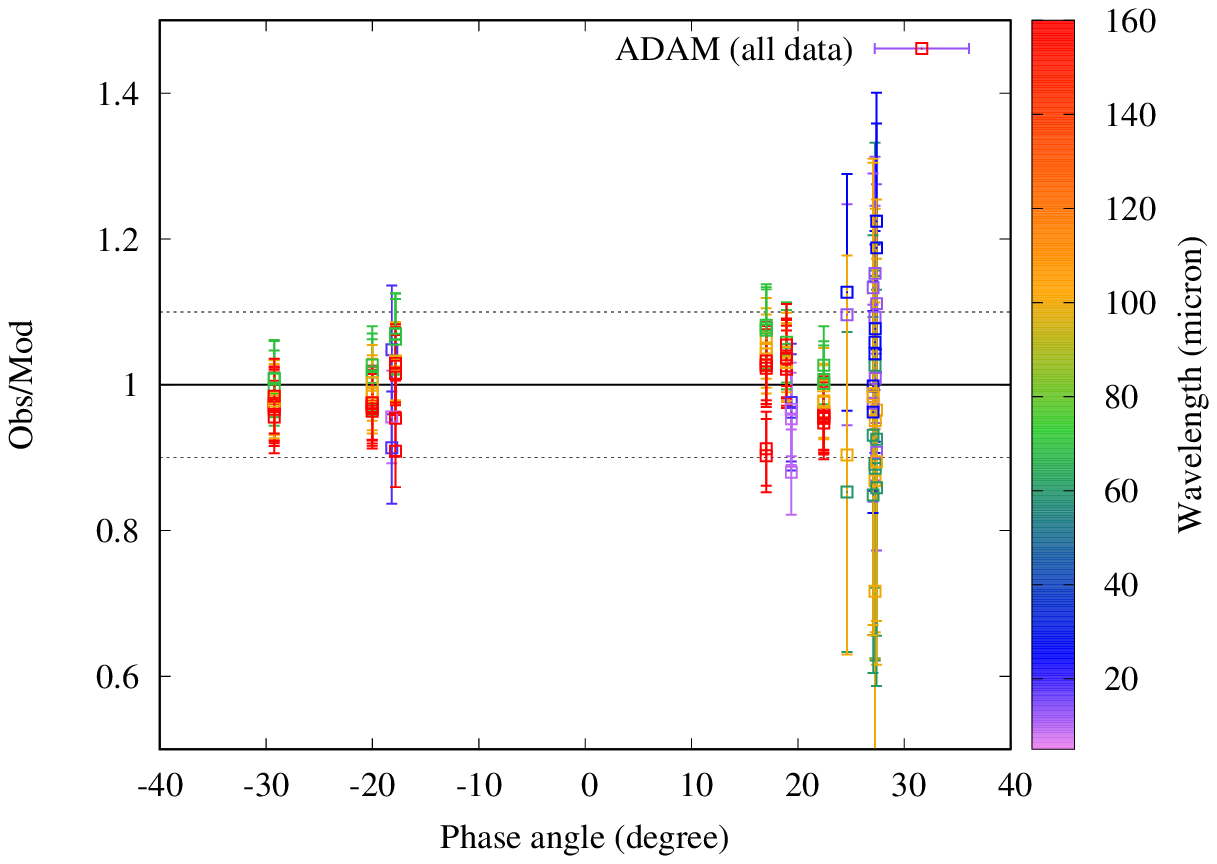}

  \caption{(3) Juno (ADAM model). See the caption in Fig.~\ref{fig:00001_OMR}. 
  }\label{fig:00003_OMR}
\end{figure}

\begin{figure}
  \centering
  \includegraphics[width=0.7\linewidth]{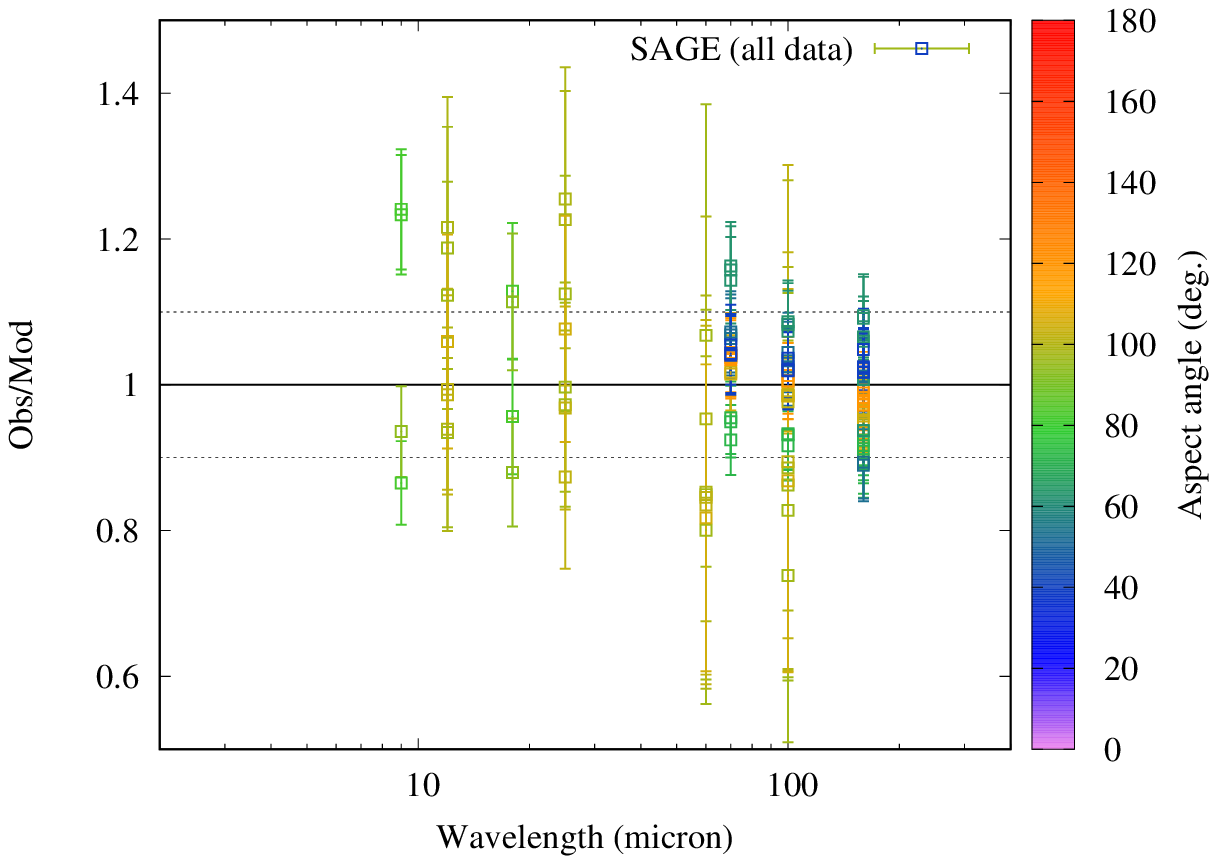}

  \includegraphics[width=0.7\linewidth]{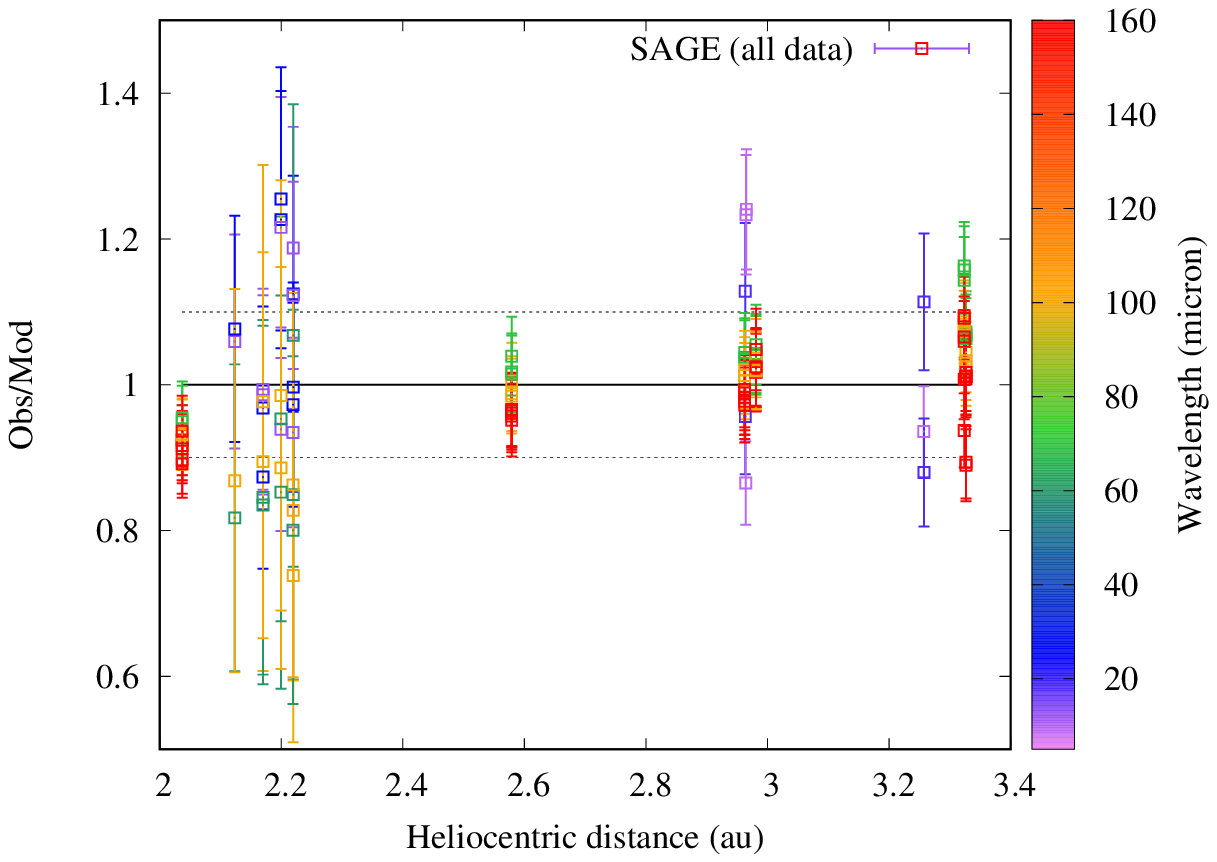}

  \includegraphics[width=0.7\linewidth]{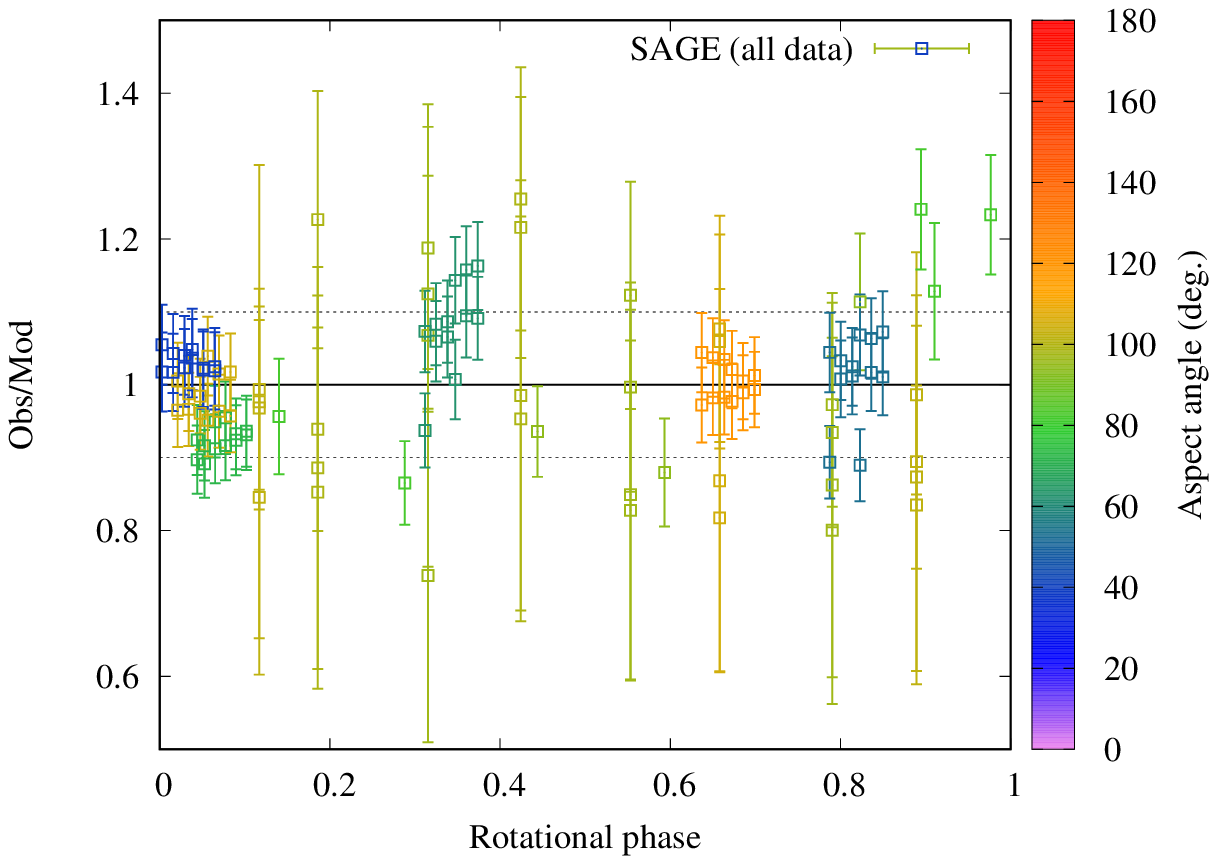}

  \includegraphics[width=0.7\linewidth]{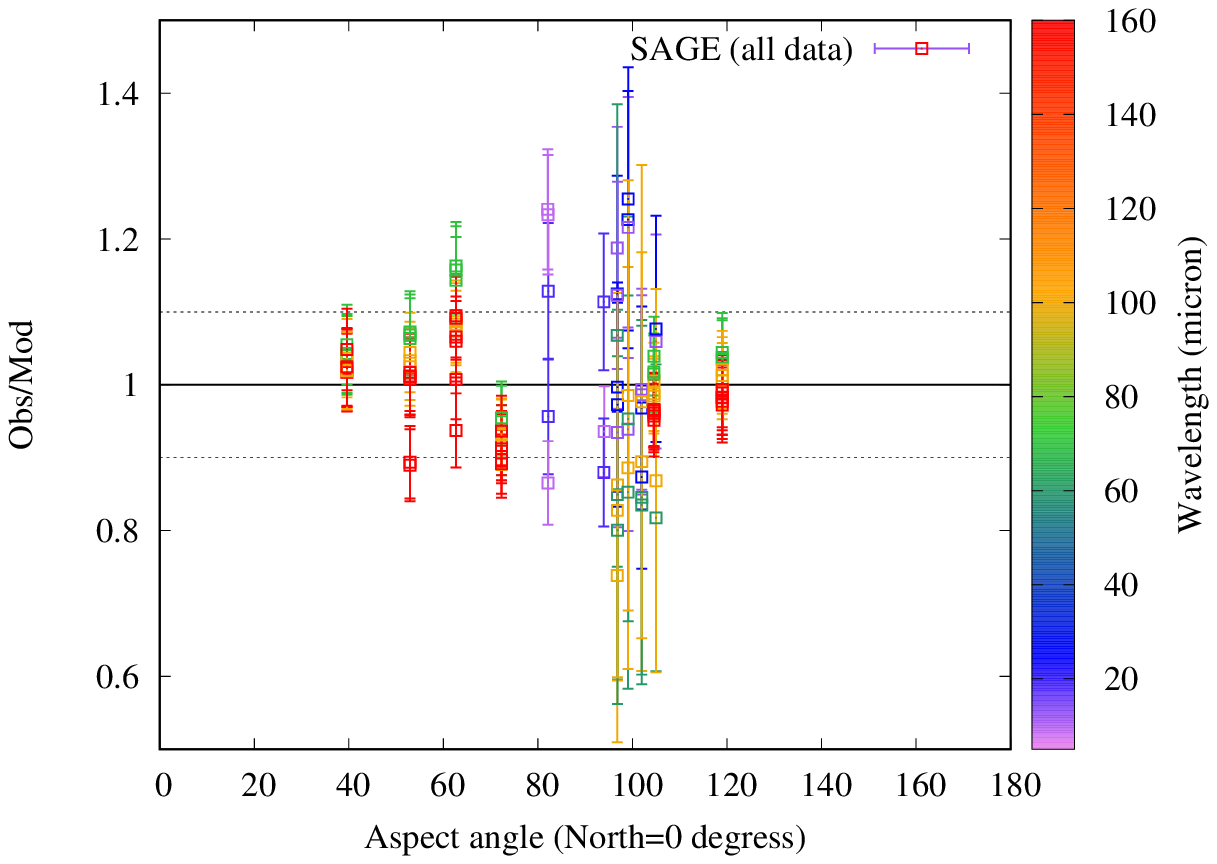}

  \includegraphics[width=0.7\linewidth]{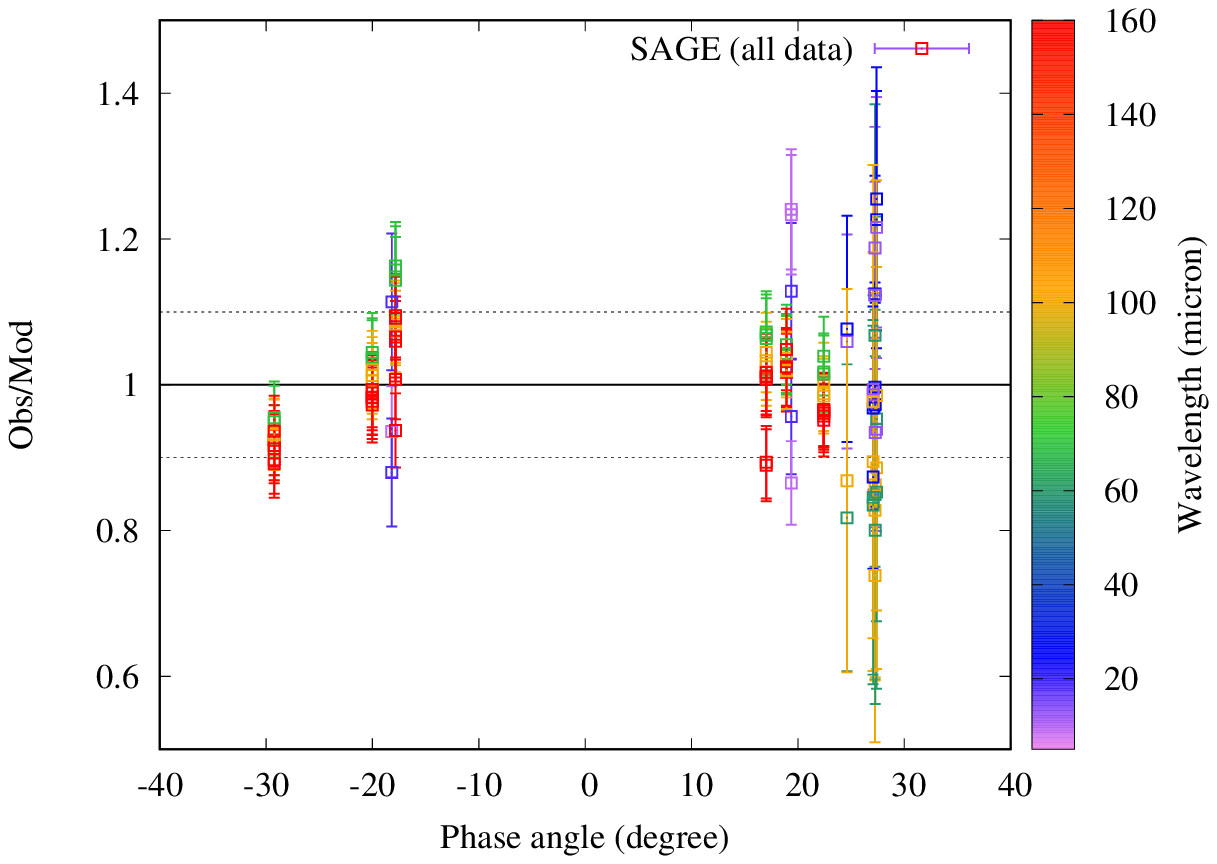}

  \caption{(3) Juno (SAGE model). See the caption in Fig.~\ref{fig:00001_OMR}. 
  }\label{fig:00003_OMRSAGE} 
\end{figure}


\begin{figure}
  \centering
  \includegraphics[width=0.7\linewidth]{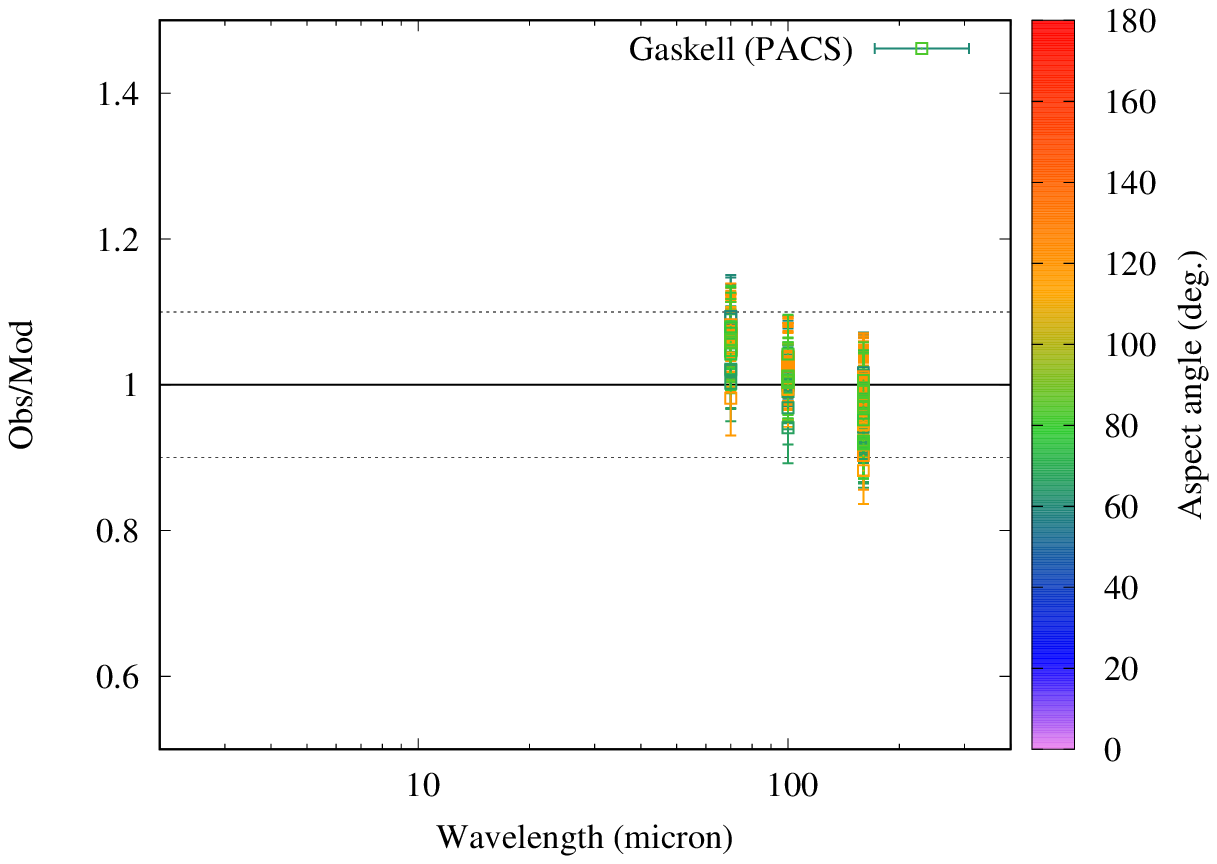}

  \includegraphics[width=0.7\linewidth]{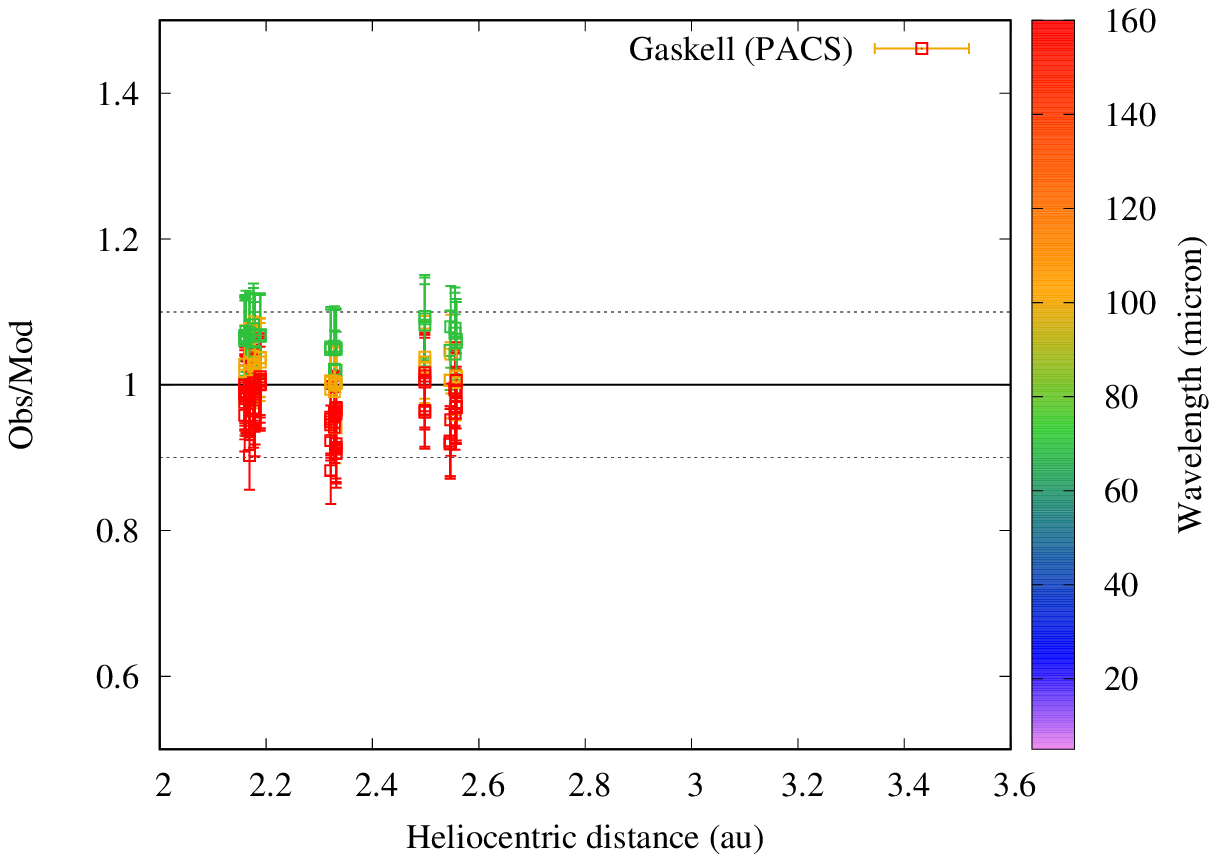}

  \includegraphics[width=0.7\linewidth]{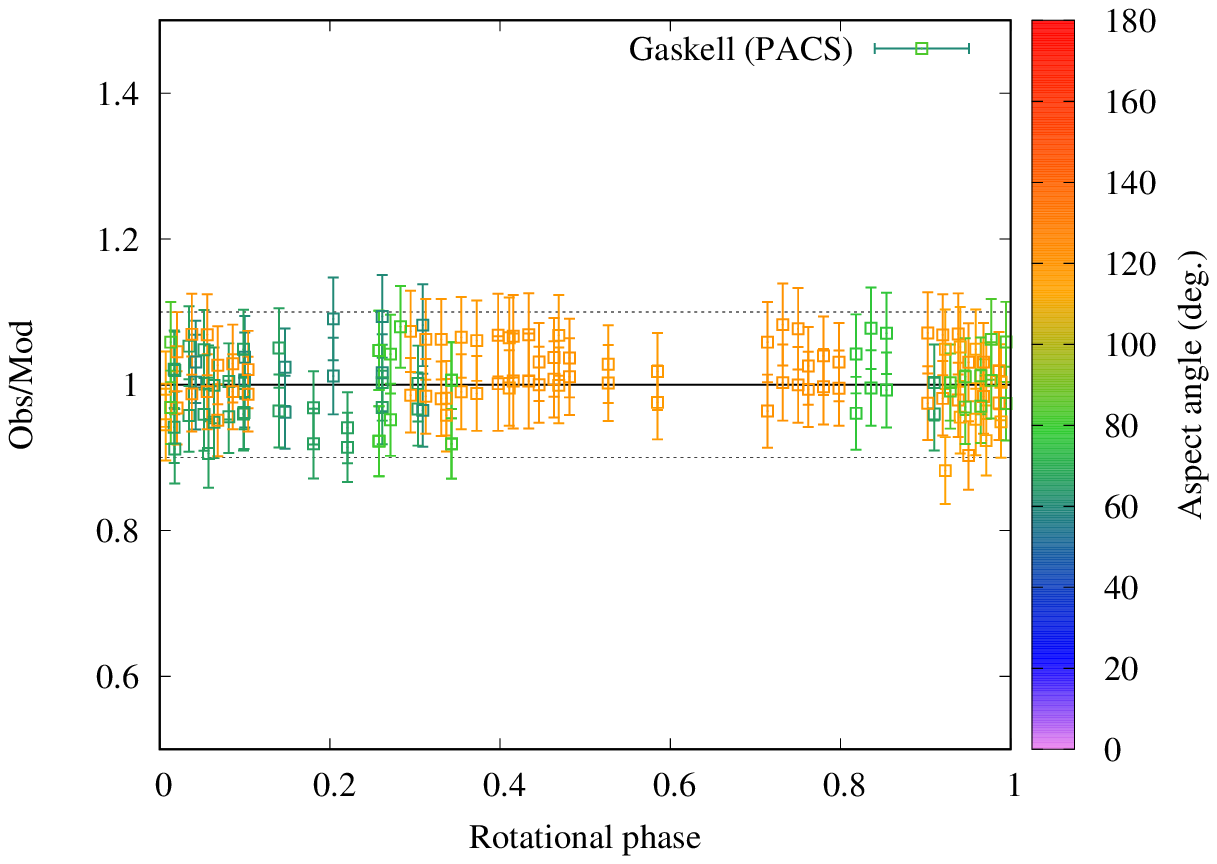}

  \includegraphics[width=0.7\linewidth]{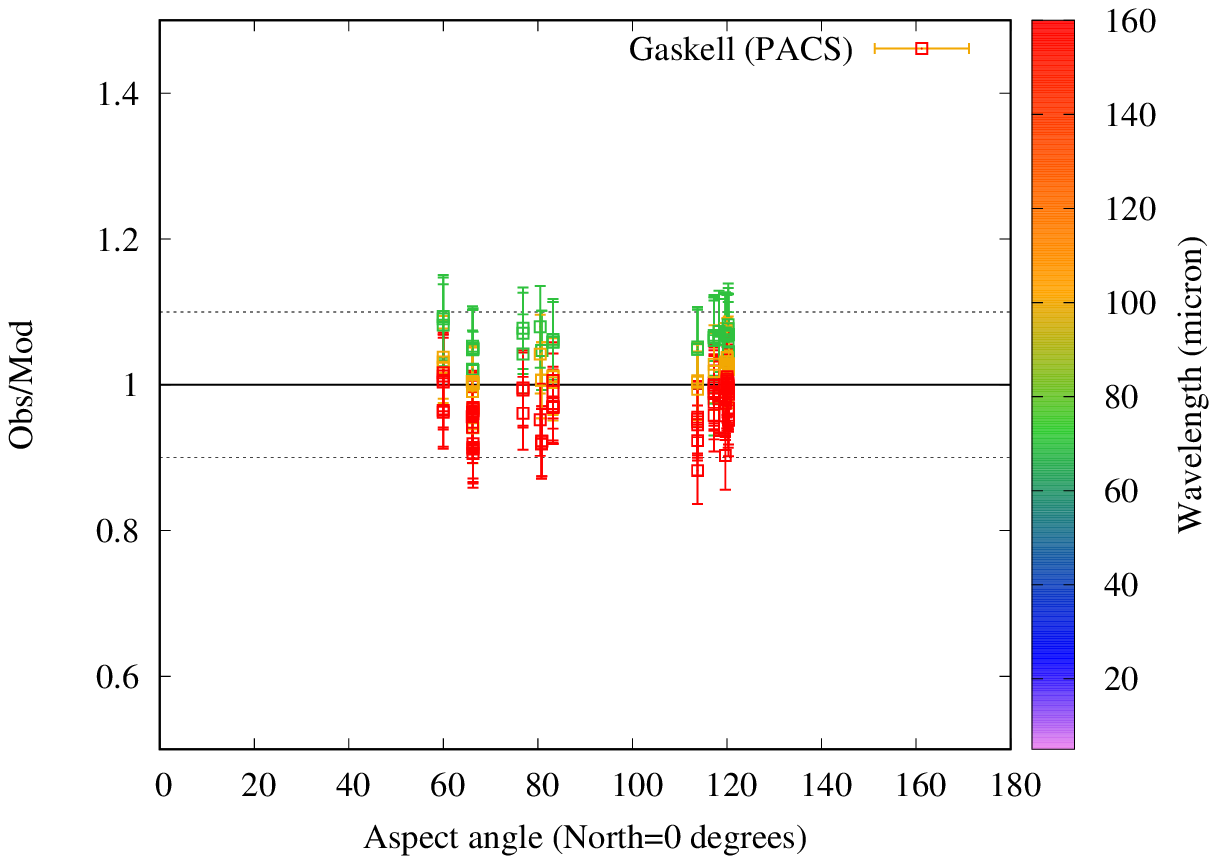}

  \includegraphics[width=0.7\linewidth]{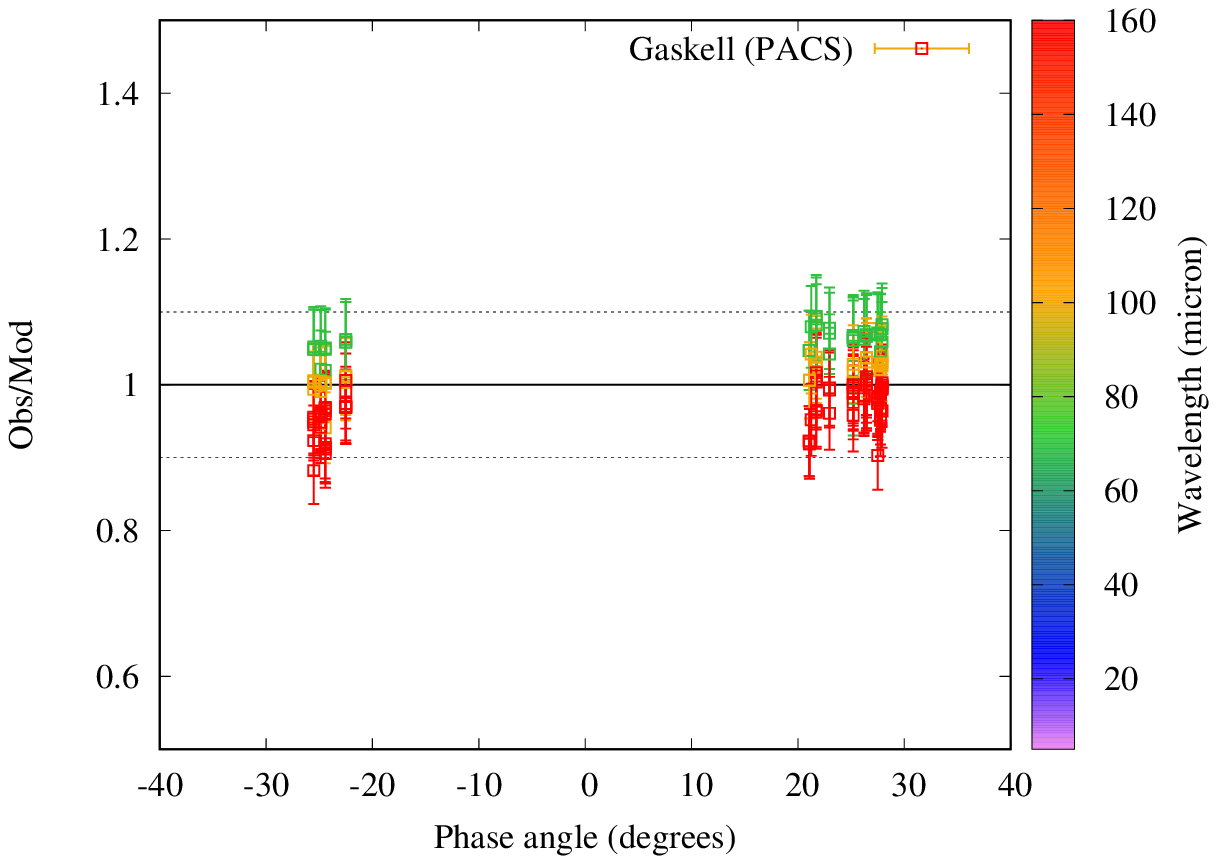}

  \caption{(4) Vesta. See the caption in Fig.~\ref{fig:00001_OMR}. 
  }\label{fig:00004_OMR} 
\end{figure}

\begin{figure}
  \centering
  \includegraphics[width=0.7\linewidth]{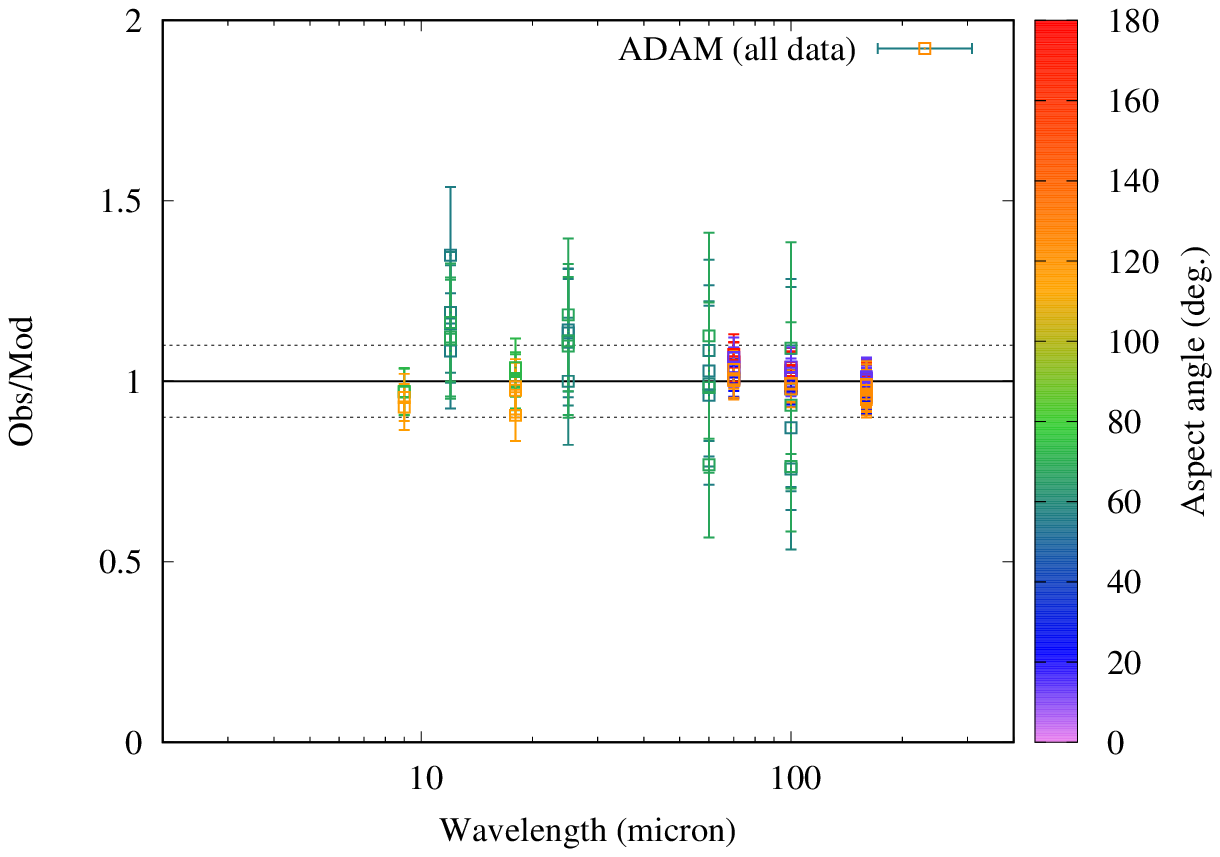}

  \includegraphics[width=0.7\linewidth]{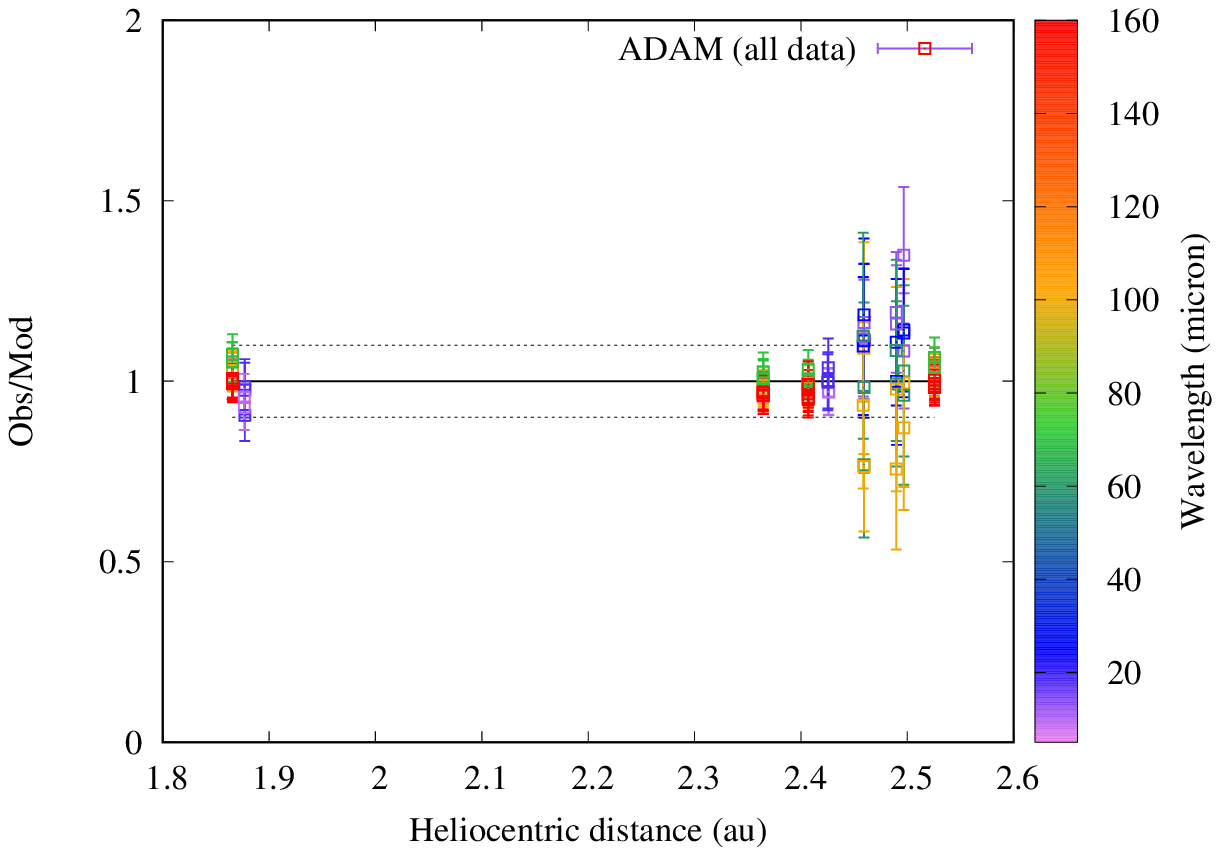}

  \includegraphics[width=0.7\linewidth]{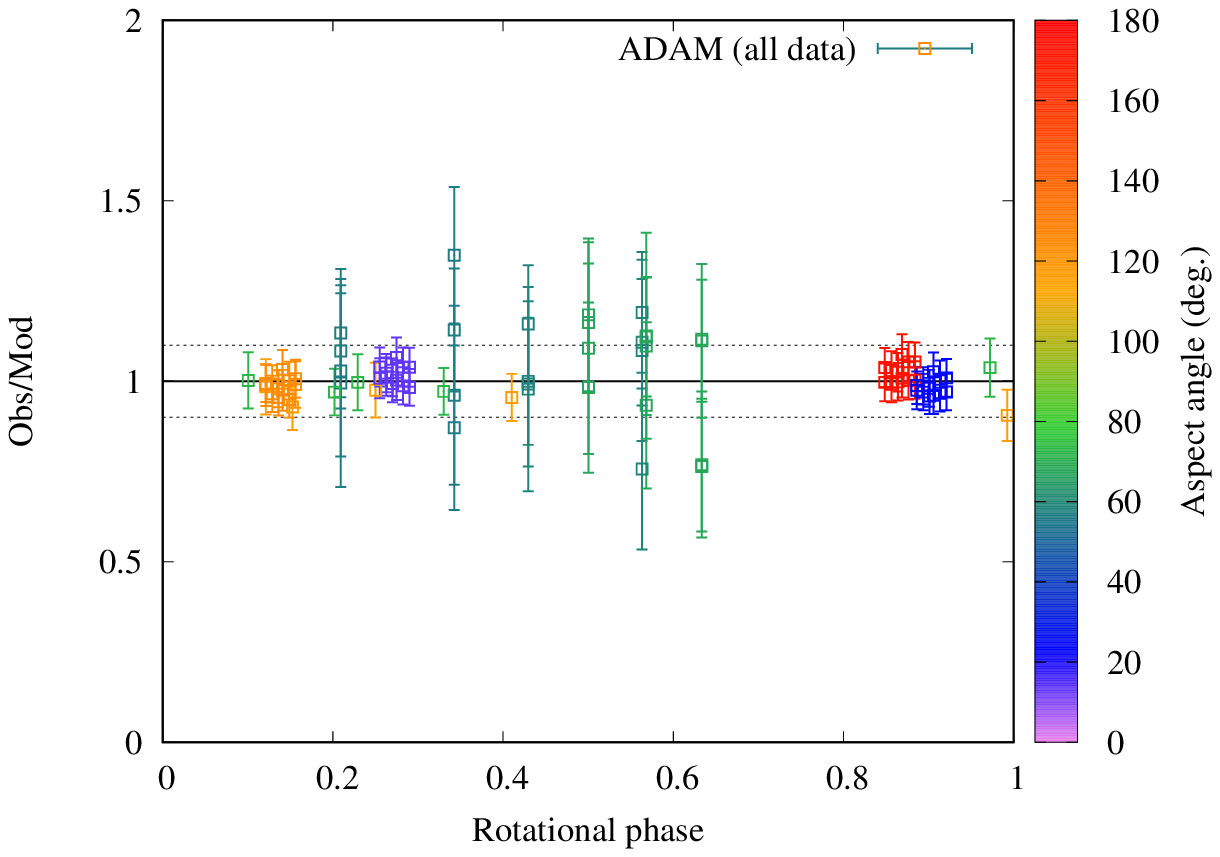}

  \includegraphics[width=0.7\linewidth]{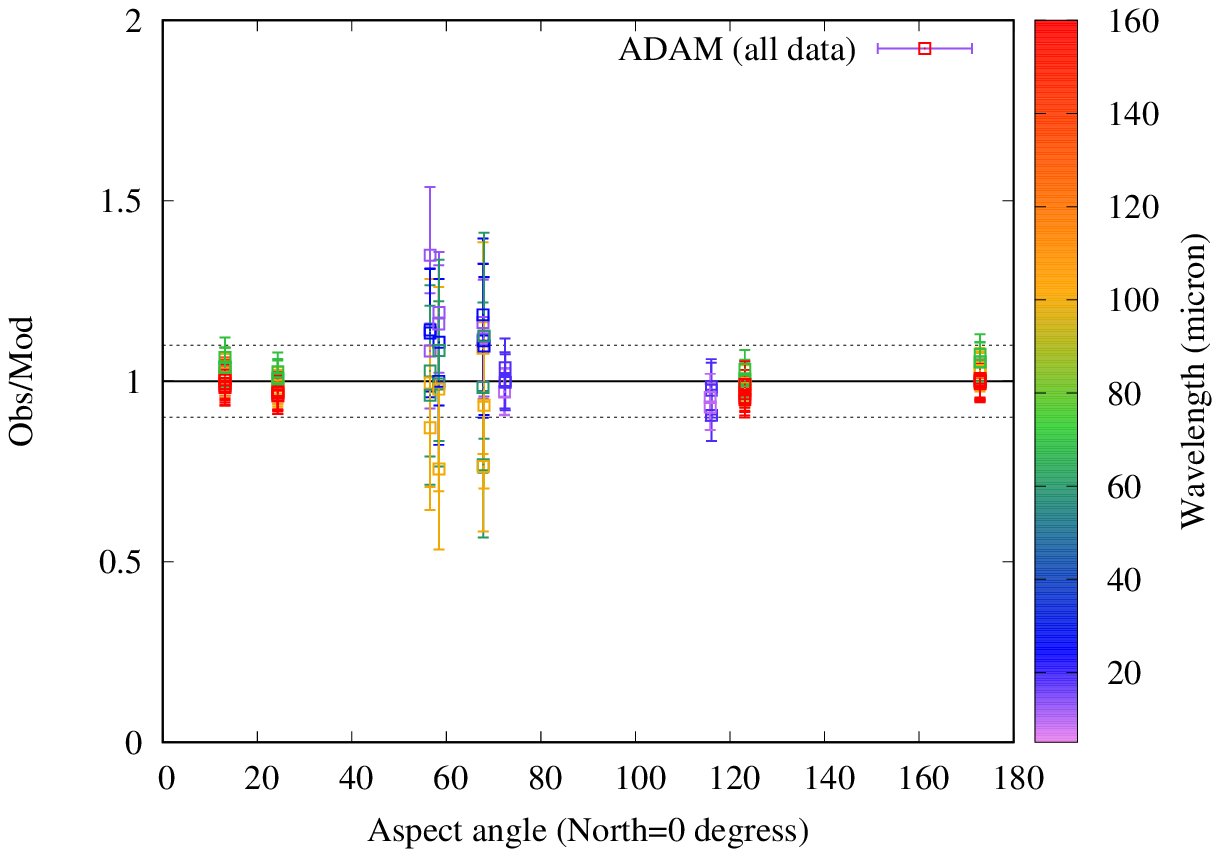}

  \includegraphics[width=0.7\linewidth]{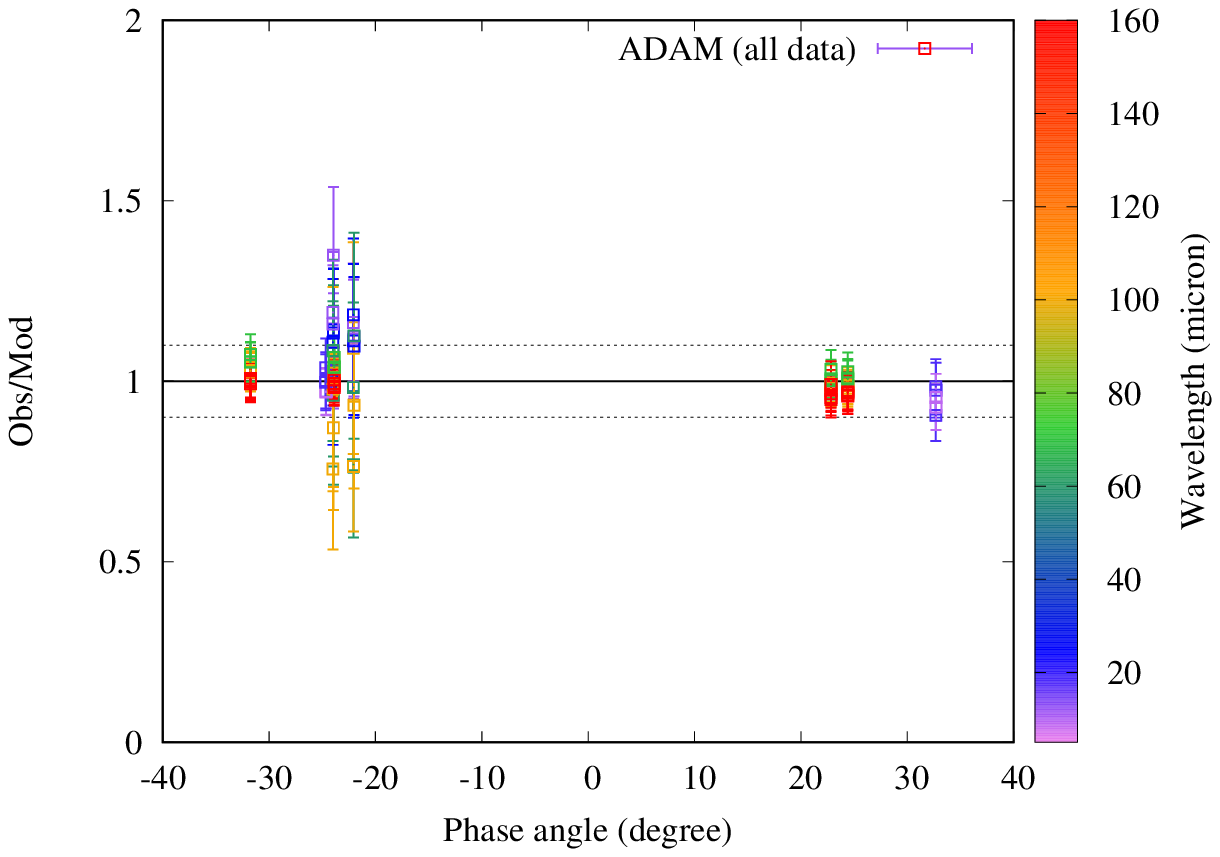}

  \caption{(8) Flora. See the caption in Fig.~\ref{fig:00001_OMR}. 
  }\label{fig:00008_OMR}
\end{figure}

\begin{figure}
  \centering
  \includegraphics[width=0.7\linewidth]{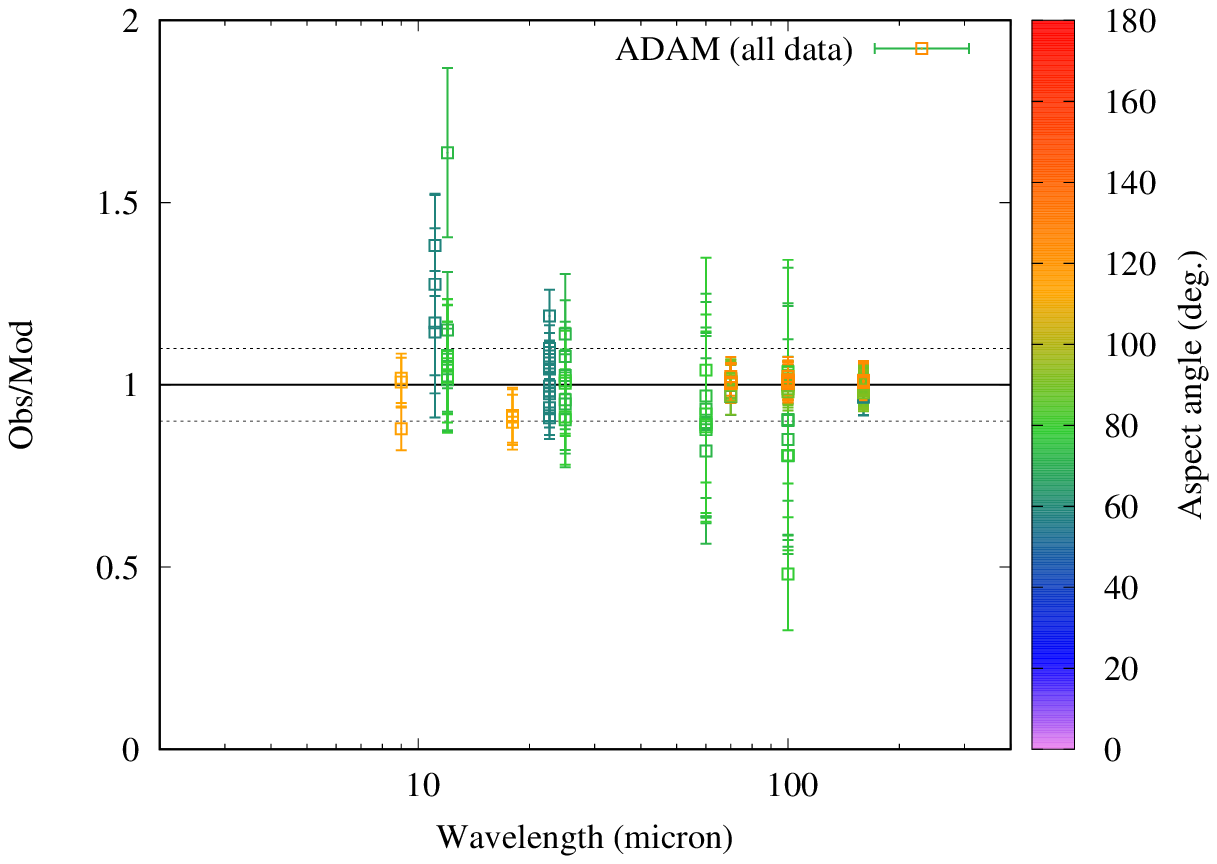}

  \includegraphics[width=0.7\linewidth]{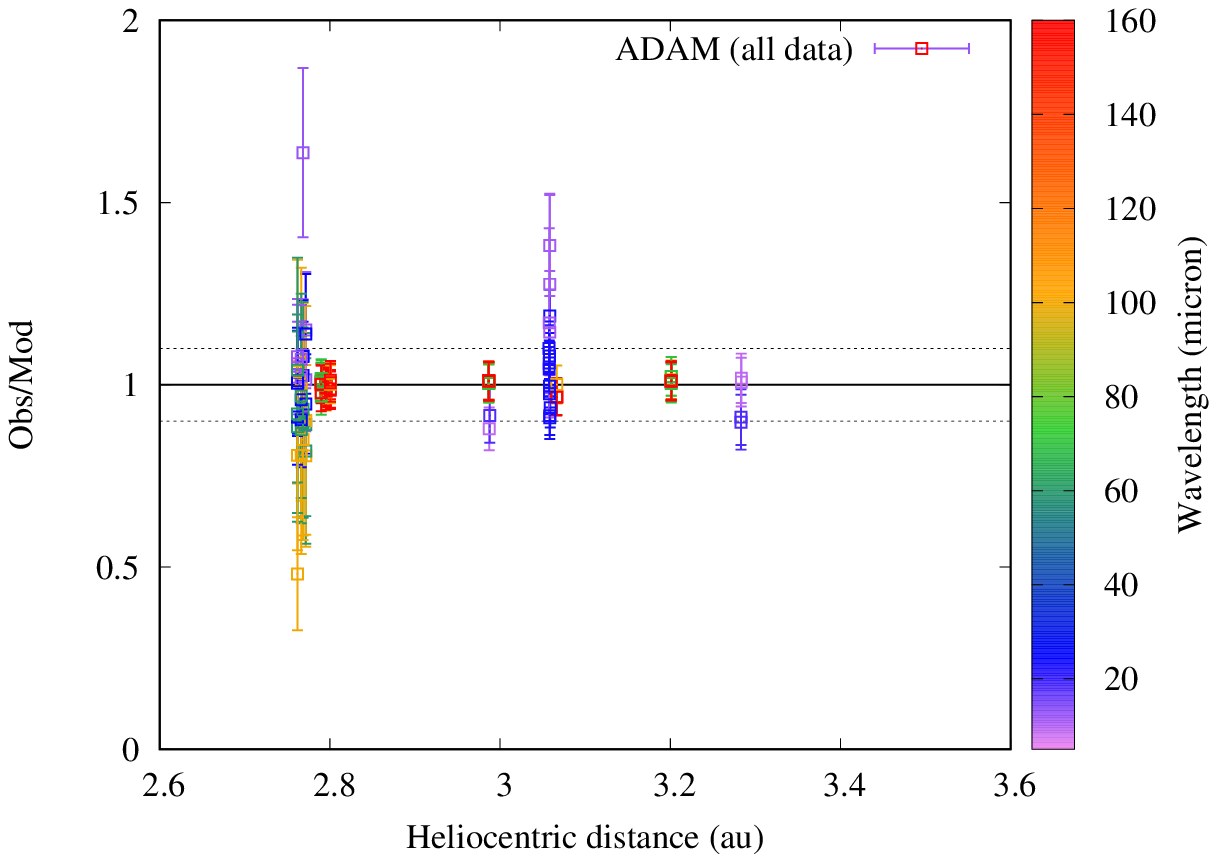}

  \includegraphics[width=0.7\linewidth]{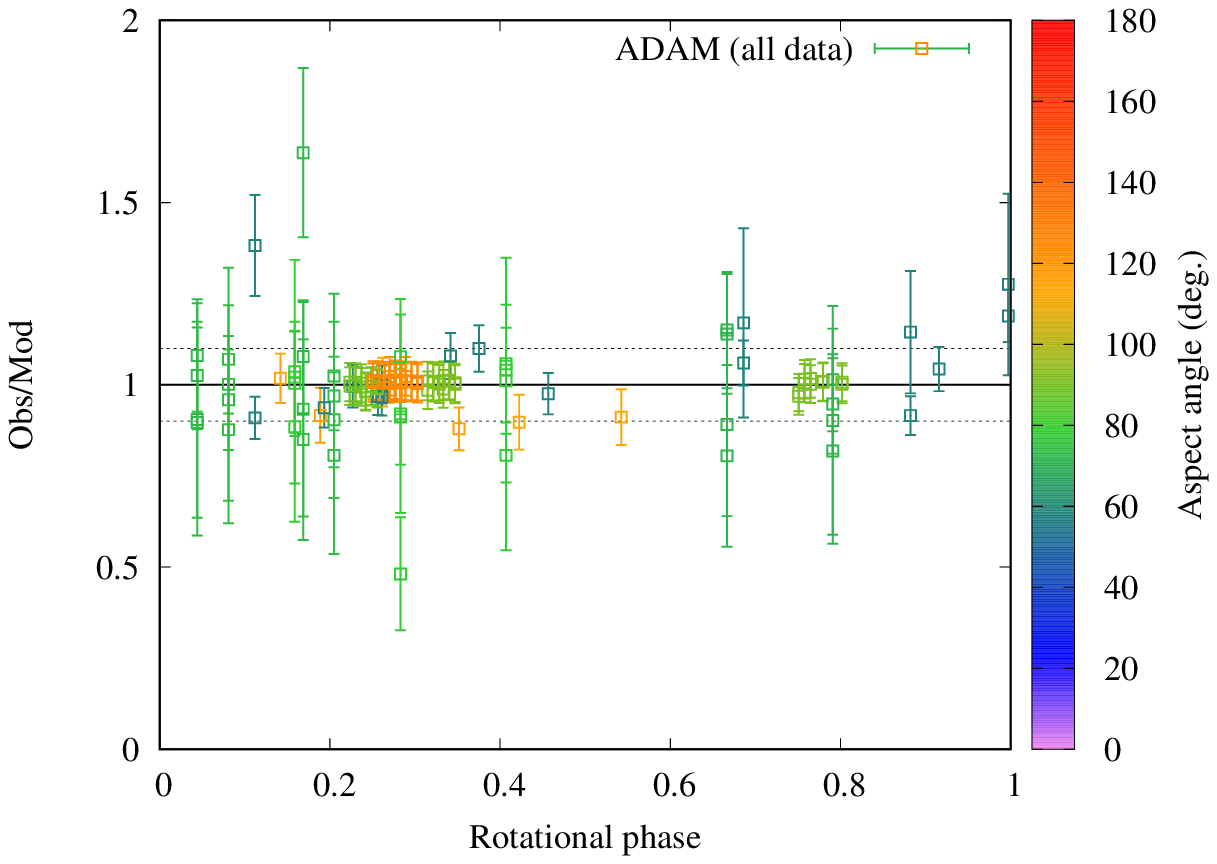}

  \includegraphics[width=0.7\linewidth]{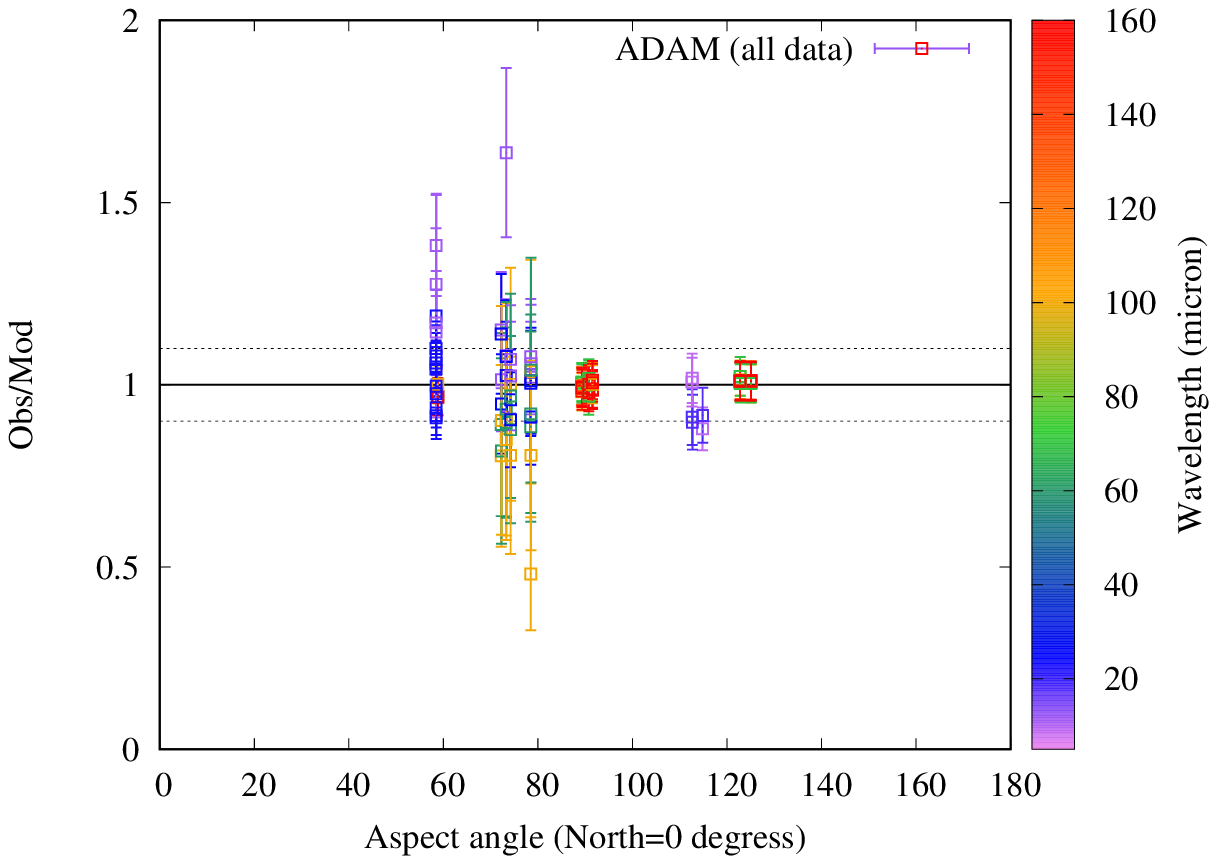}

  \includegraphics[width=0.7\linewidth]{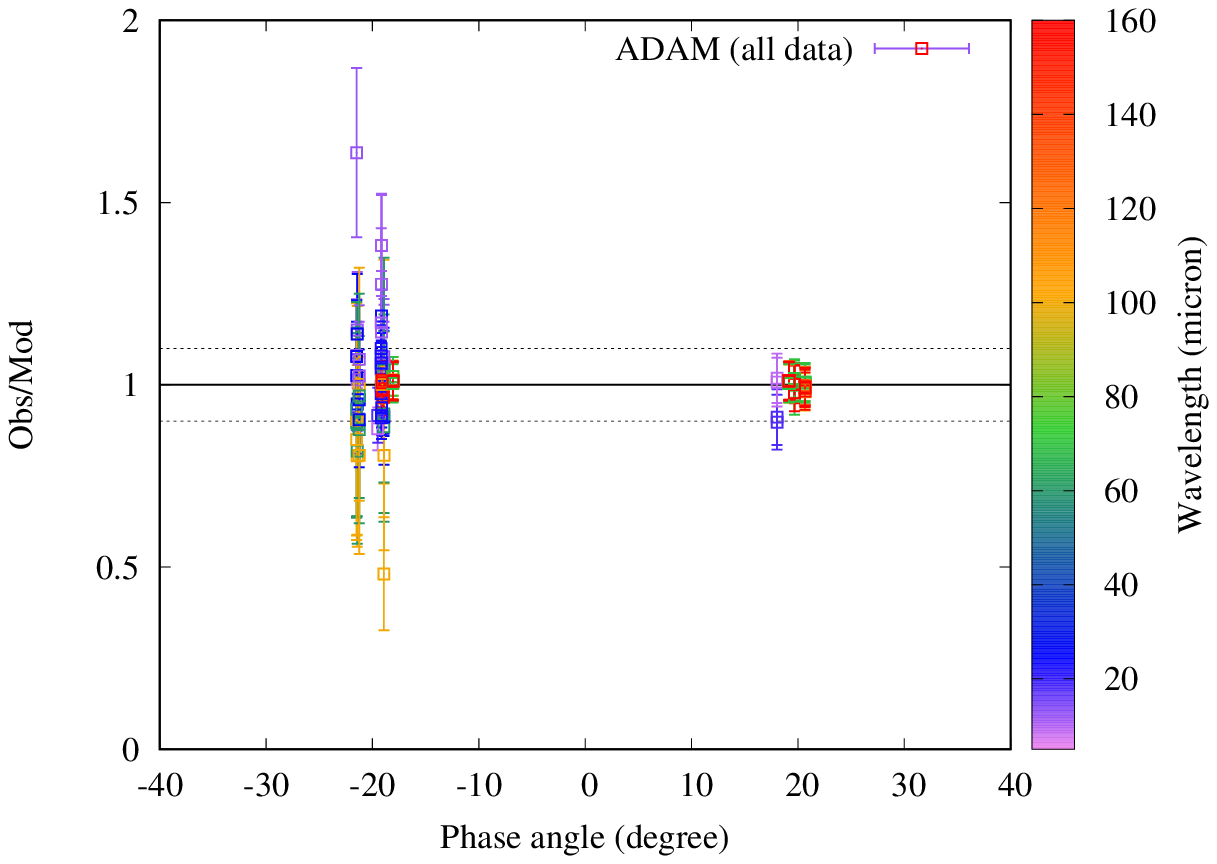}

  \caption{(10) Hygiea. See the caption in Fig.~\ref{fig:00001_OMR}. 
  }\label{fig:00010_OMR}
\end{figure}

\begin{figure}
  \centering
  \includegraphics[width=0.7\linewidth]{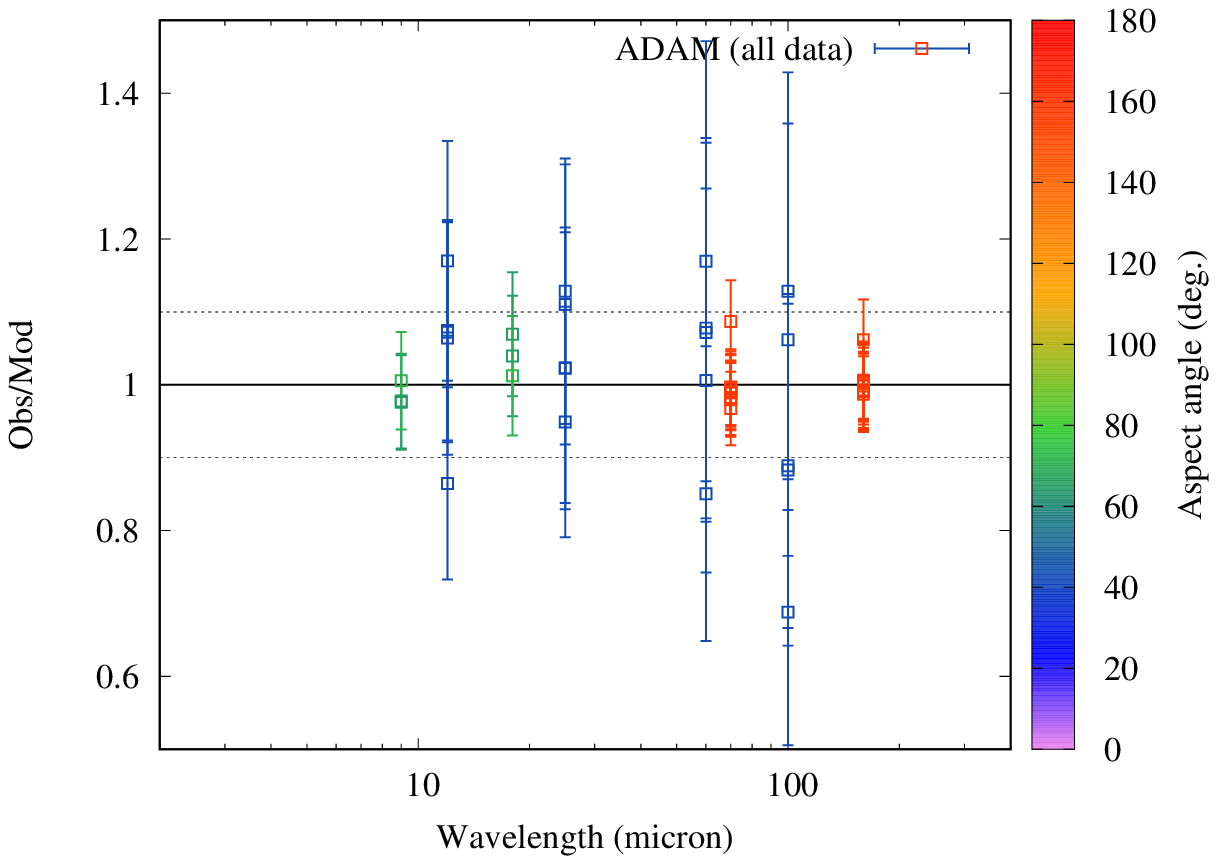}

  \includegraphics[width=0.7\linewidth]{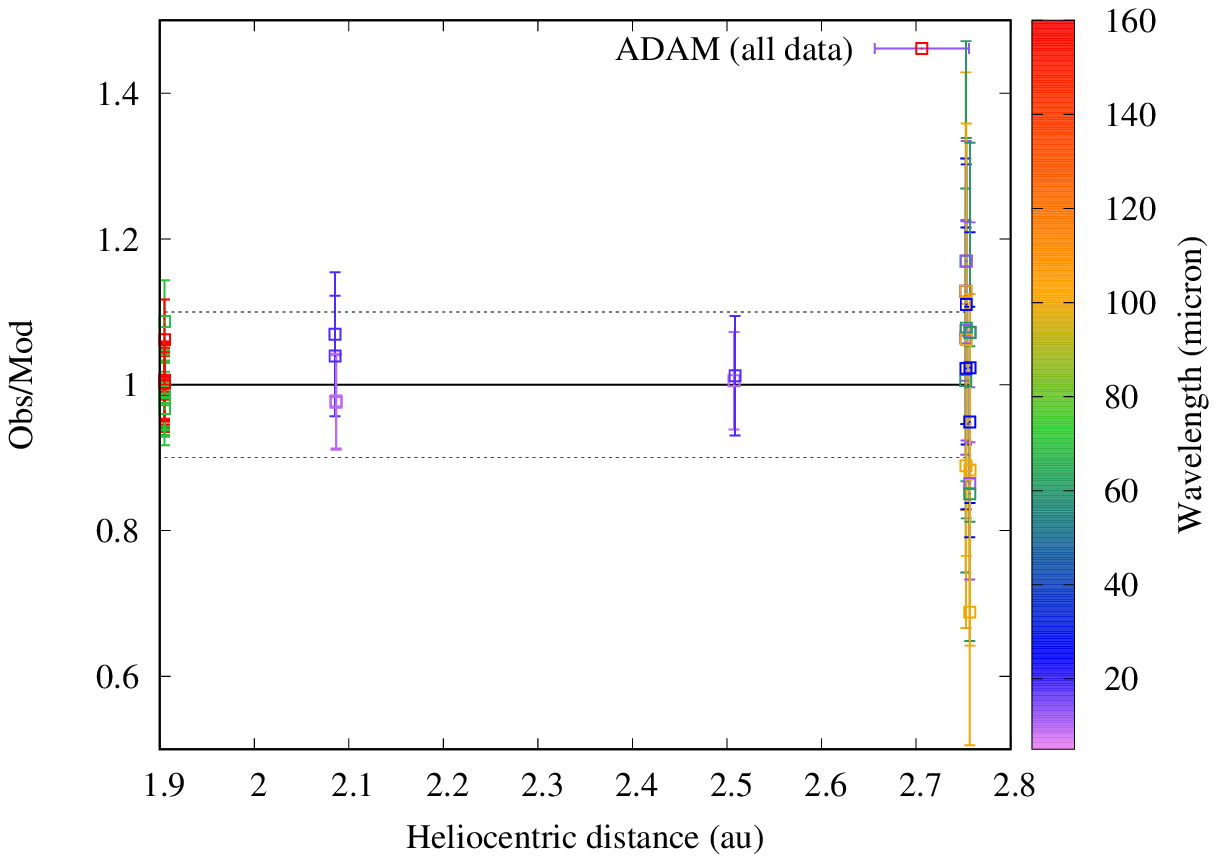}

  \includegraphics[width=0.7\linewidth]{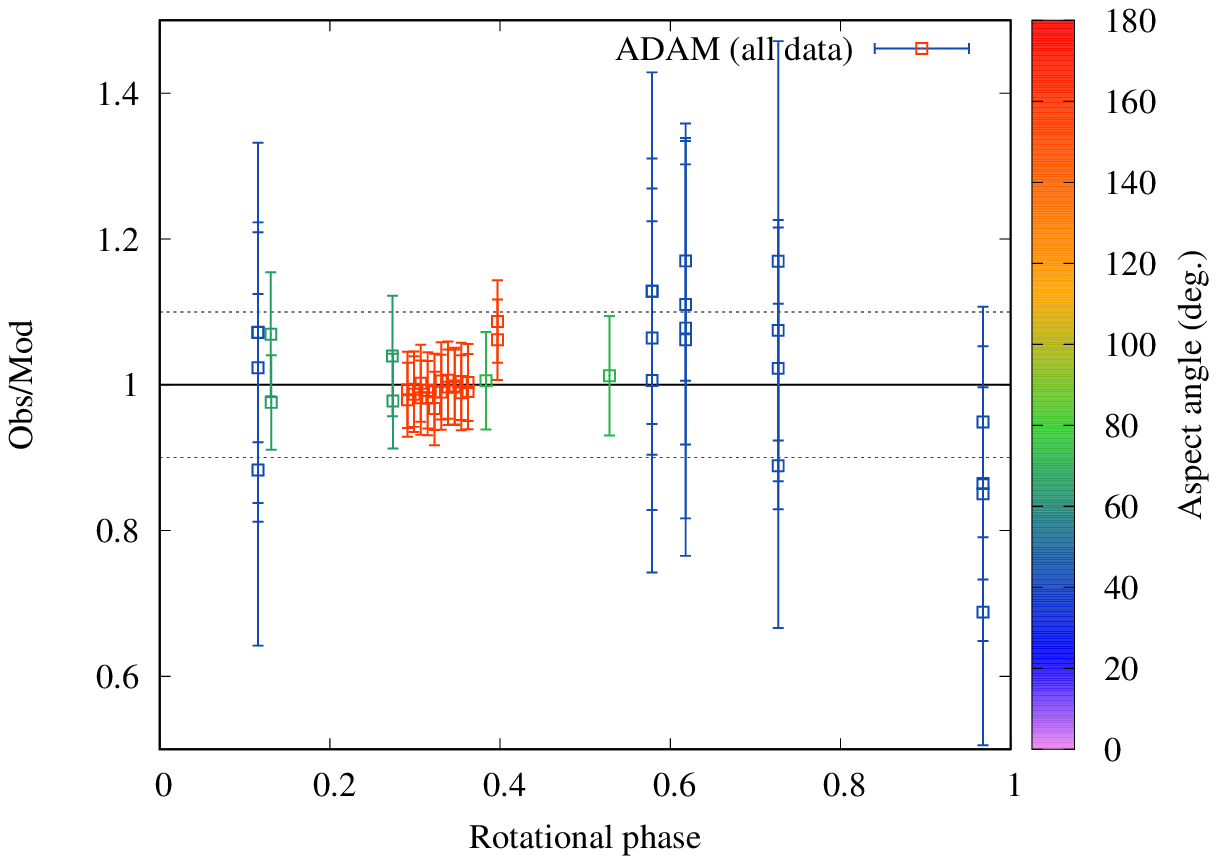}

  \includegraphics[width=0.7\linewidth]{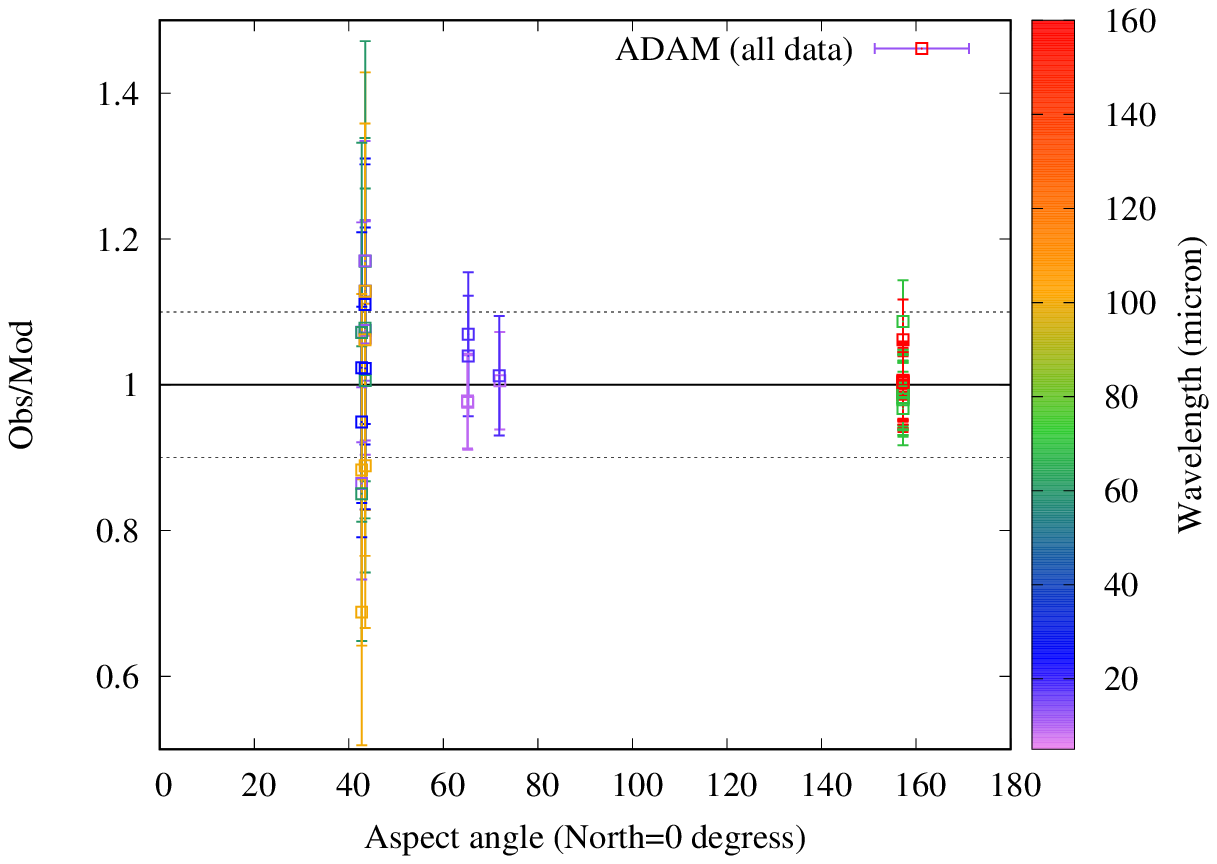}

  \includegraphics[width=0.7\linewidth]{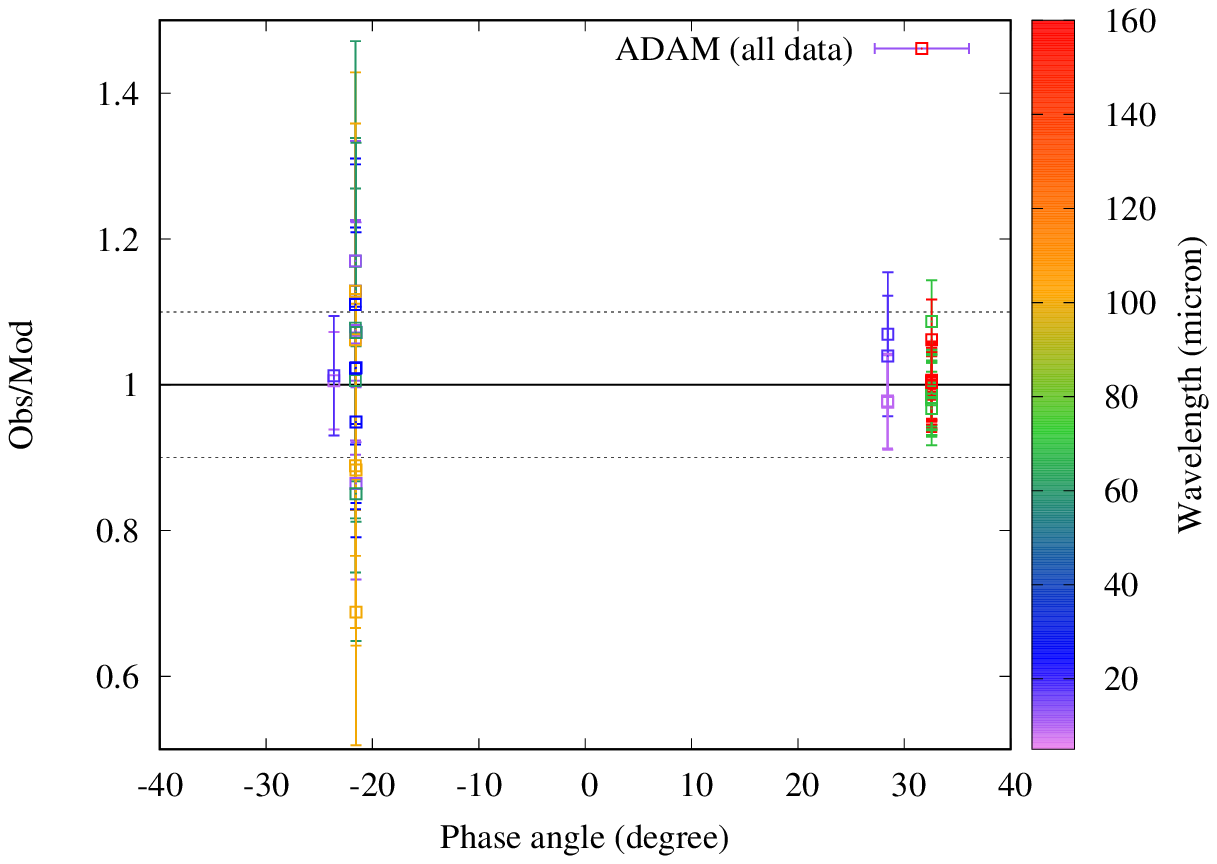}

  \caption{(18) Melpomene. See the caption in Fig.~\ref{fig:00001_OMR}. 
  }\label{fig:00018_OMR}
\end{figure}

\begin{figure}
  \centering
  \includegraphics[width=0.7\linewidth]{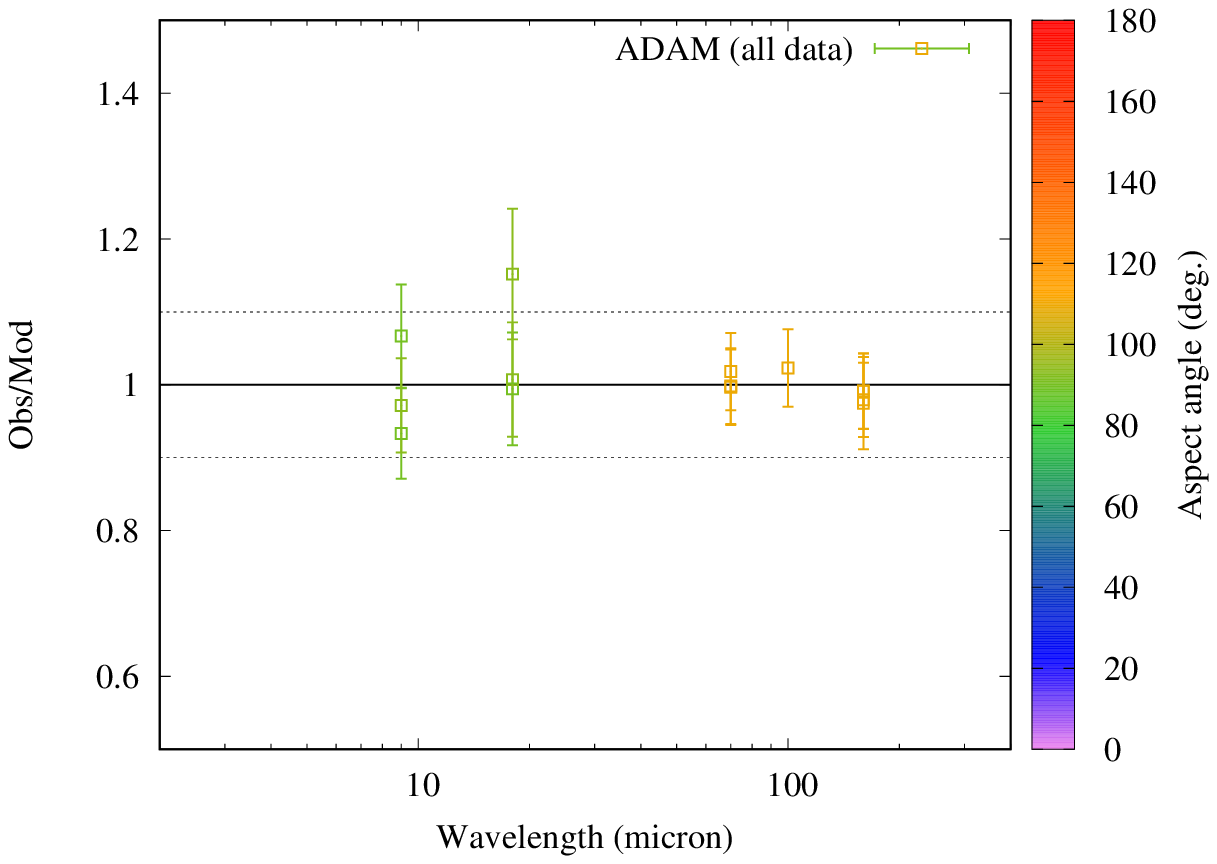}

  \includegraphics[width=0.7\linewidth]{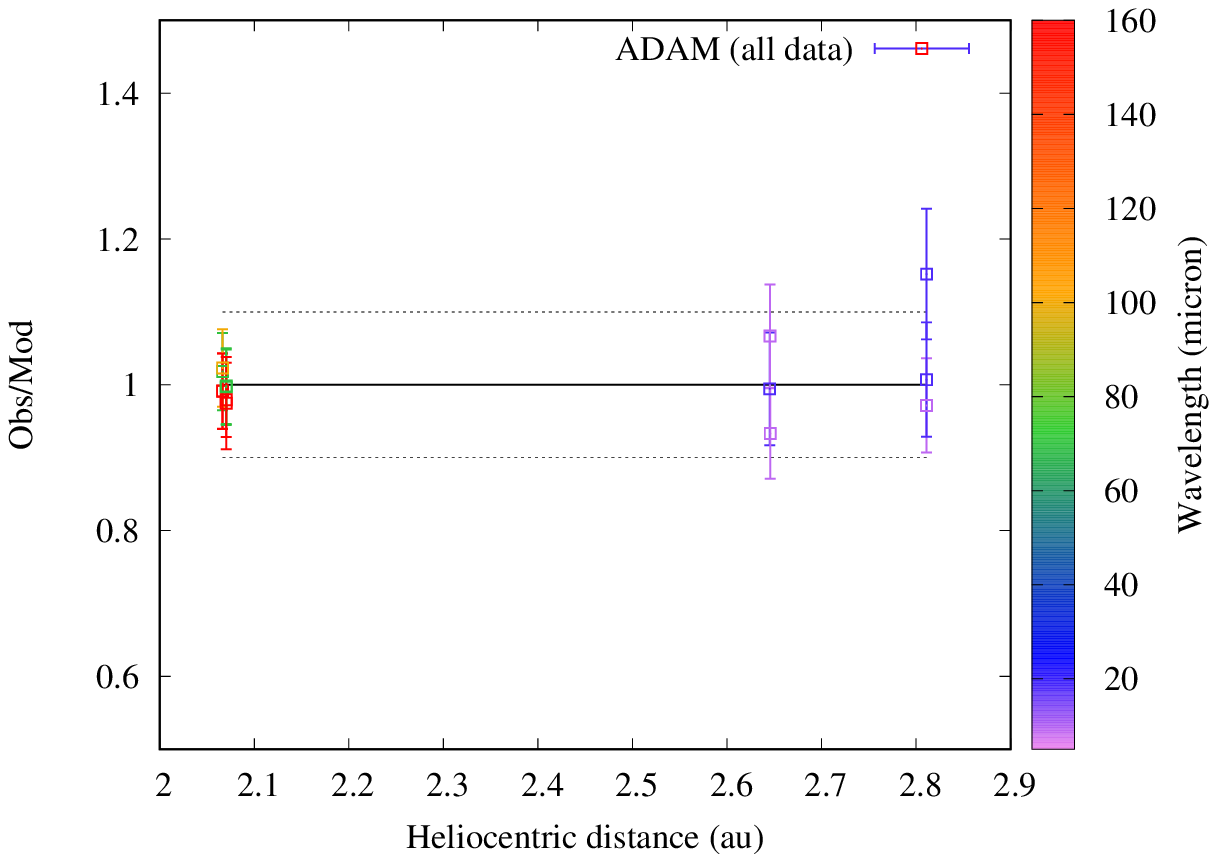}

  \includegraphics[width=0.7\linewidth]{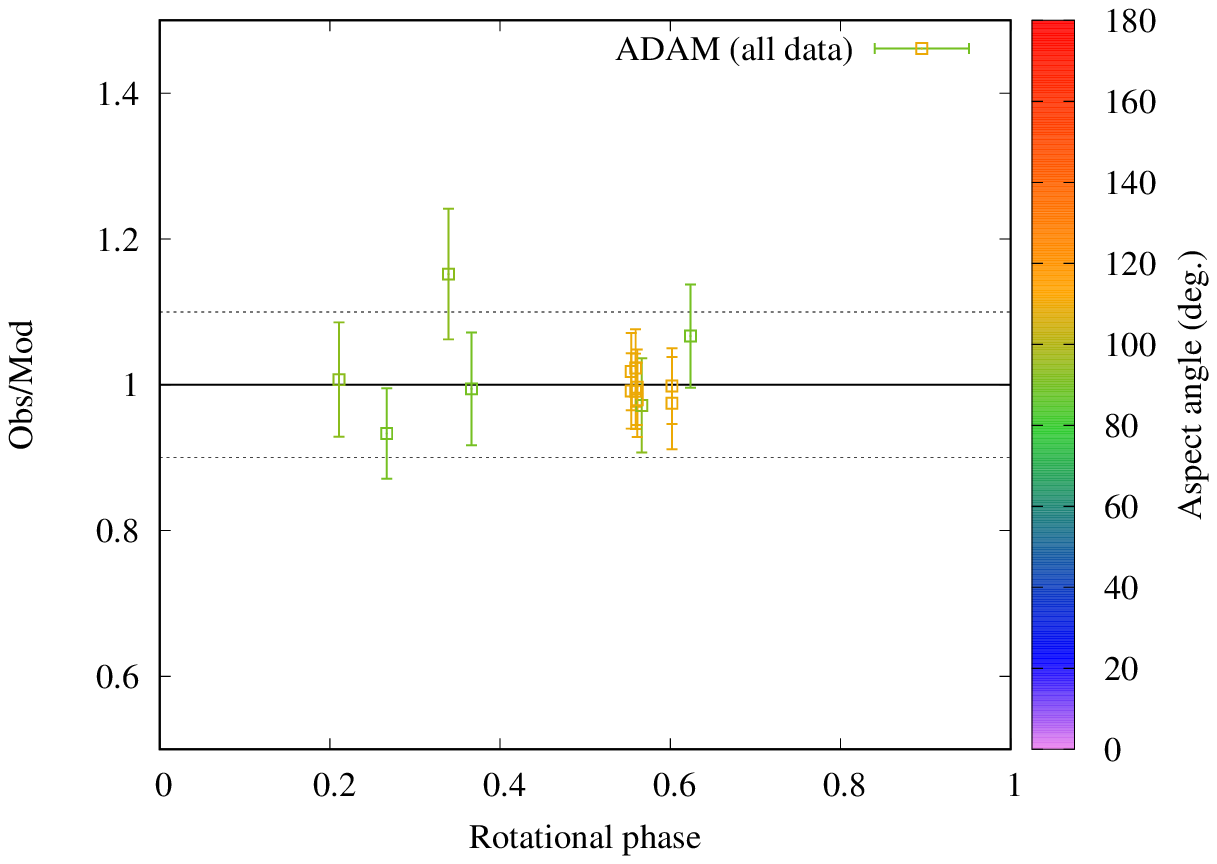}

  \includegraphics[width=0.7\linewidth]{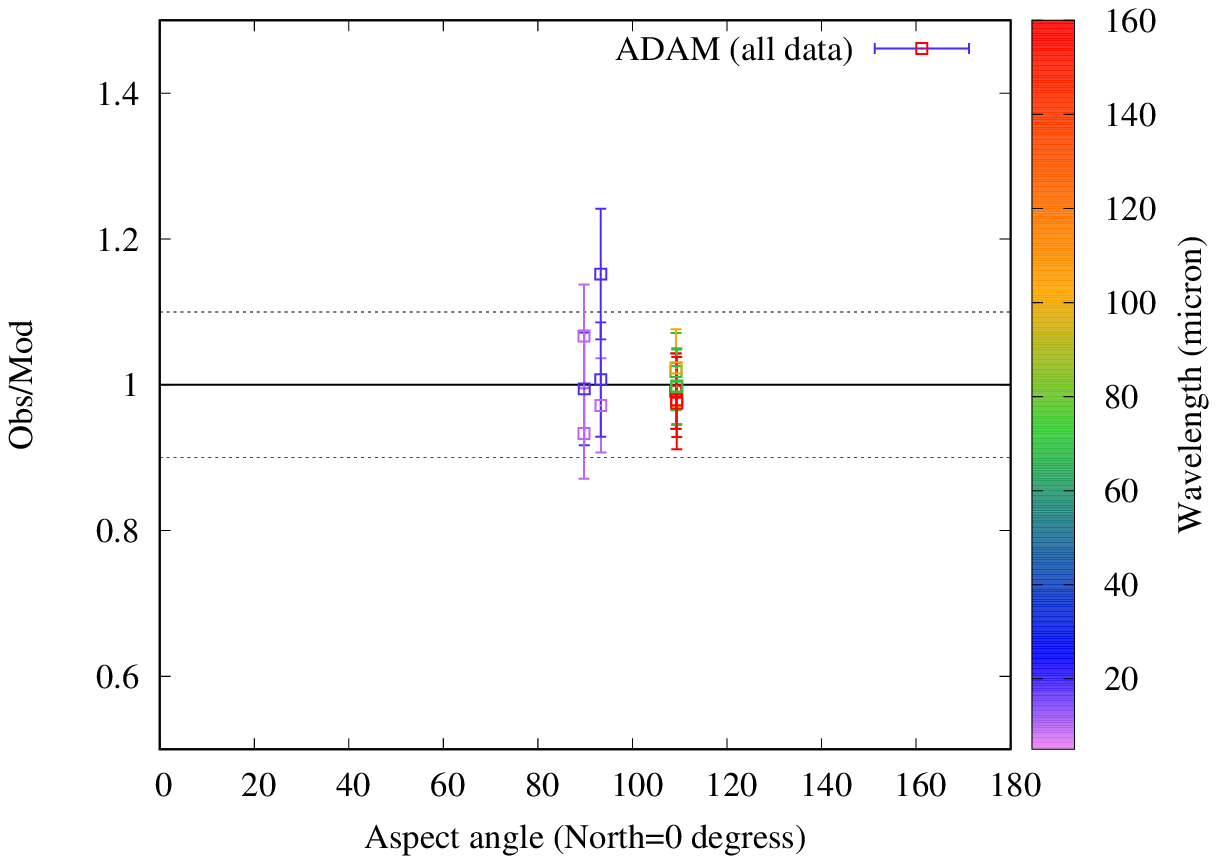}

  \includegraphics[width=0.7\linewidth]{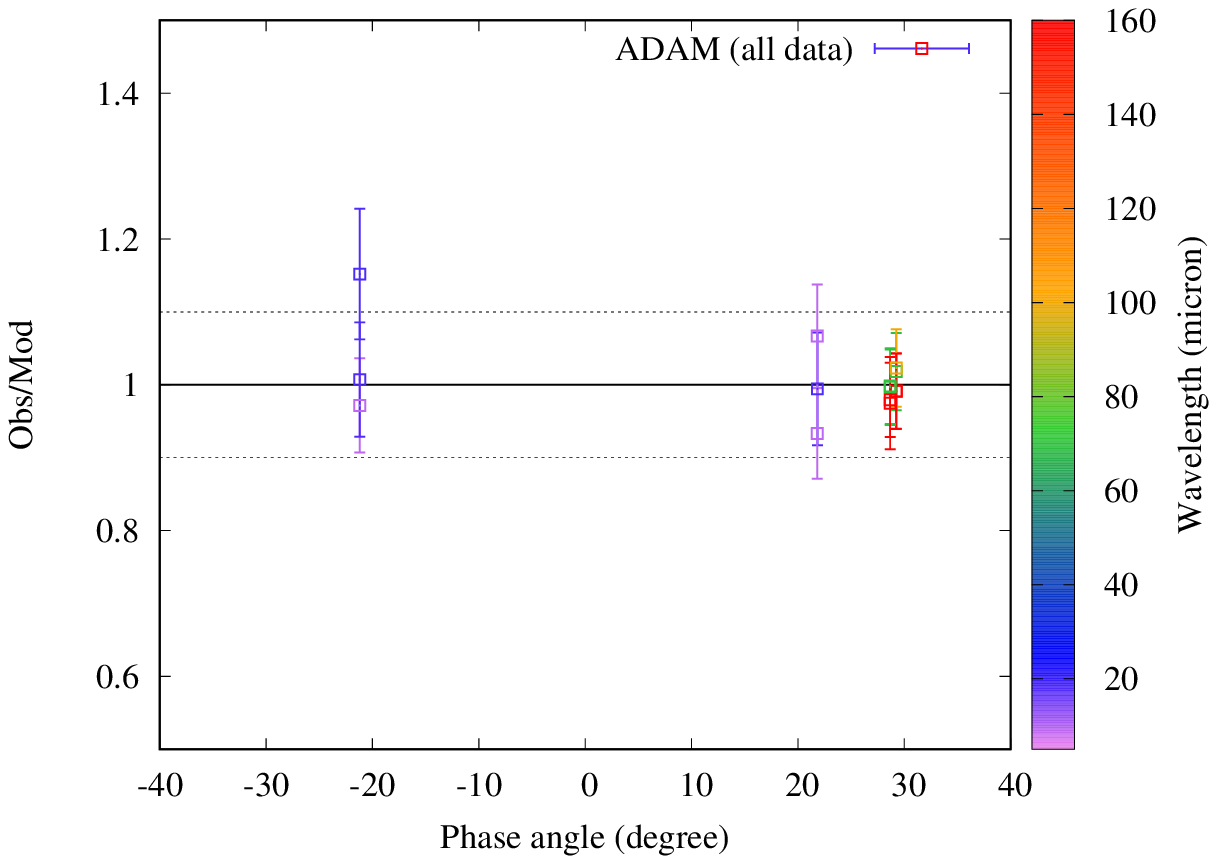}

  \caption{(19) Fortuna. See the caption in Fig.~\ref{fig:00001_OMR}. 
  }\label{fig:00019_OMR}
\end{figure}

\begin{figure}
  \centering
  \includegraphics[width=0.7\linewidth]{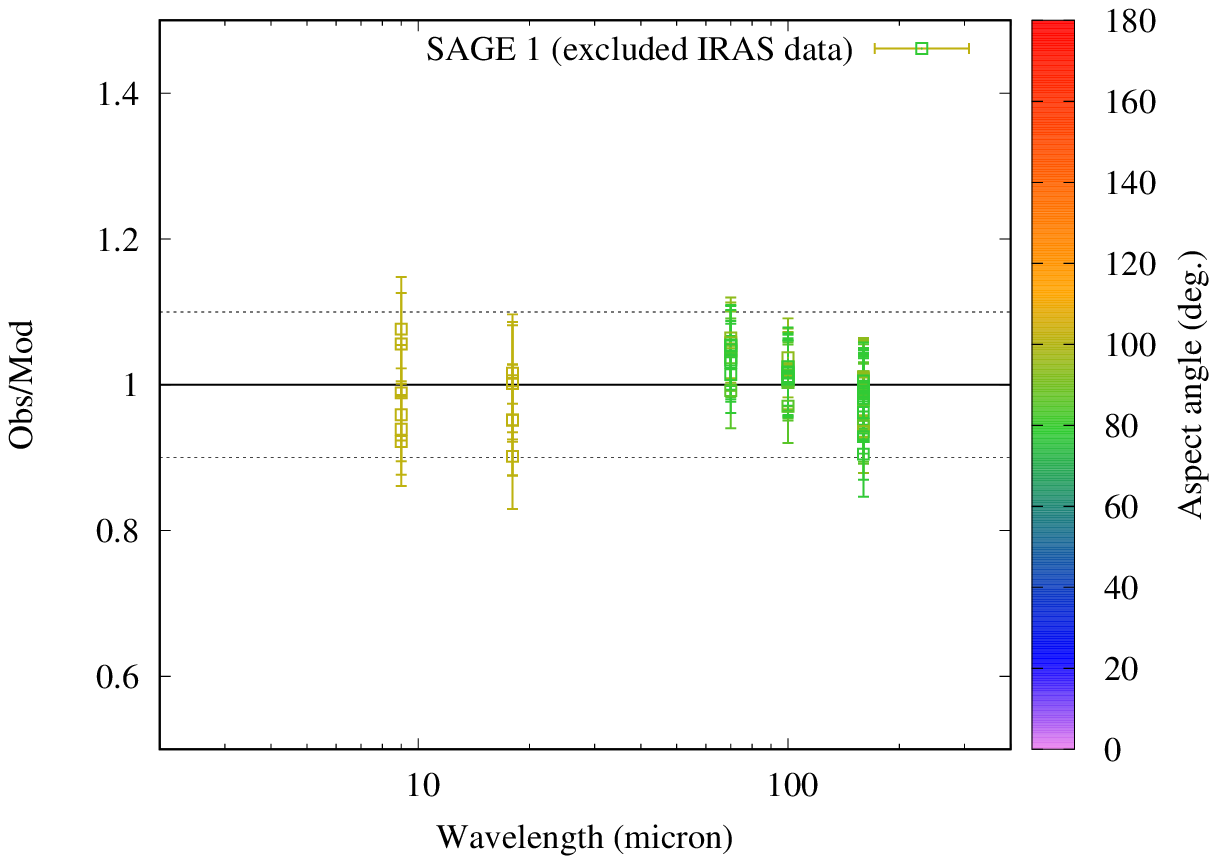}

  \includegraphics[width=0.7\linewidth]{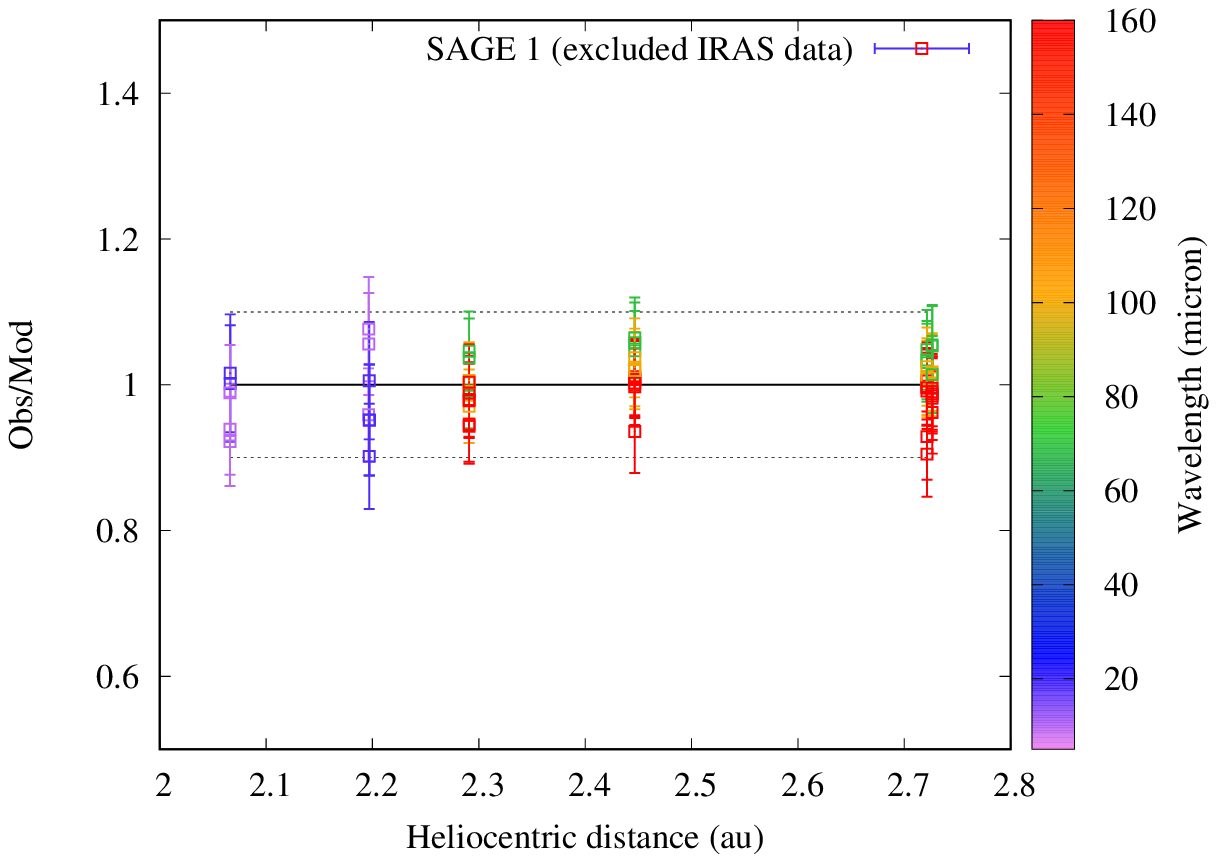}

  \includegraphics[width=0.7\linewidth]{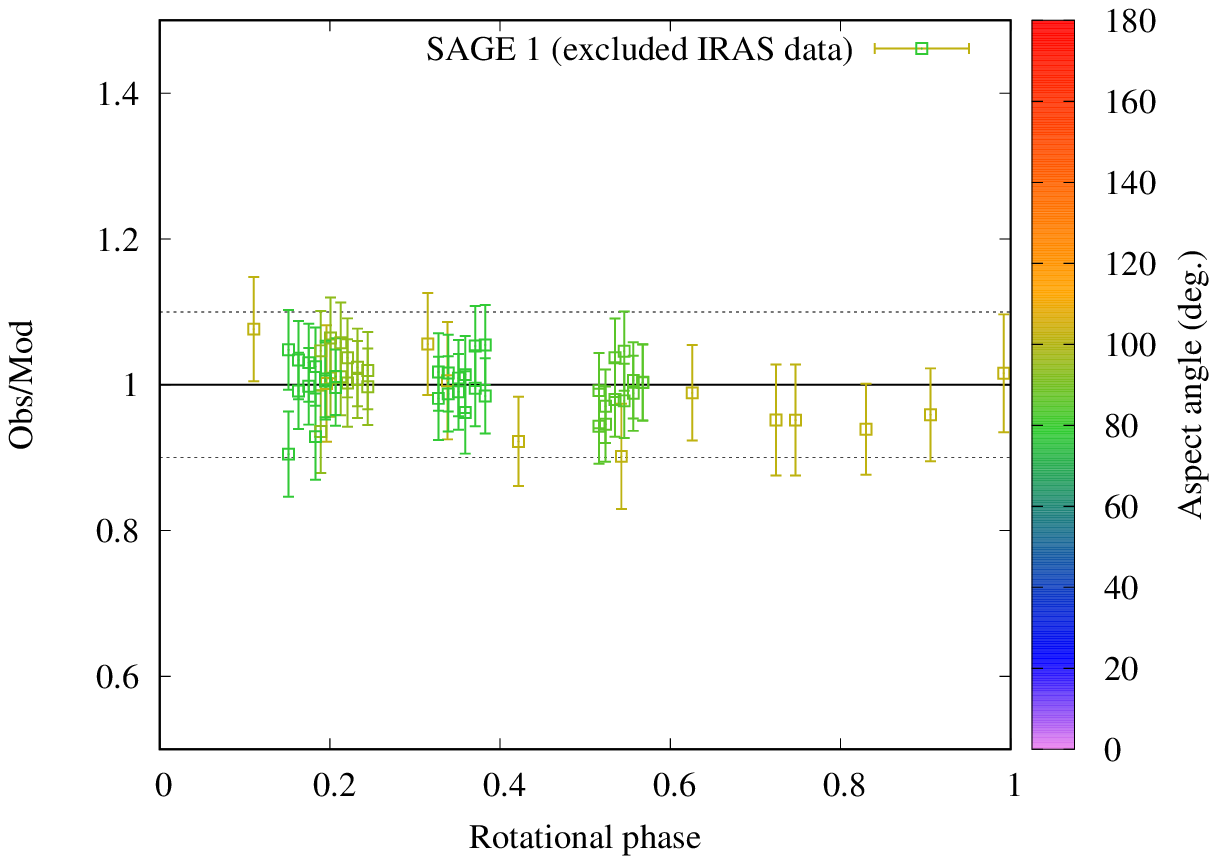}

  \includegraphics[width=0.7\linewidth]{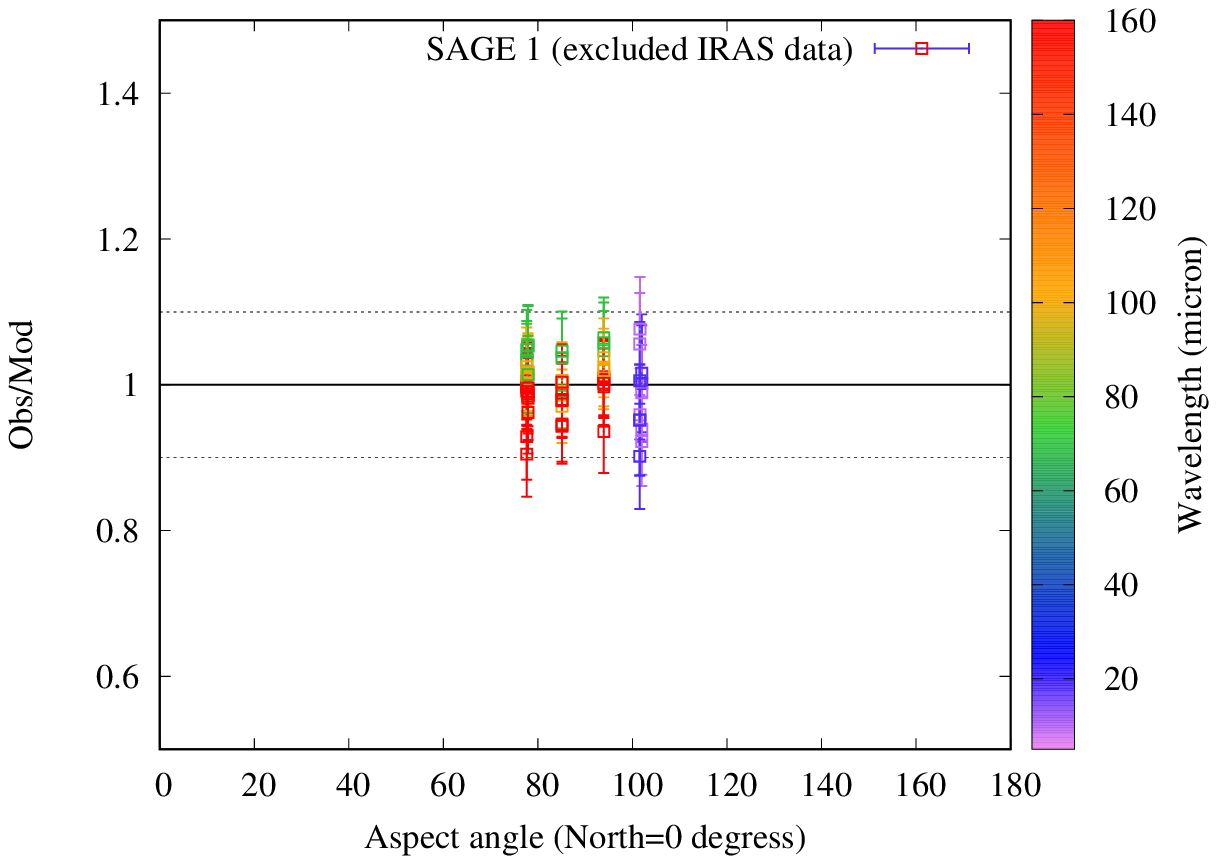}

  \includegraphics[width=0.7\linewidth]{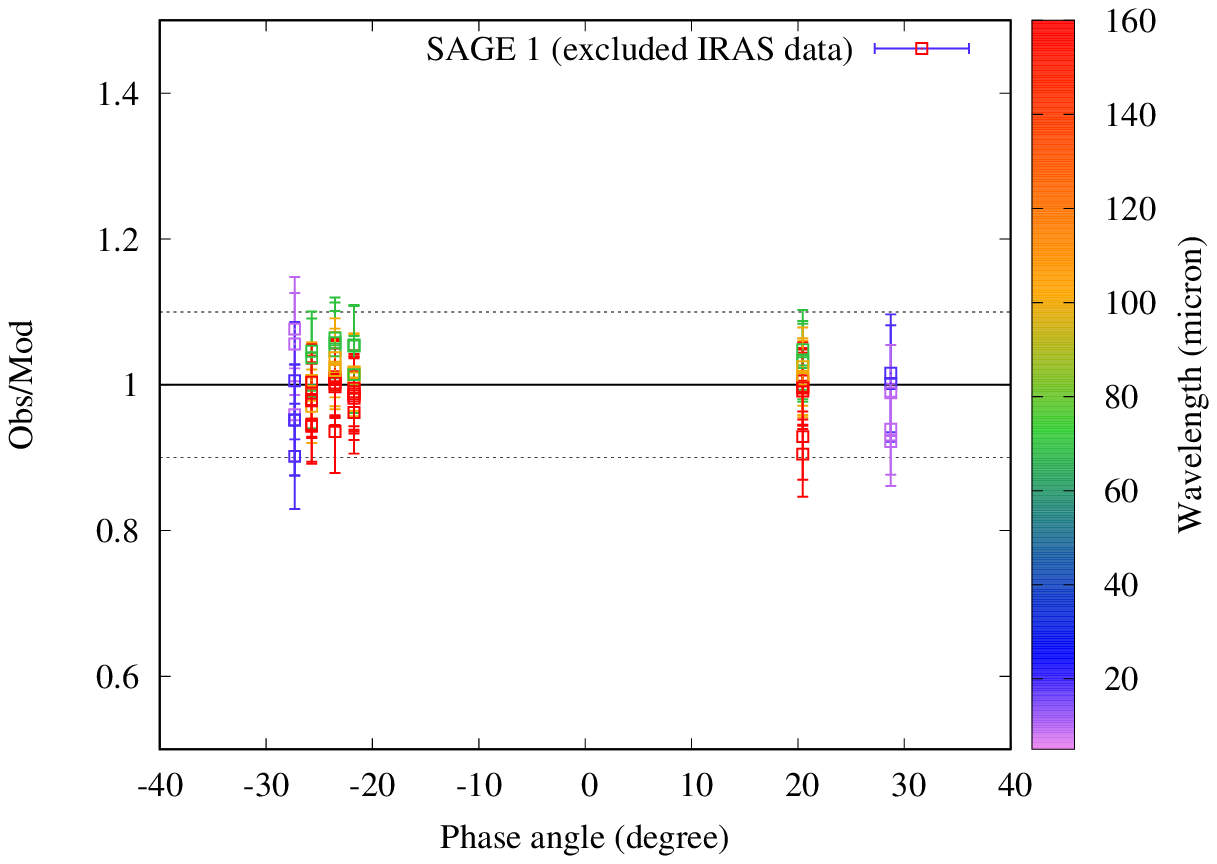}

  \caption{(20) Massalia (SAGE model 1). Similar plots were obtained for the second  mirror solution. 
  }\label{fig:00020_OMR}
\end{figure}

\subsection{(21) Lutetia}\label{sec:lutetia}

\begin{figure}
  \centering
  \includegraphics[width=0.7\linewidth]{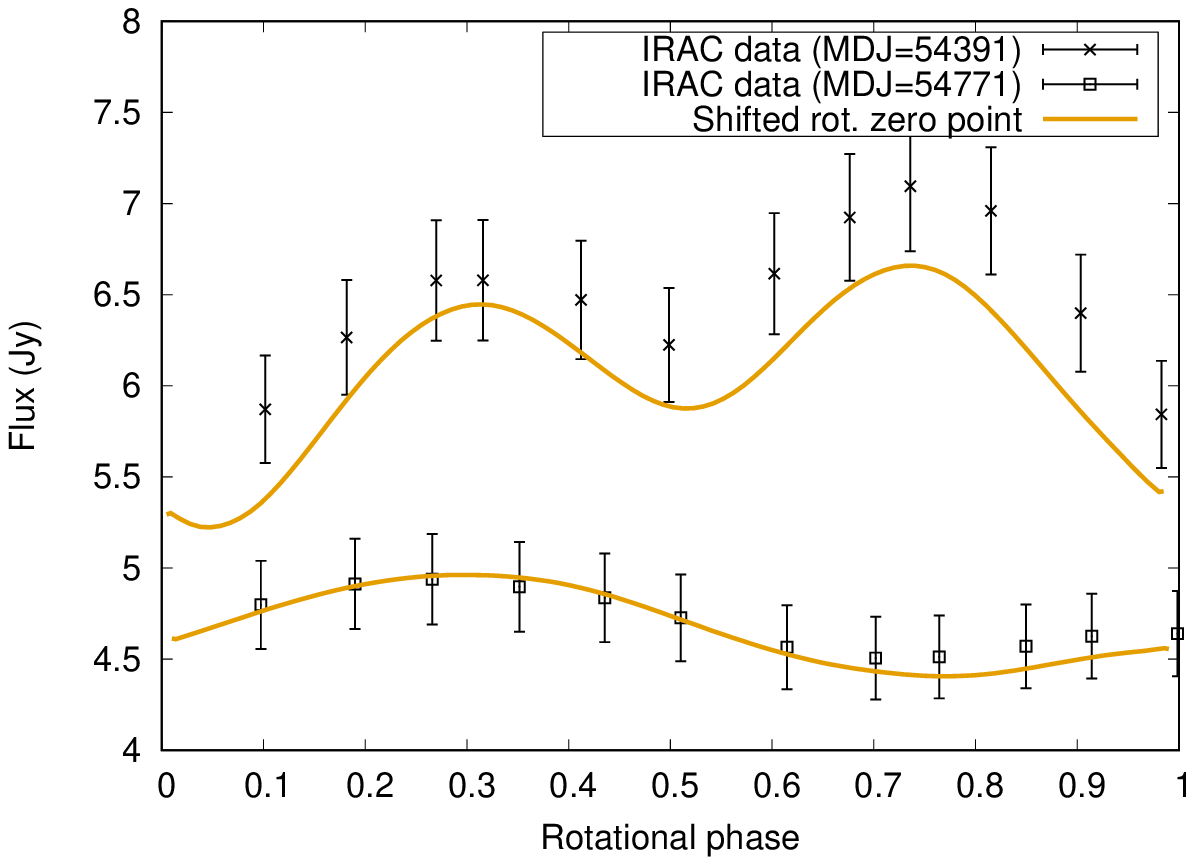}
  \caption{Spitzer IRAC thermal light curve of (21) Lutetia ($\lambda=$ 7.872 $\mu$m) and our best-fitting model fluxes. \citet{ORourke2012} provide details of the observations. MDJ is the modified Julian date, the first one (crosses) corresponding to October 17 2007 15:13 UT, the second one (empty squares) to October 31 2008 19:58 UT. }
  \label{fig:ThLC}
\end{figure}

\begin{figure}
  \centering
  \includegraphics[width=0.7\linewidth]{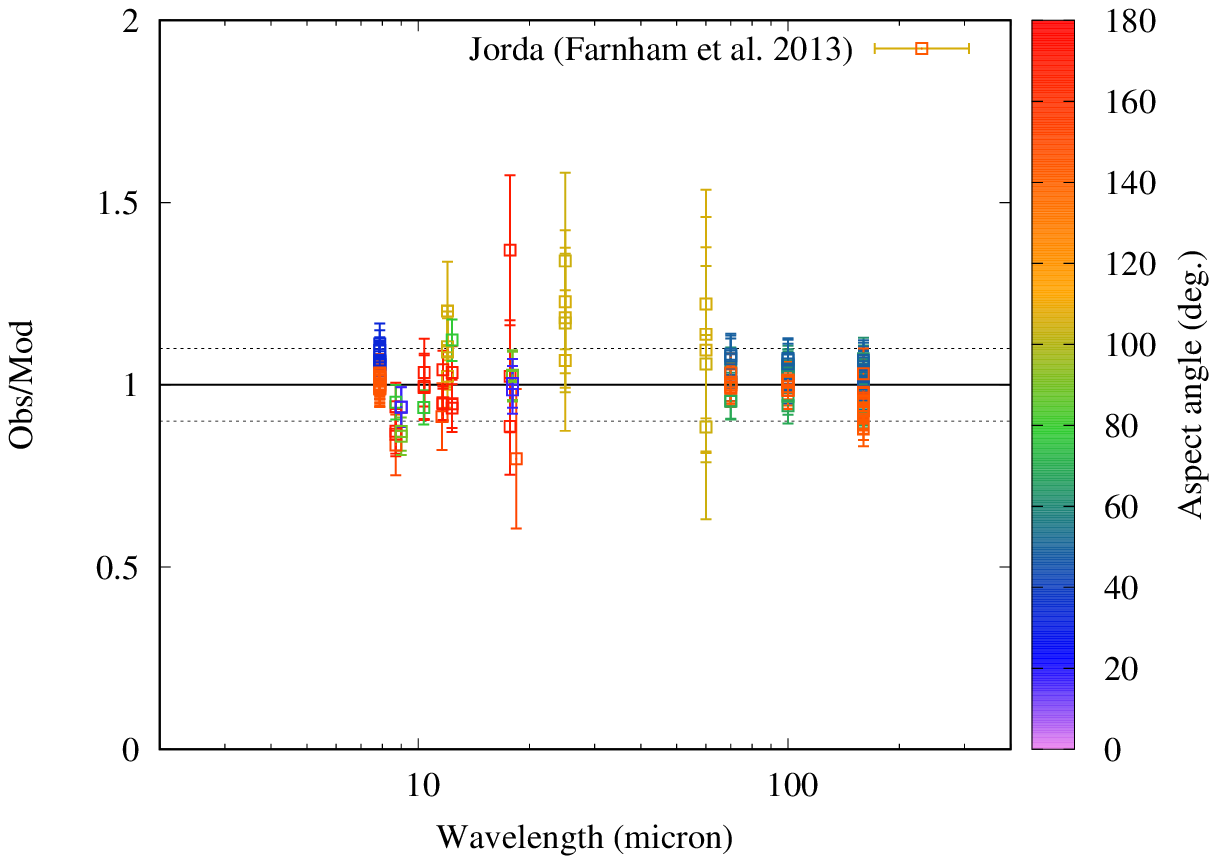}

  \includegraphics[width=0.7\linewidth]{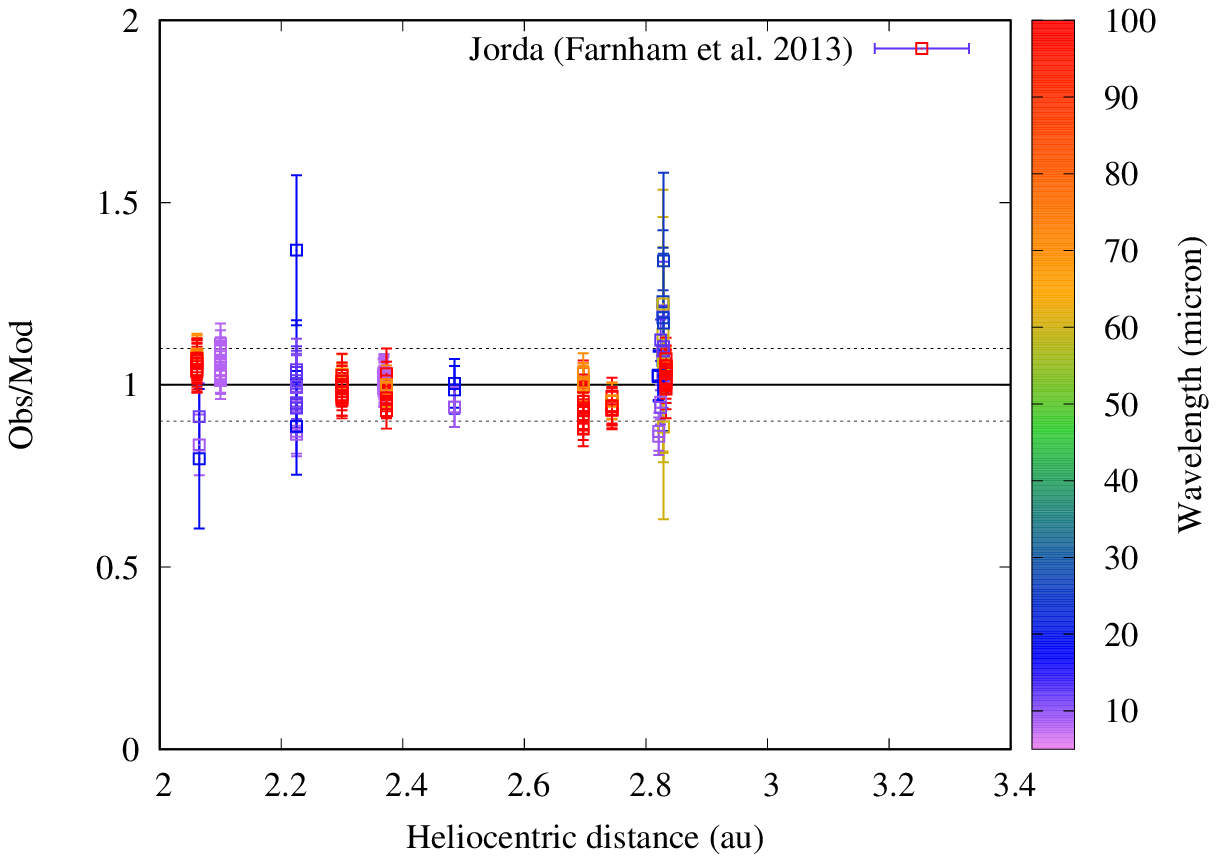}

  \includegraphics[width=0.7\linewidth]{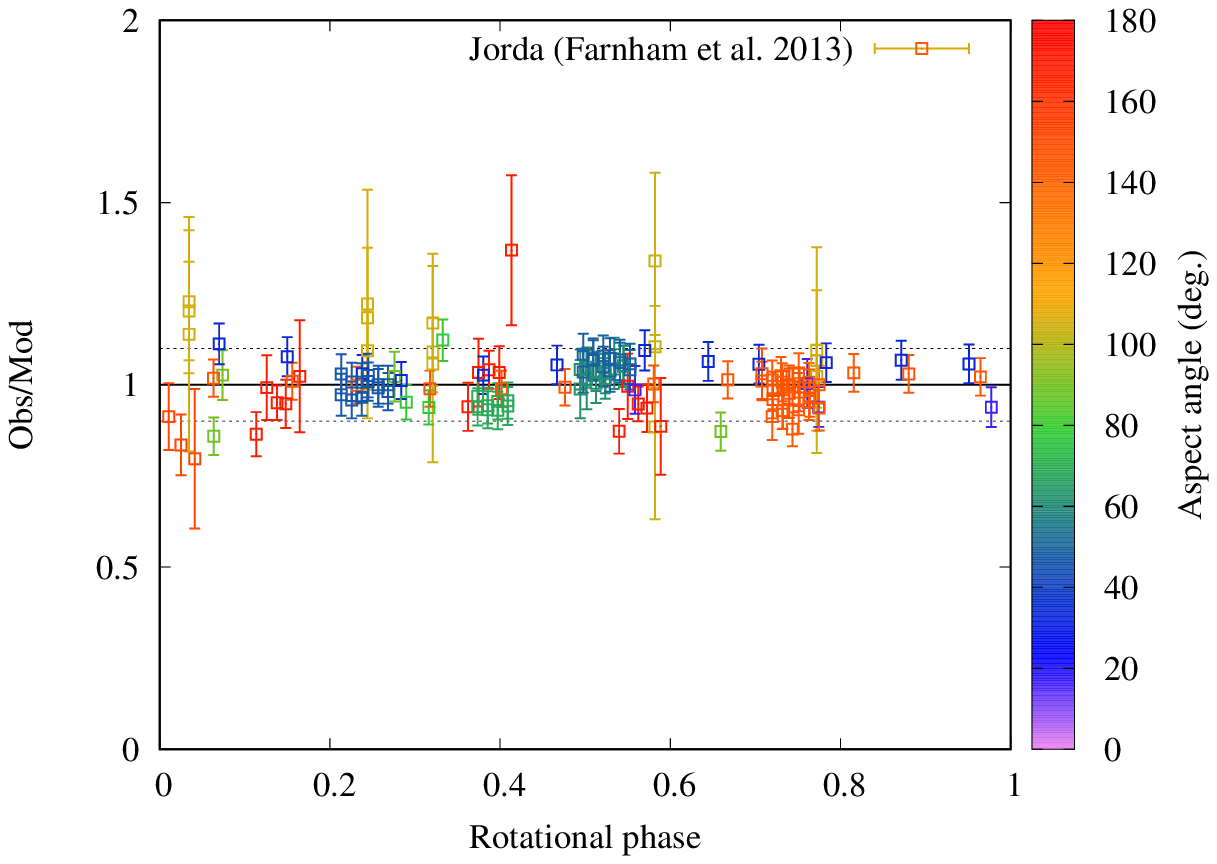}

  \includegraphics[width=0.7\linewidth]{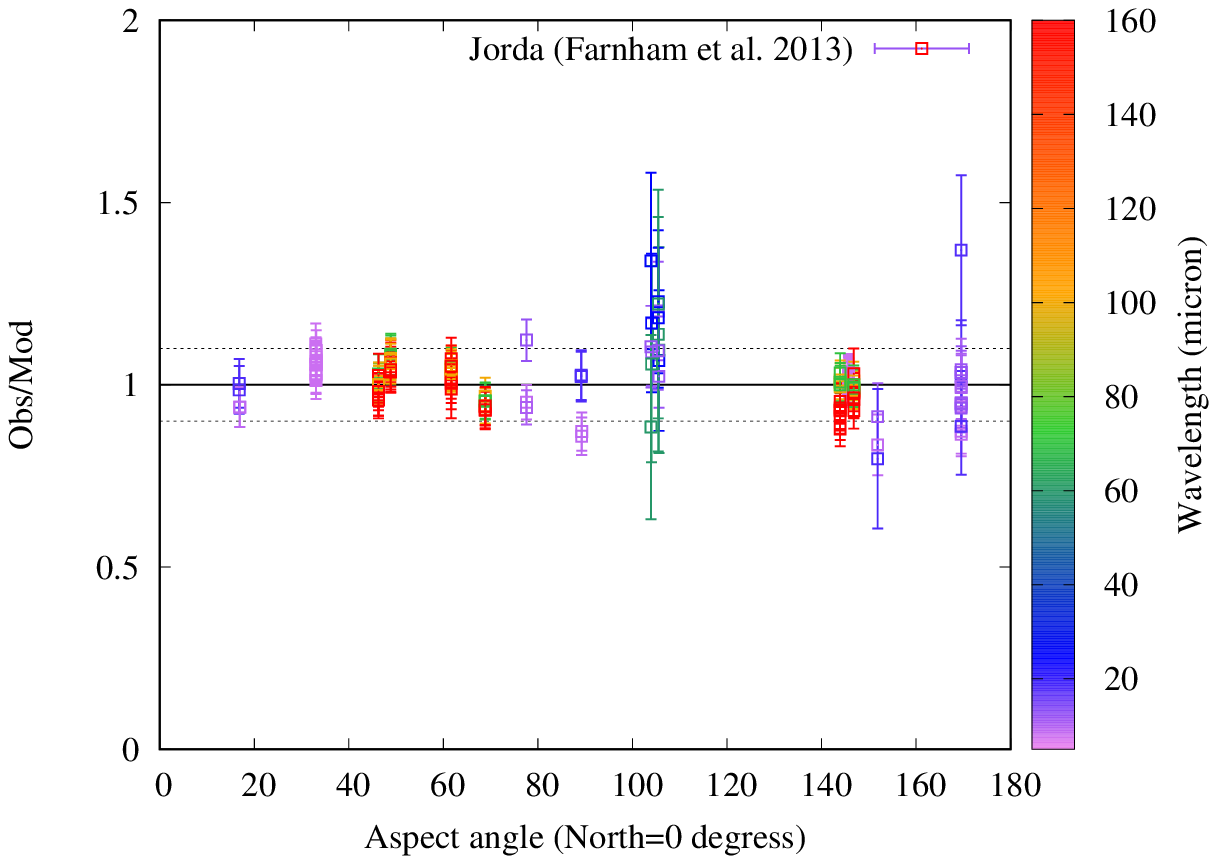}

  \includegraphics[width=0.7\linewidth]{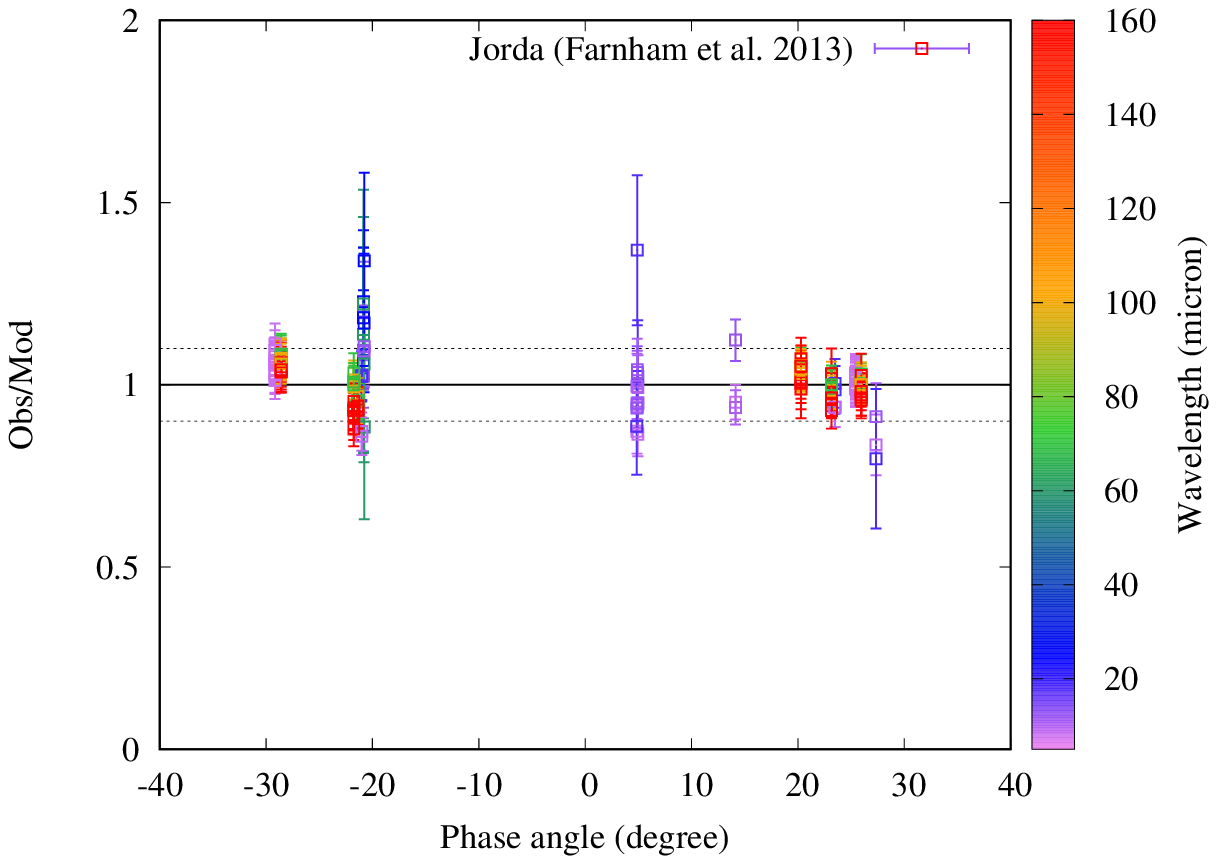}

  \caption{(21) Lutetia. See the caption in Fig.~\ref{fig:00001_OMR}. 
  }\label{fig:00021_OMR}
\end{figure}

For this target we incorporated most of the data featured in
\citet{ORourke2012} in our analysis, for example Spitzer Infrared Array Camera (IRAC) fluxes, with the notable exception of Herschel
SPIRE fluxes, which we did not model in this work. As expected, our results for Lutetia
are fully consistent with the previous work. 
O'Rourke et al. suggest that a localised inaccuracy in the shape model, which is after
all based on data that did not cover 100\% of Lutetia's surface, is responsible for the
inability of the TPM model to fully reproduce the October 17 IRAC thermal light curve
(see their Fig. 4). 
Here we also considered an offset in the rotational
zero-phase as a possible explanation, so we repeated the analysis with a delay of
20 degrees. 
As Fig.~\ref{fig:ThLC} shows, the maxima and minima of the model were better aligned with both IRAC light curves but the overall flux levels of the October 17 one were still not matched. 
Nevertheless, this mismatch is small, on average 5\% (pink OMRs at aspect angle close to 40 degrees in the fourth panel of Fig.~\ref{fig:00021_OMR}), and a possible localised inaccuracy in the shape-model still cannot be ruled out.

\begin{figure}
  \centering
  \includegraphics[width=0.7\linewidth]{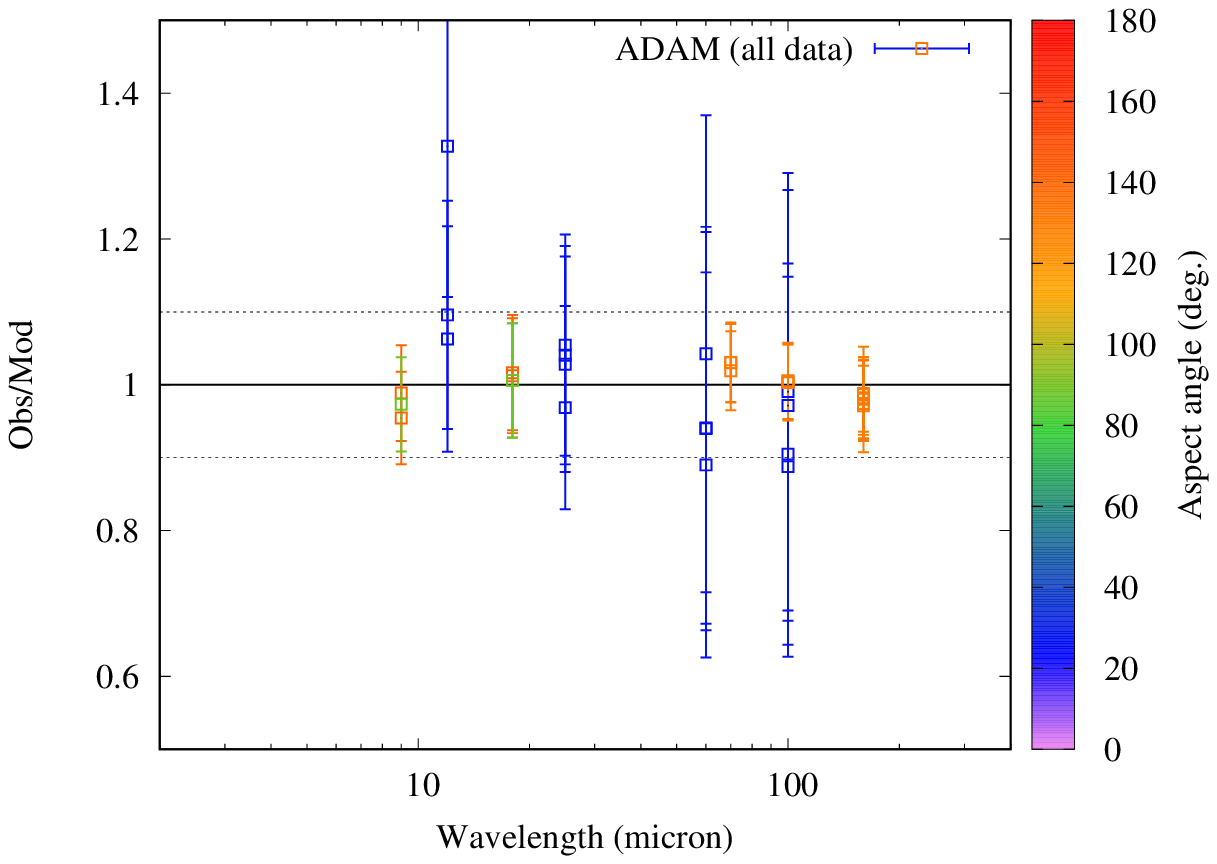}

  \includegraphics[width=0.7\linewidth]{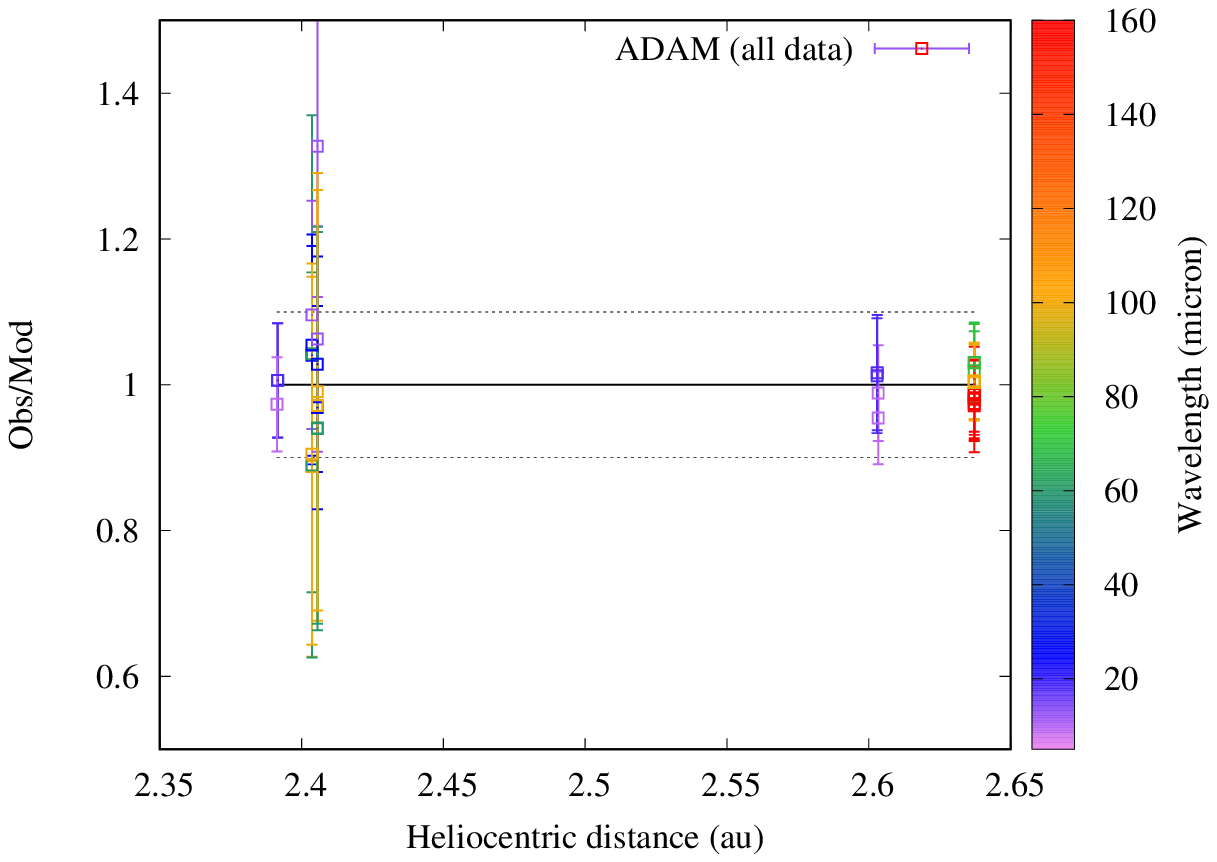}

  \includegraphics[width=0.7\linewidth]{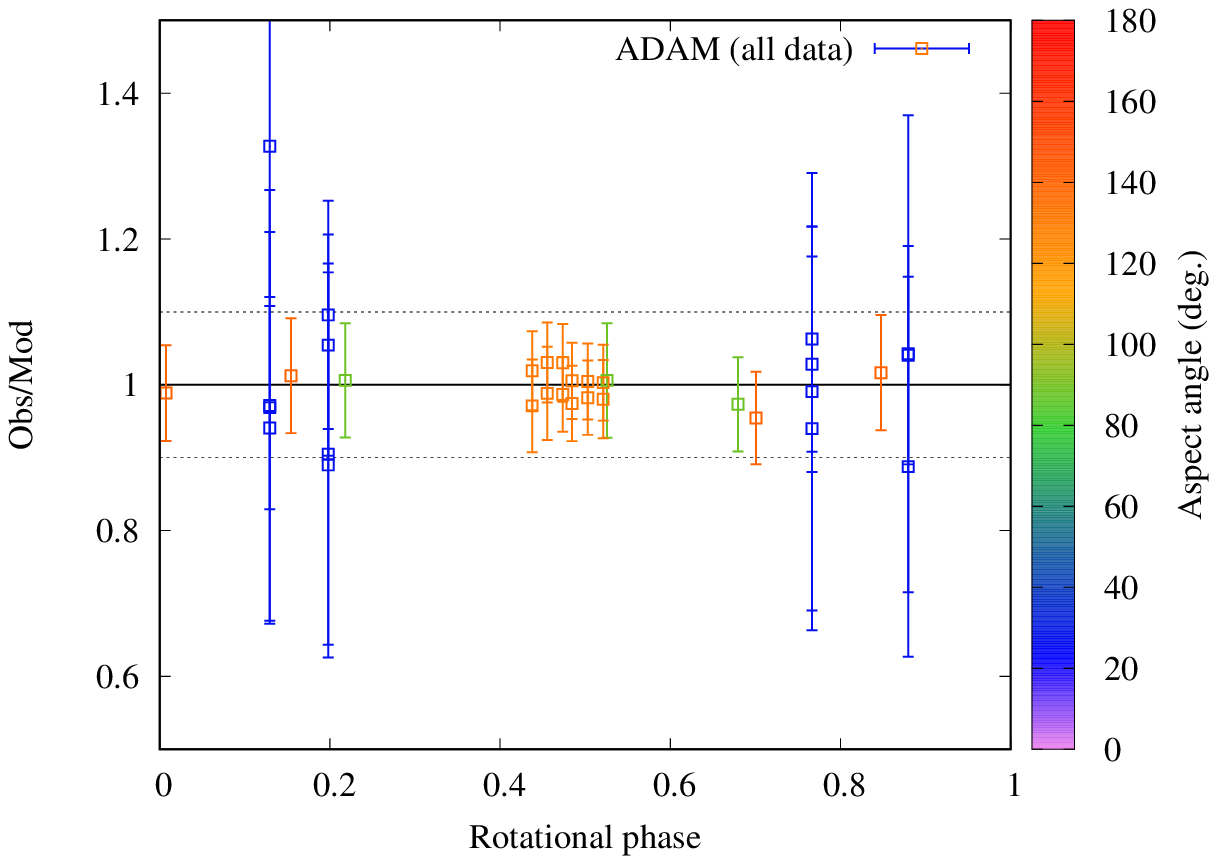}

  \includegraphics[width=0.7\linewidth]{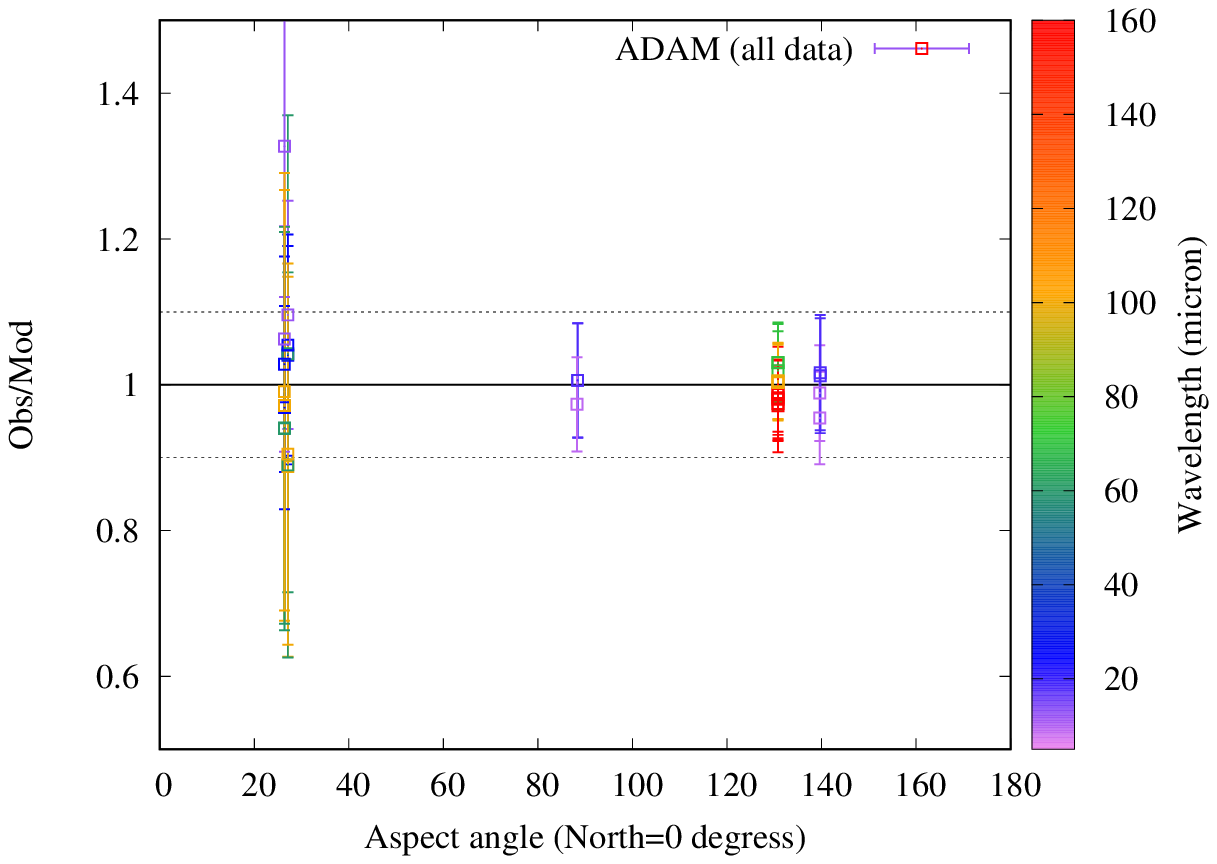}

  \includegraphics[width=0.7\linewidth]{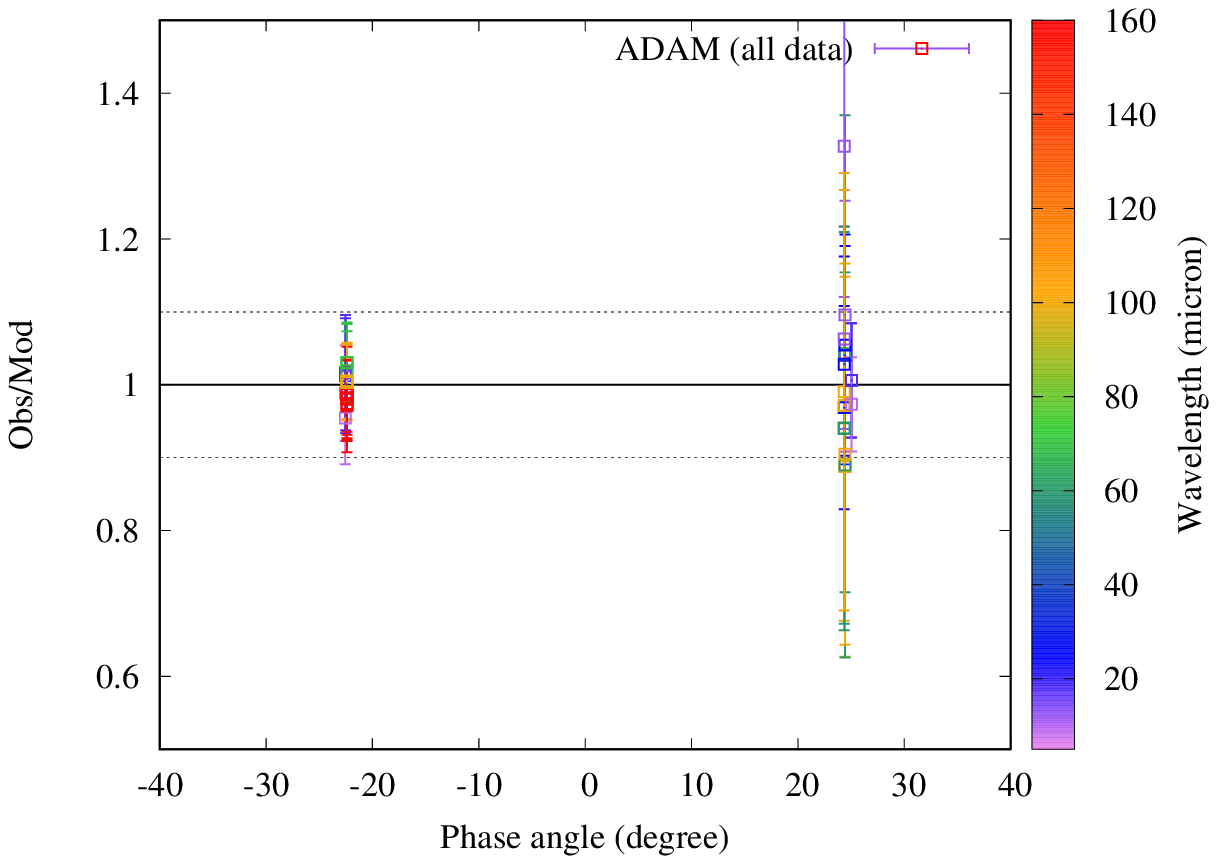}

  \caption{(29) Amphitrite. See the caption in Fig.~\ref{fig:00001_OMR}. 
  }\label{fig:00029_OMR}
\end{figure}

\begin{figure}
  \centering
  \includegraphics[width=0.7\linewidth]{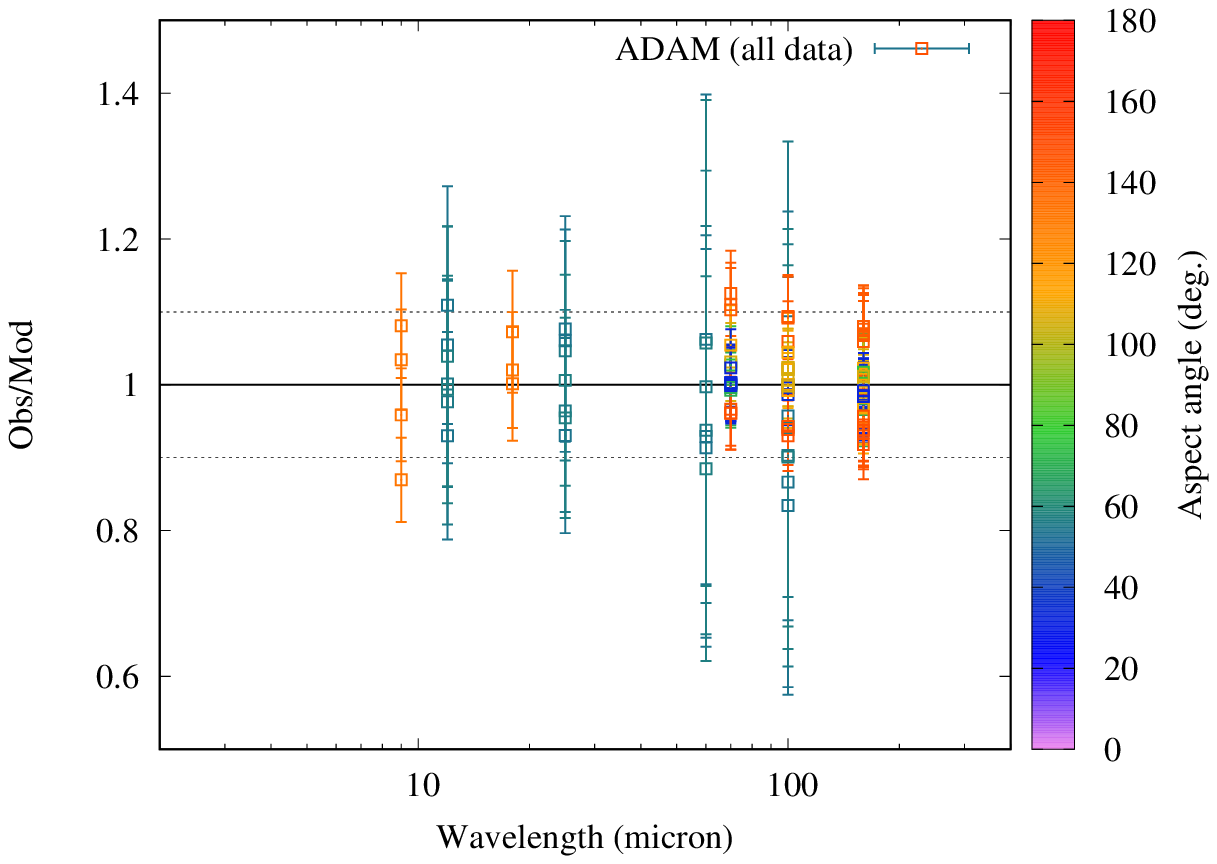}

  \includegraphics[width=0.7\linewidth]{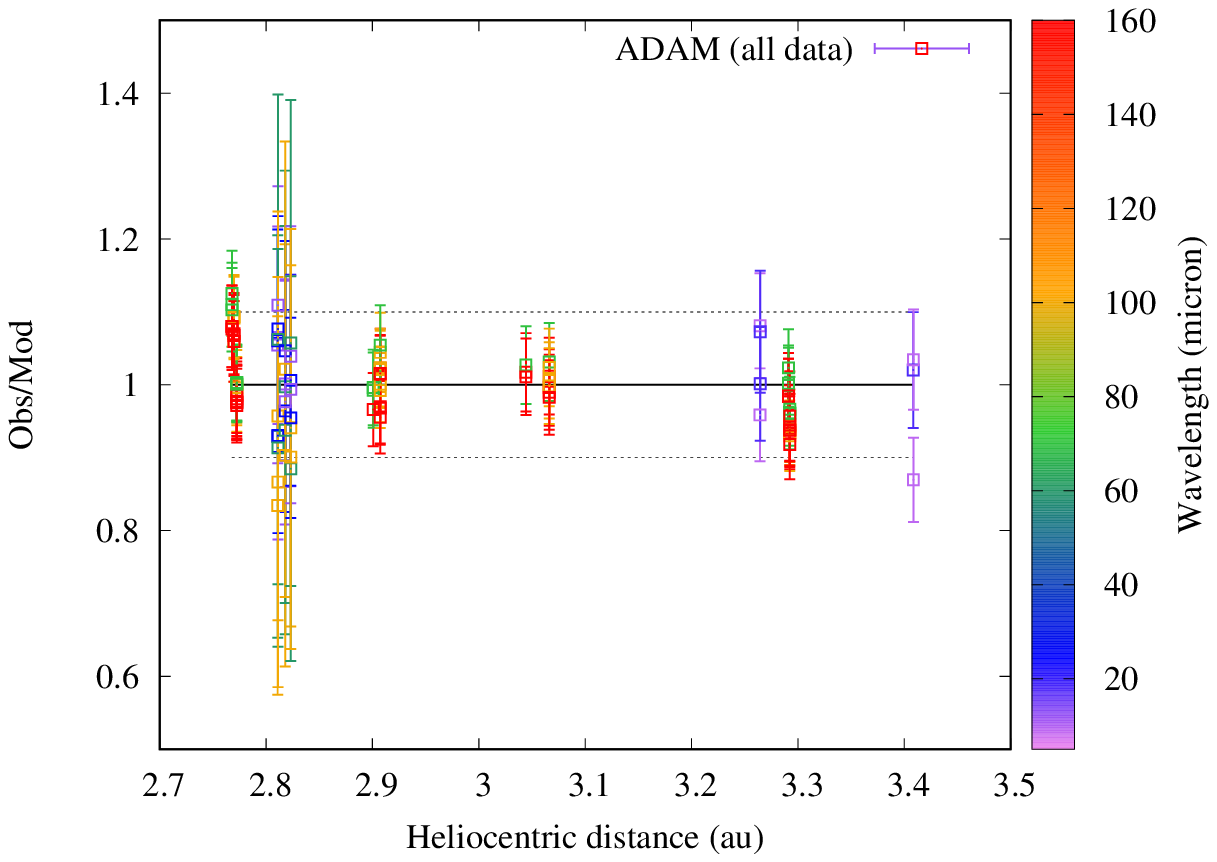}

  \includegraphics[width=0.7\linewidth]{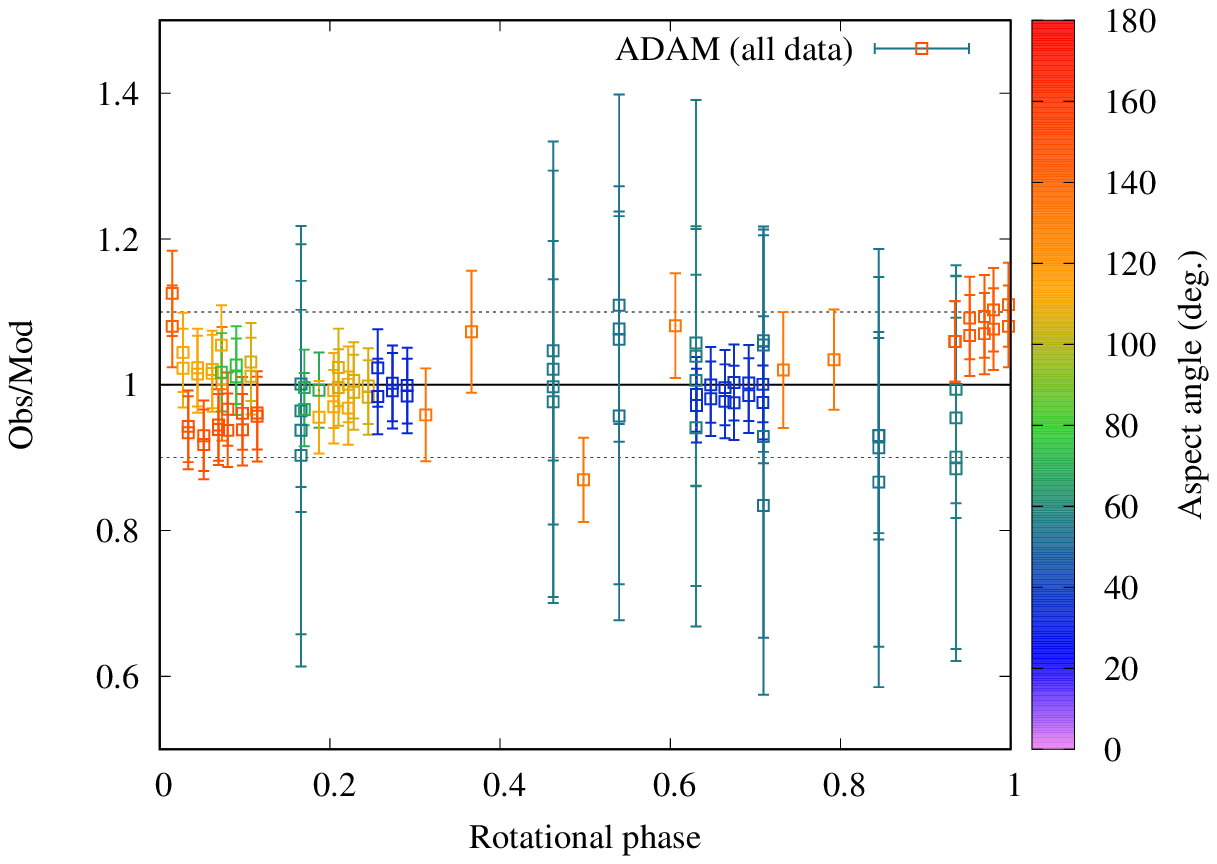}

  \includegraphics[width=0.7\linewidth]{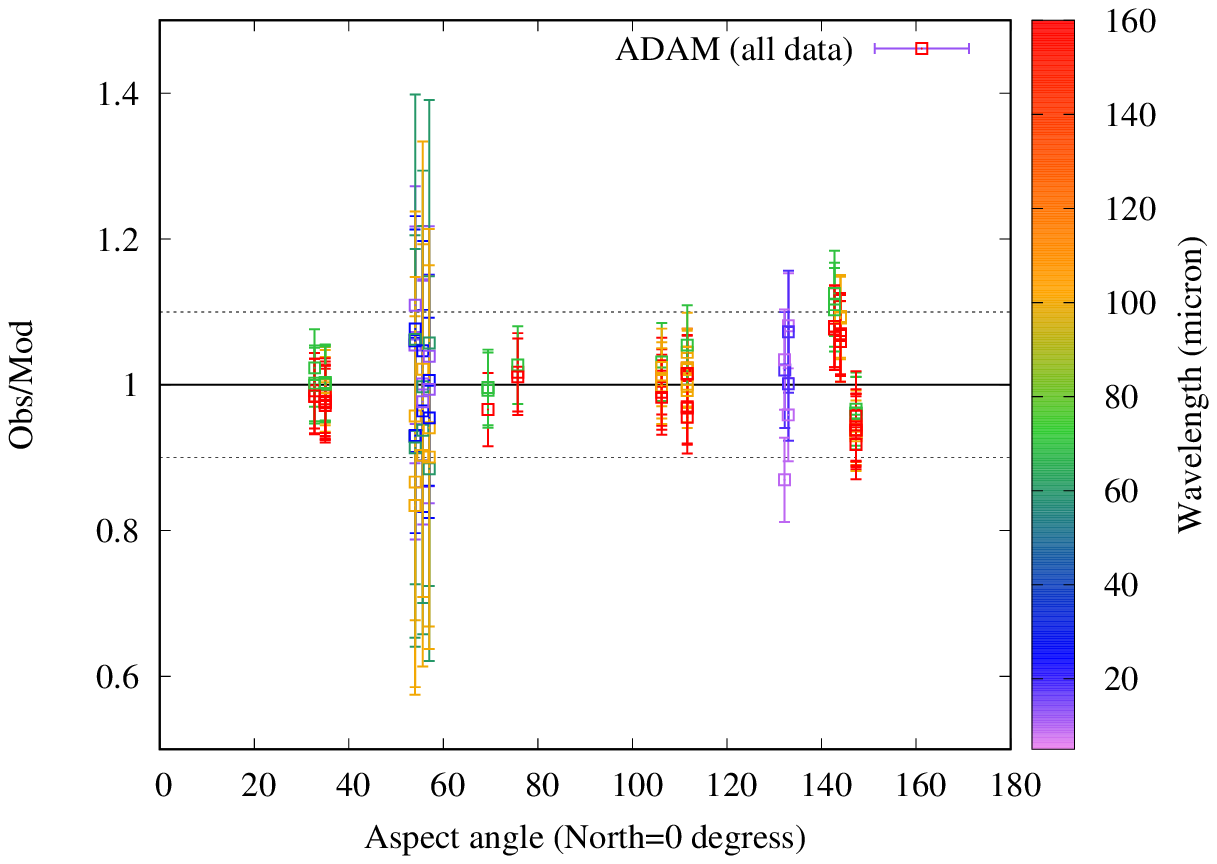}

  \includegraphics[width=0.7\linewidth]{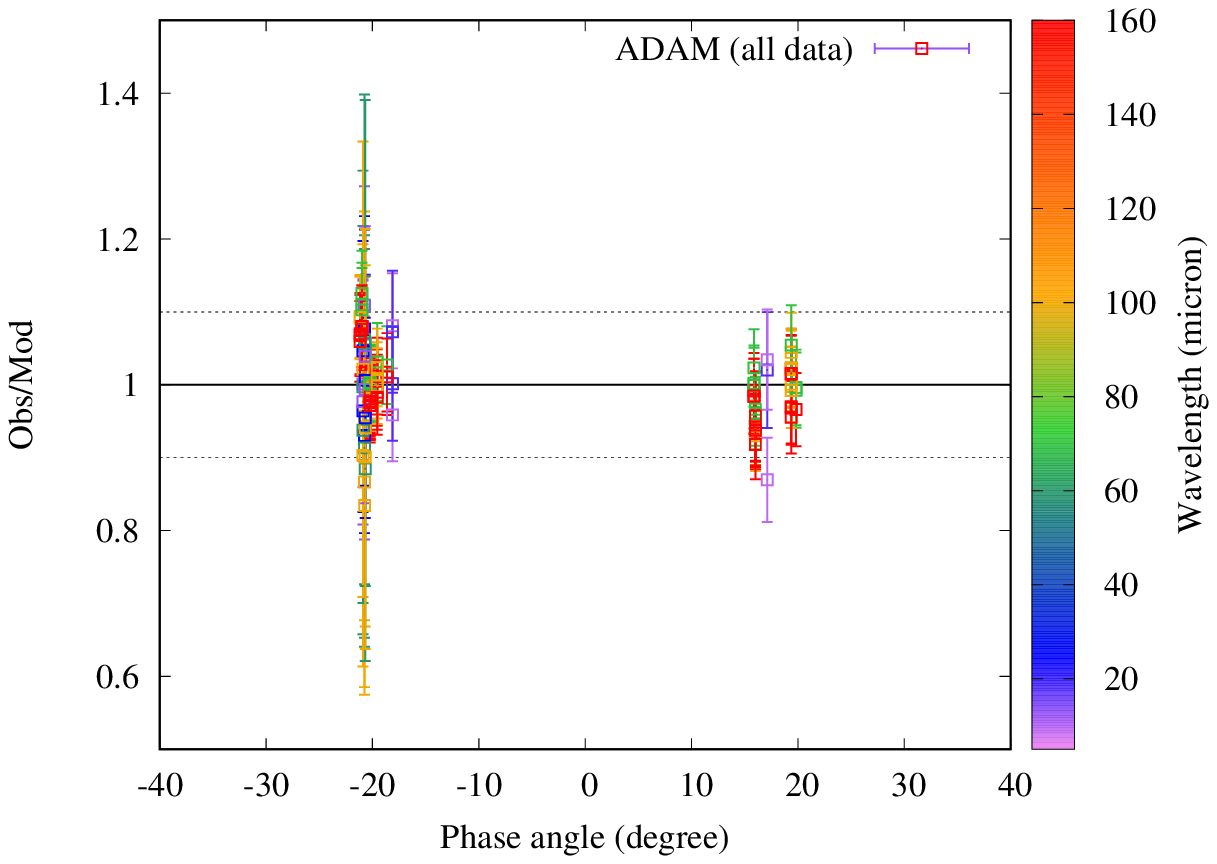}

  \caption{(52) Europa. See the caption in Fig.~\ref{fig:00001_OMR}. 
  }\label{fig:00052_OMR}
\end{figure}

\begin{figure}
  \centering
  \includegraphics[width=0.7\linewidth]{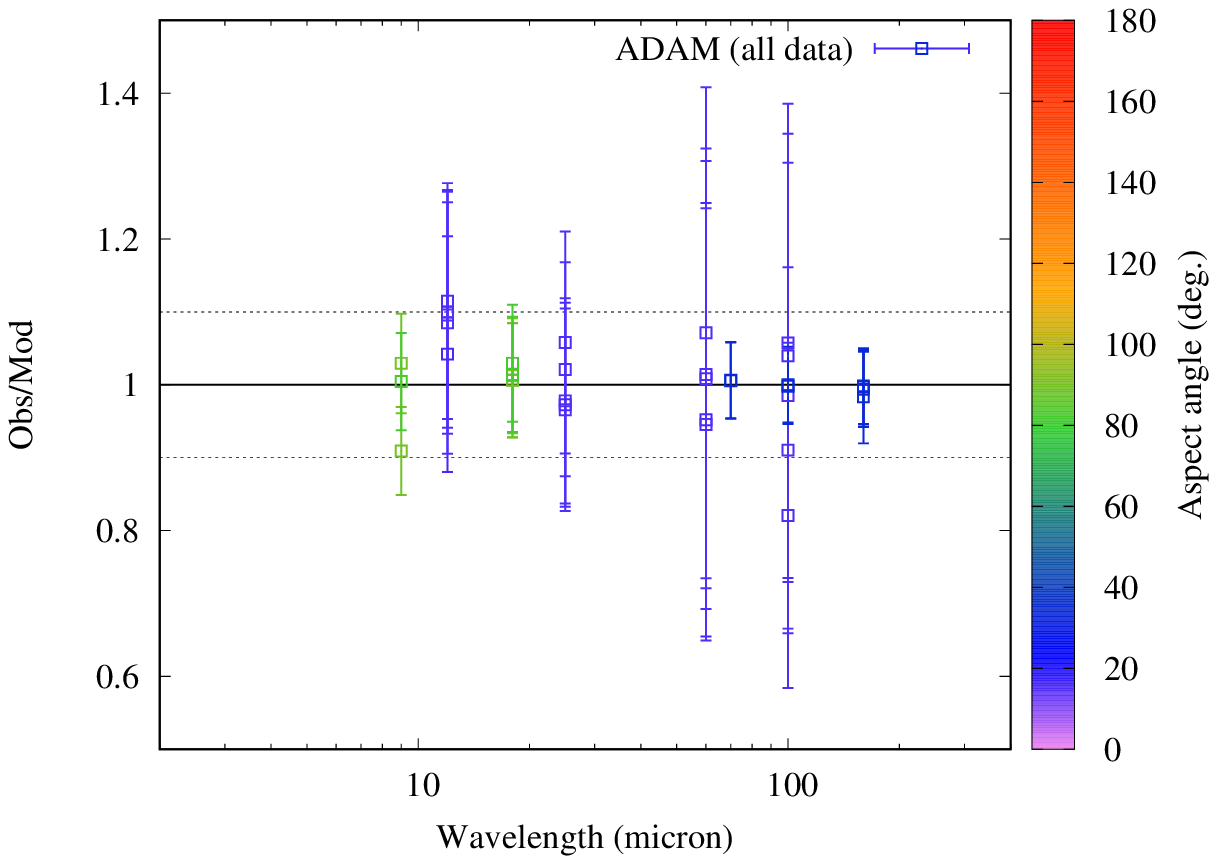}

  \includegraphics[width=0.7\linewidth]{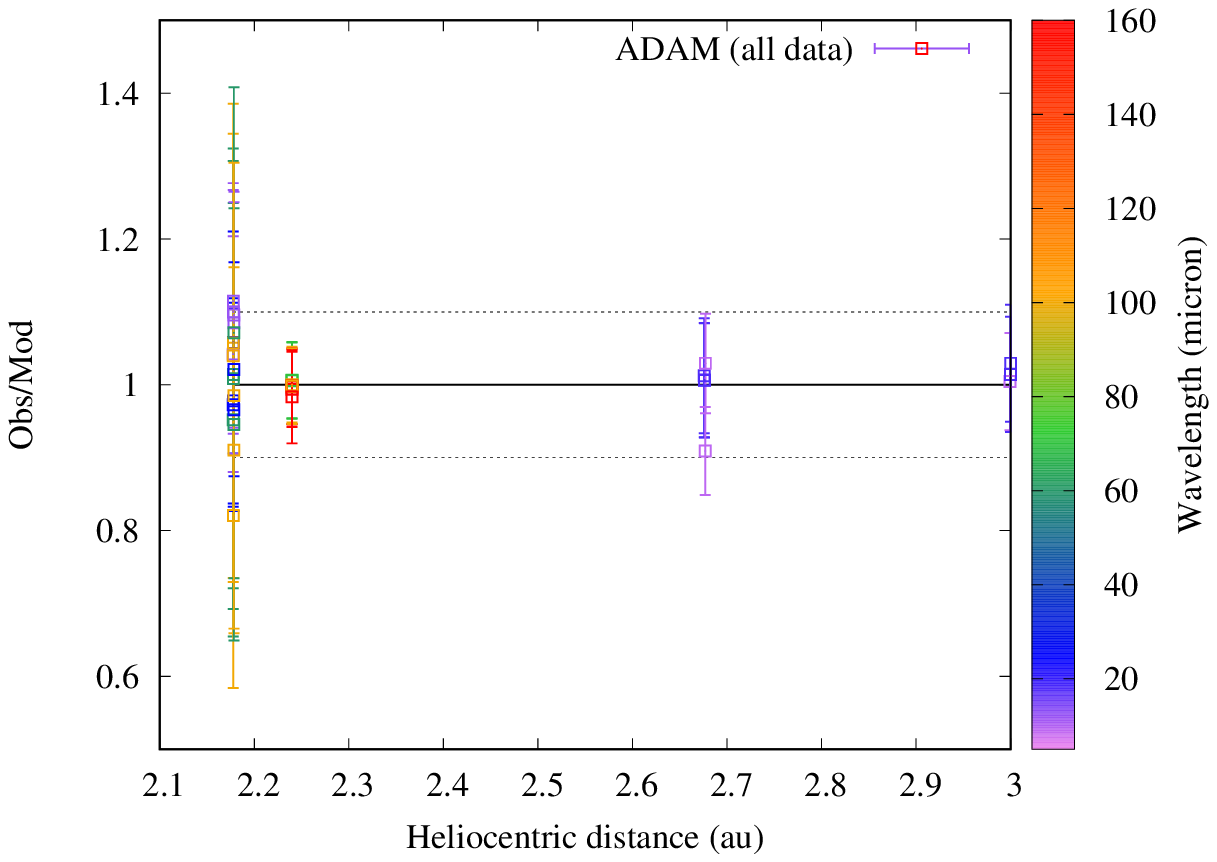}

  \includegraphics[width=0.7\linewidth]{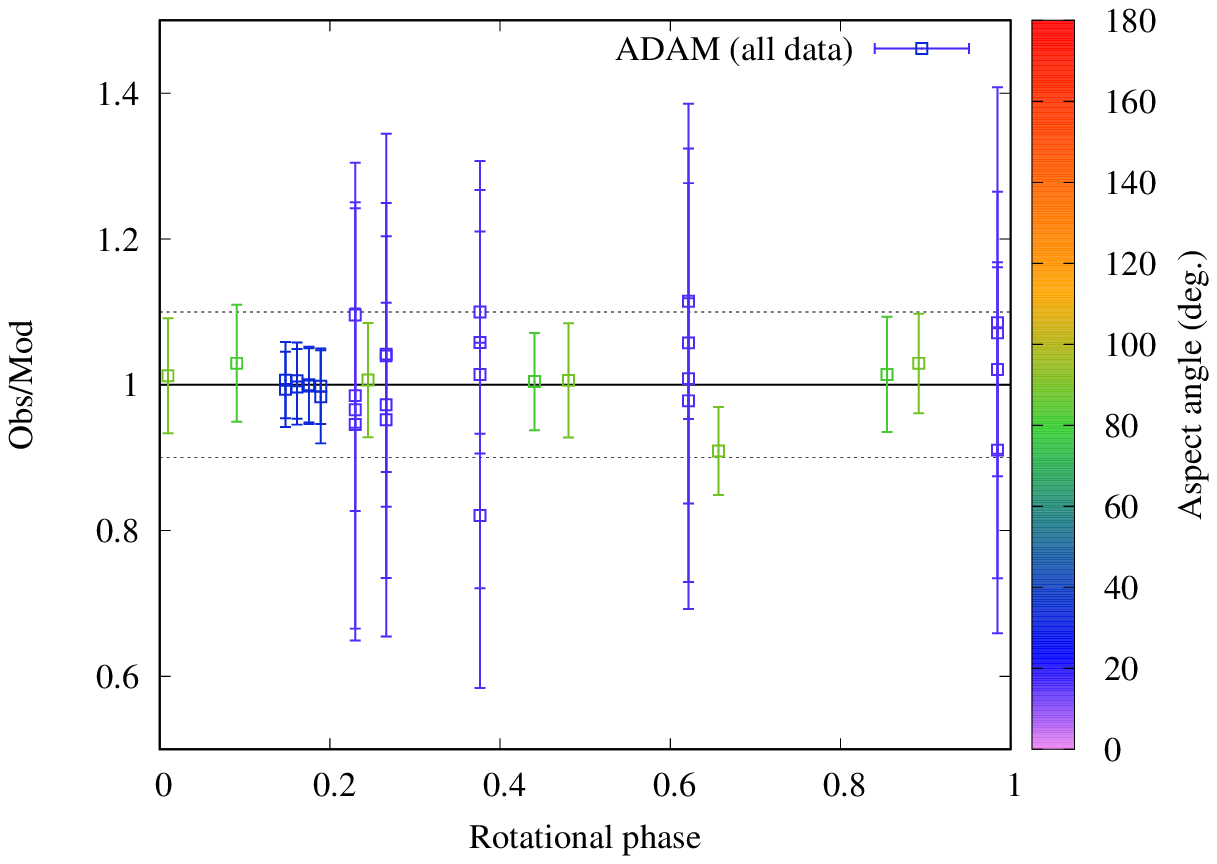}

  \includegraphics[width=0.7\linewidth]{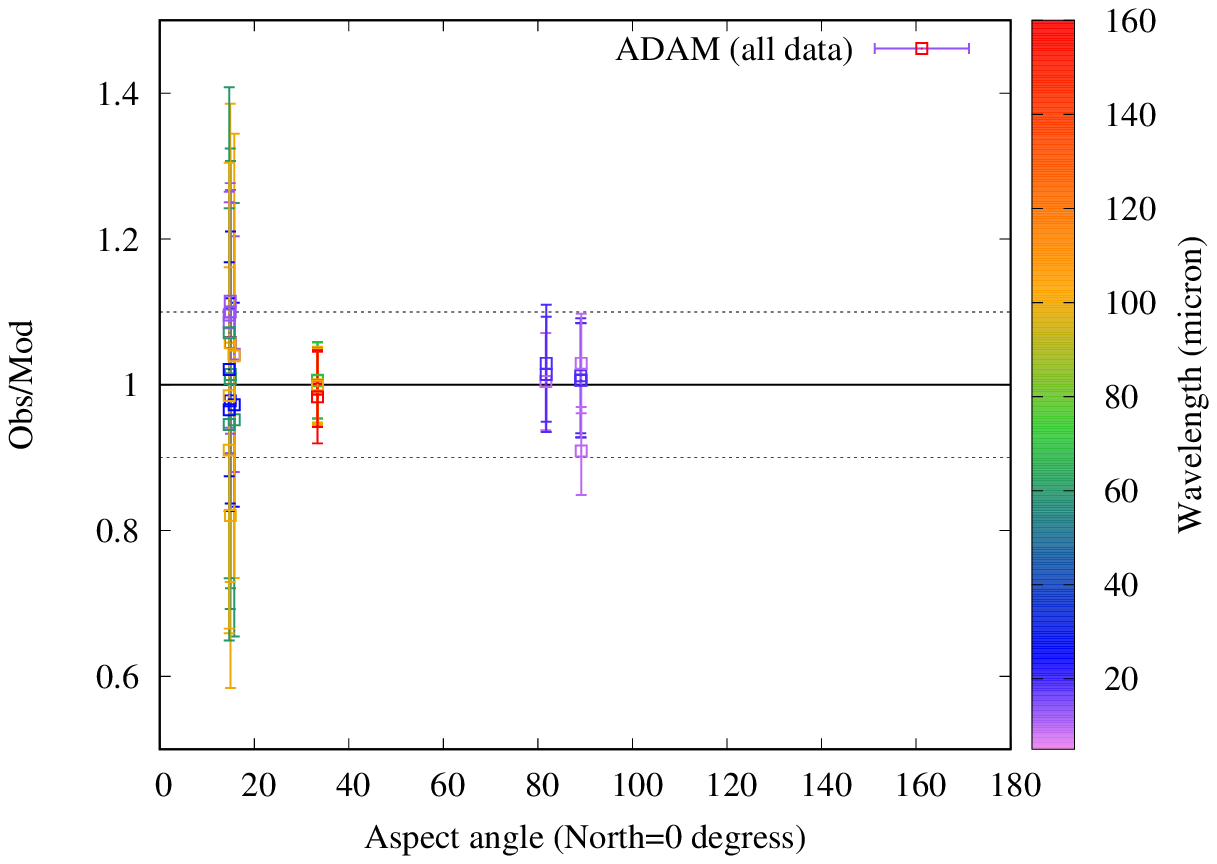}

  \includegraphics[width=0.7\linewidth]{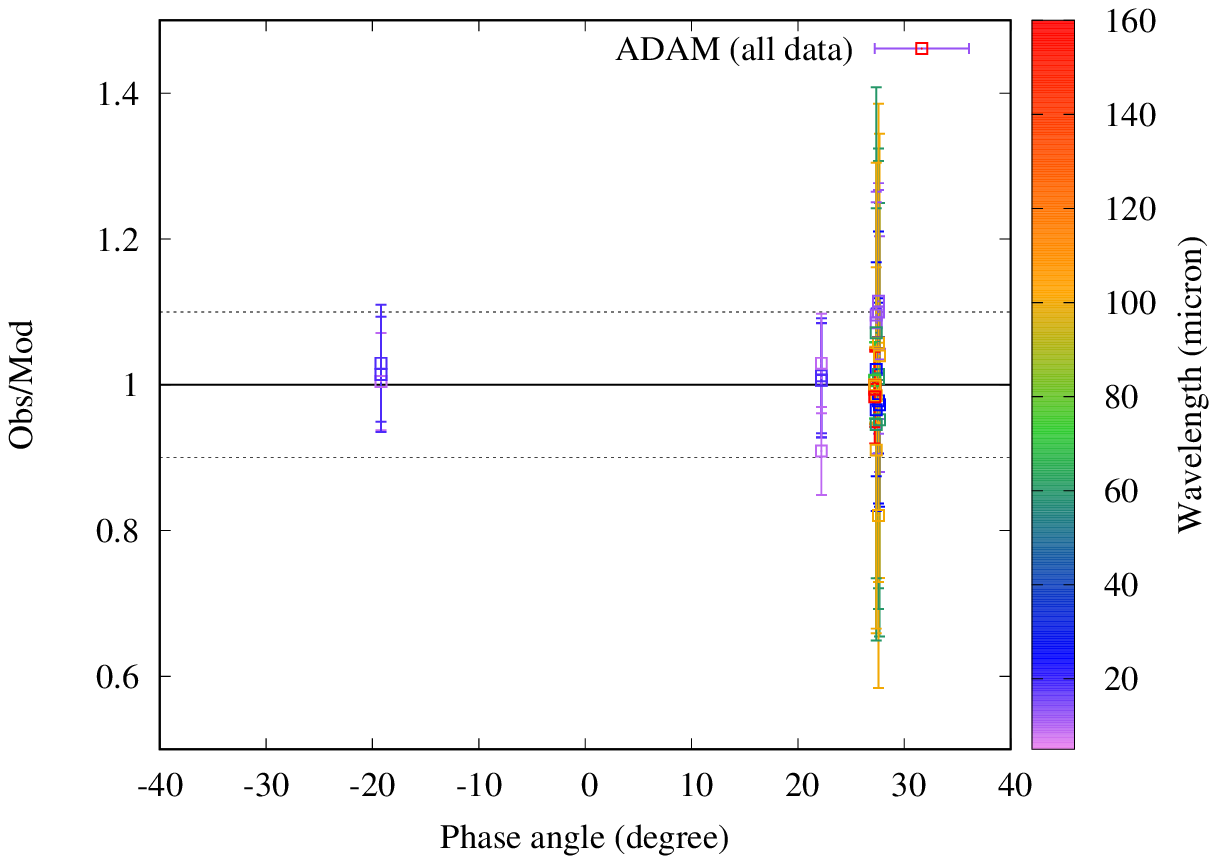}

  \caption{(54) Alexandra. See the caption in Fig.~\ref{fig:00001_OMR}. 
  }\label{fig:00054_OMR}
\end{figure}

\begin{figure}
  \centering
  \includegraphics[width=0.7\linewidth]{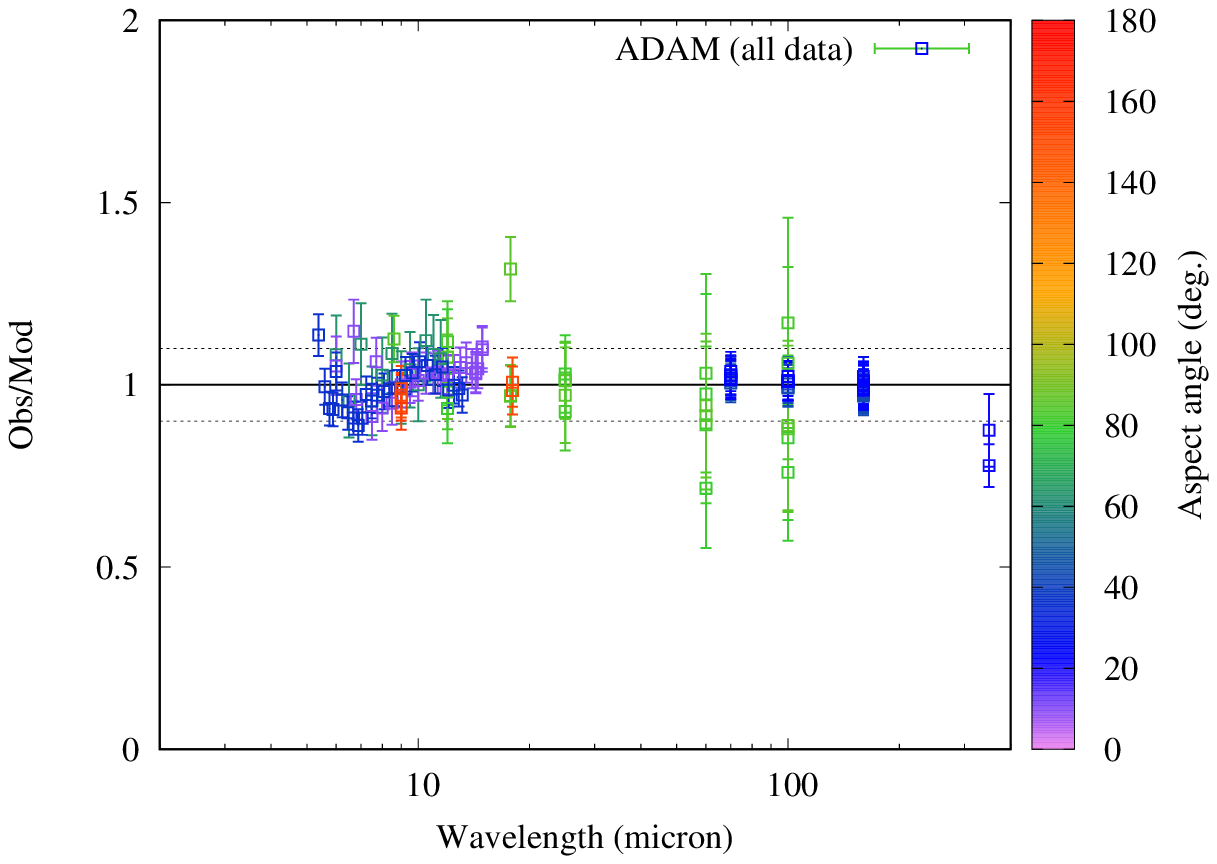}

  \includegraphics[width=0.7\linewidth]{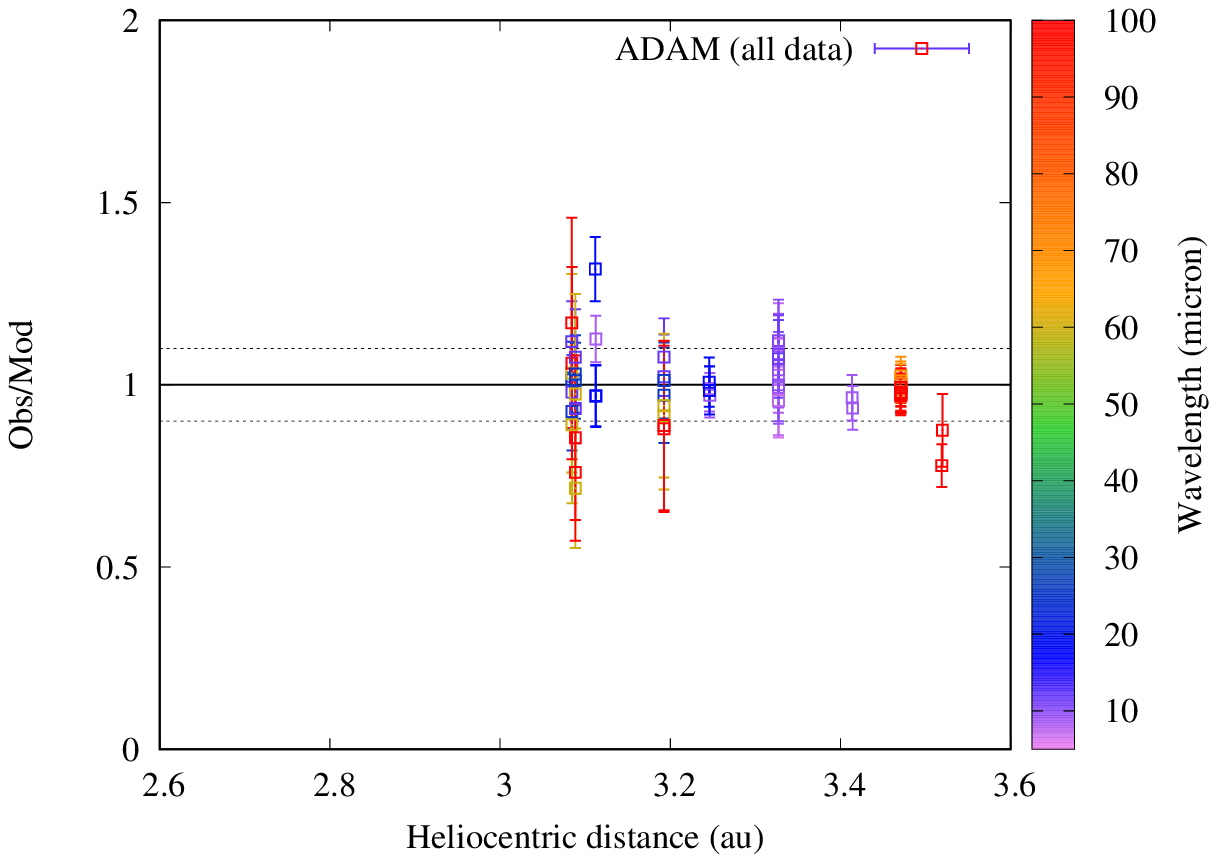}

  \includegraphics[width=0.7\linewidth]{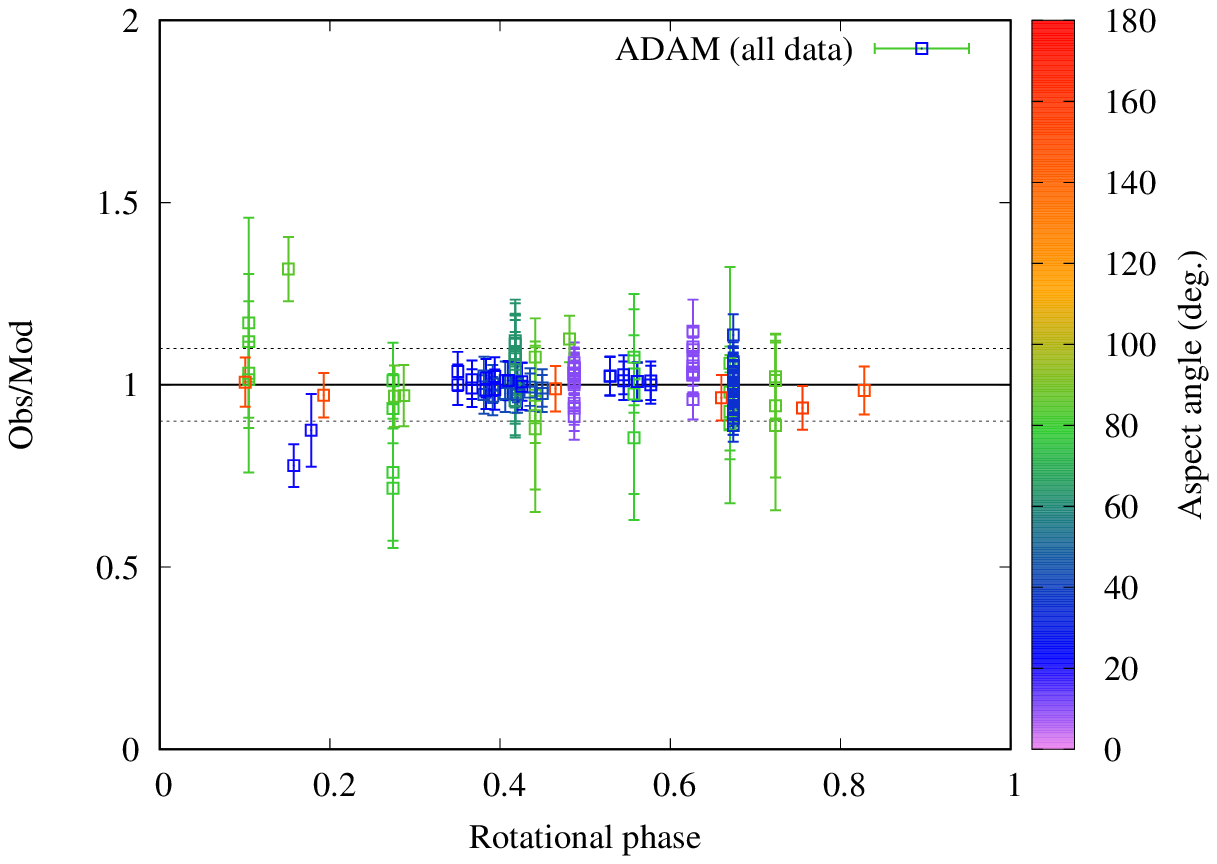}

  \includegraphics[width=0.7\linewidth]{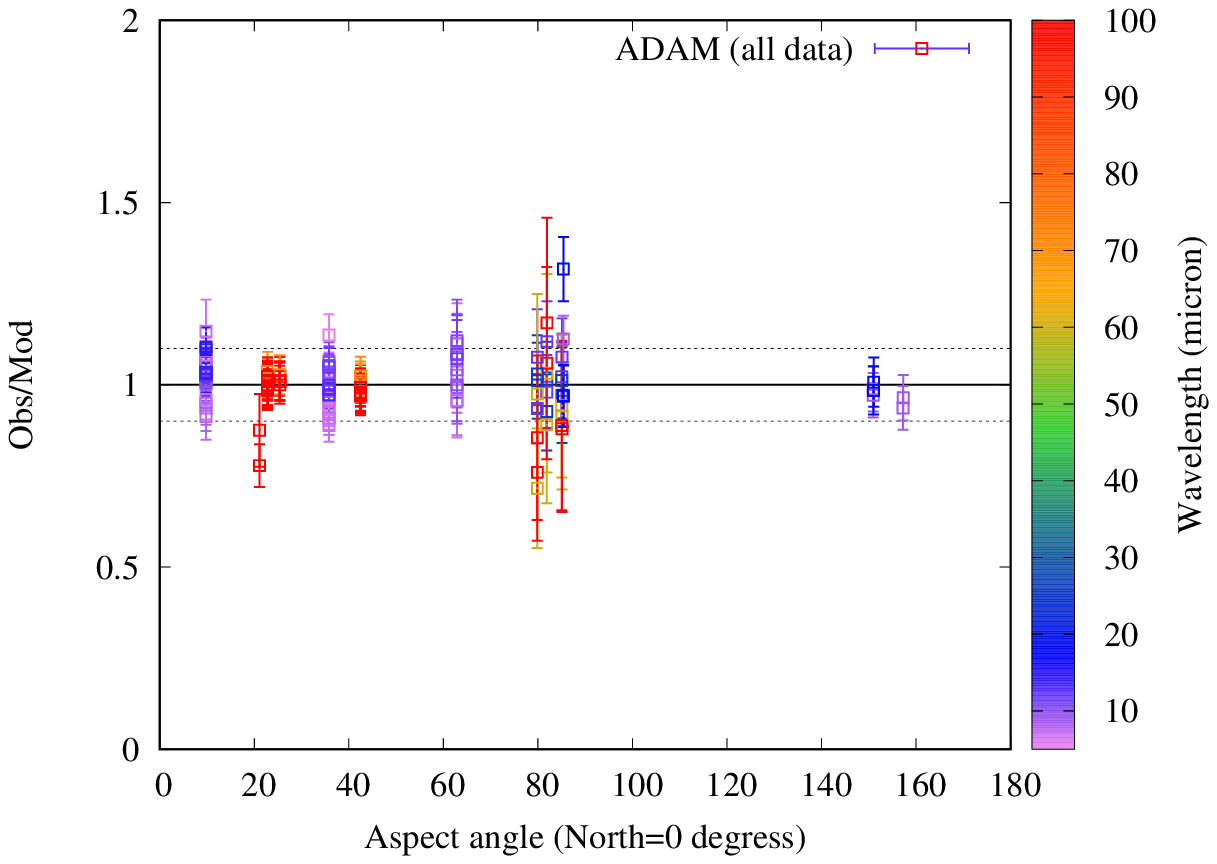}

  \includegraphics[width=0.7\linewidth]{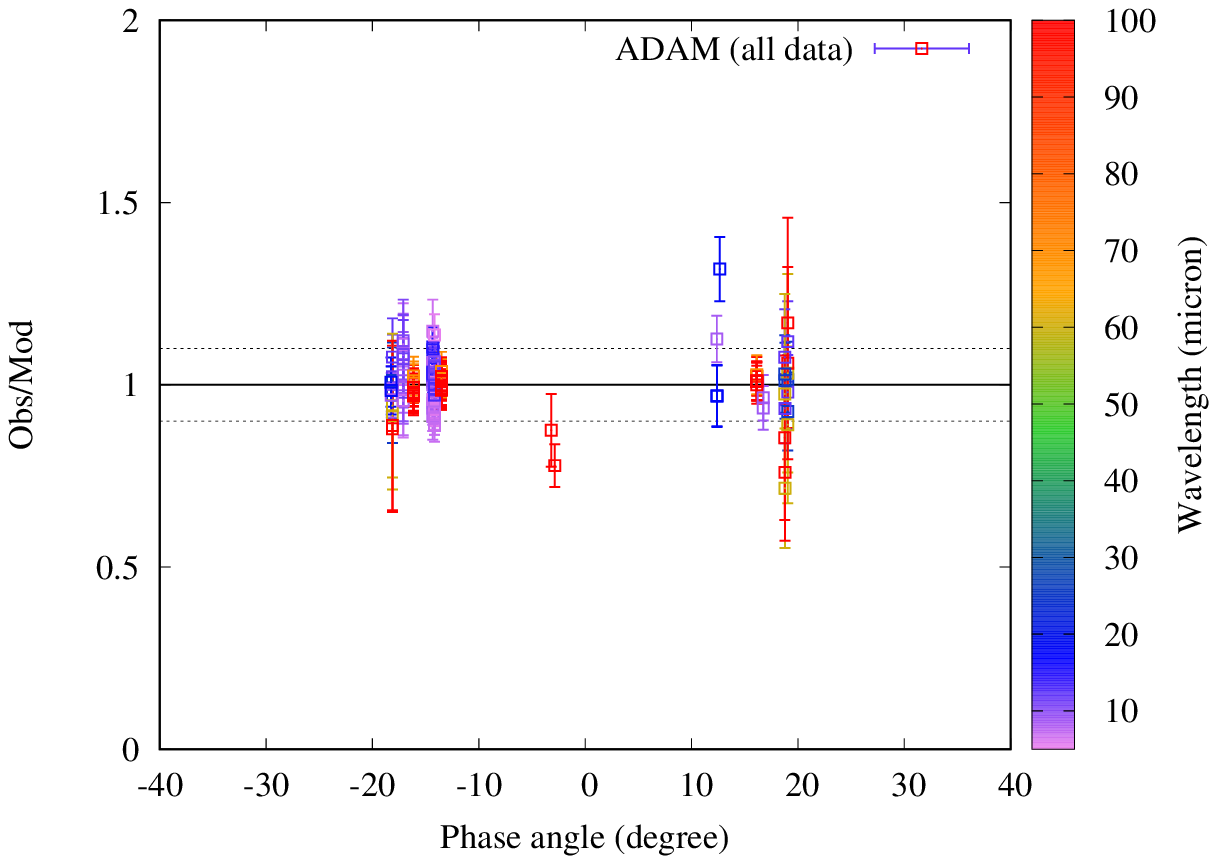}

  \caption{(65) Cybele. See the caption in Fig.~\ref{fig:00001_OMR}. 
  }\label{fig:00065_OMR}
\end{figure}

\begin{figure}
  \centering
  \includegraphics[width=0.7\linewidth]{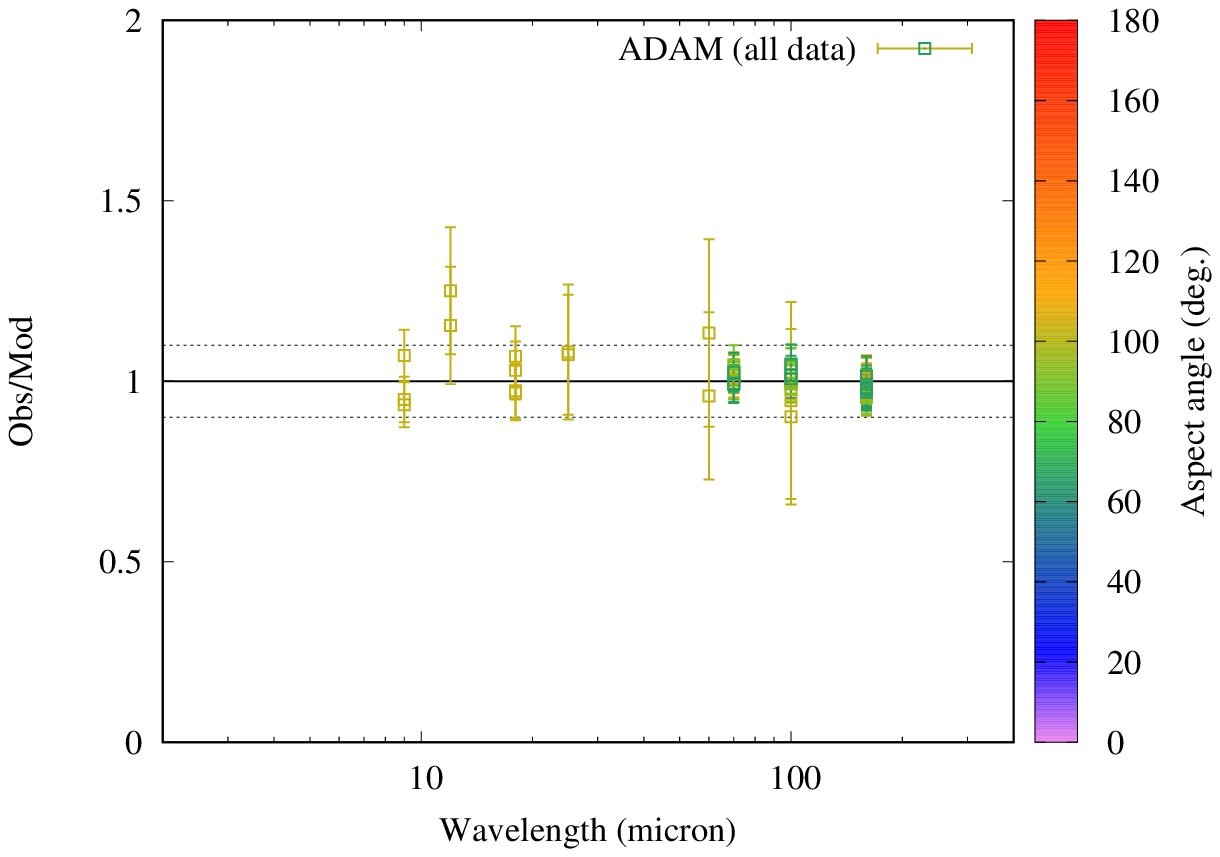}

  \includegraphics[width=0.7\linewidth]{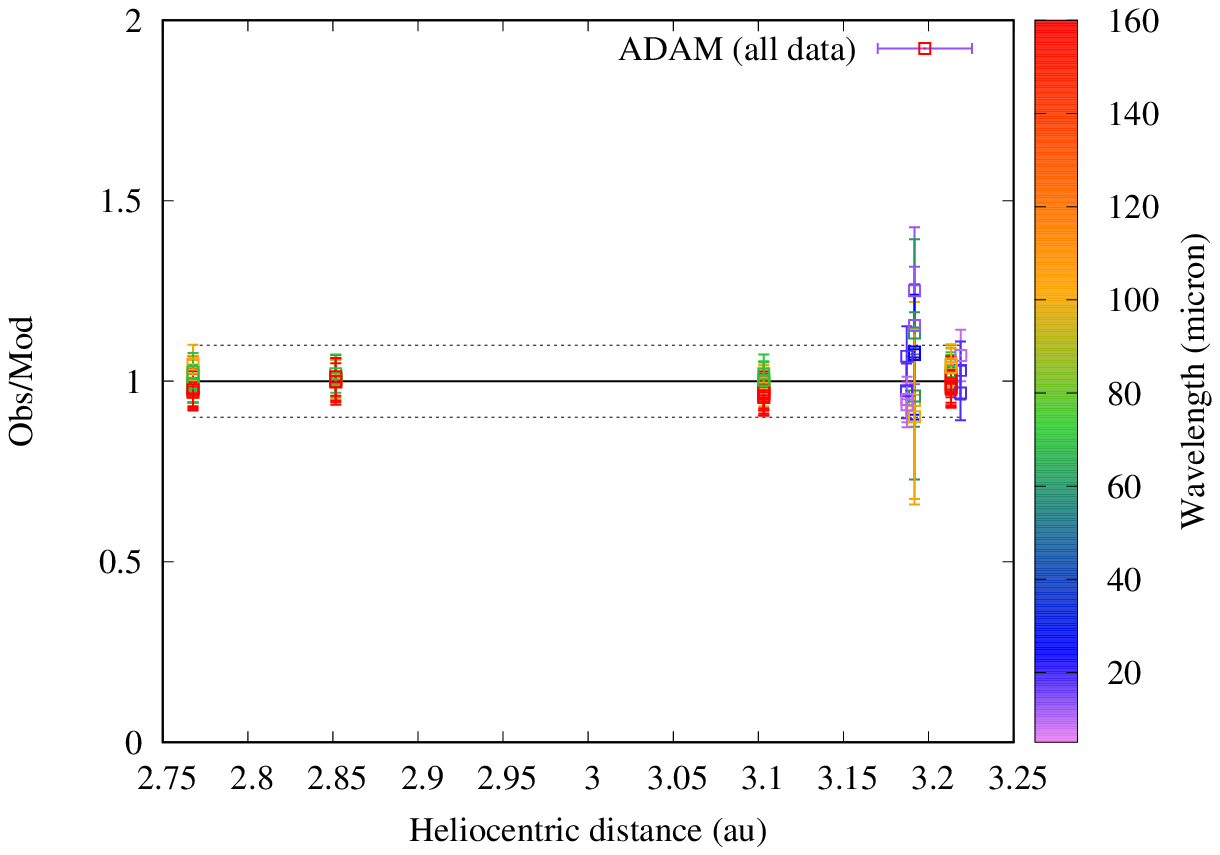}

  \includegraphics[width=0.7\linewidth]{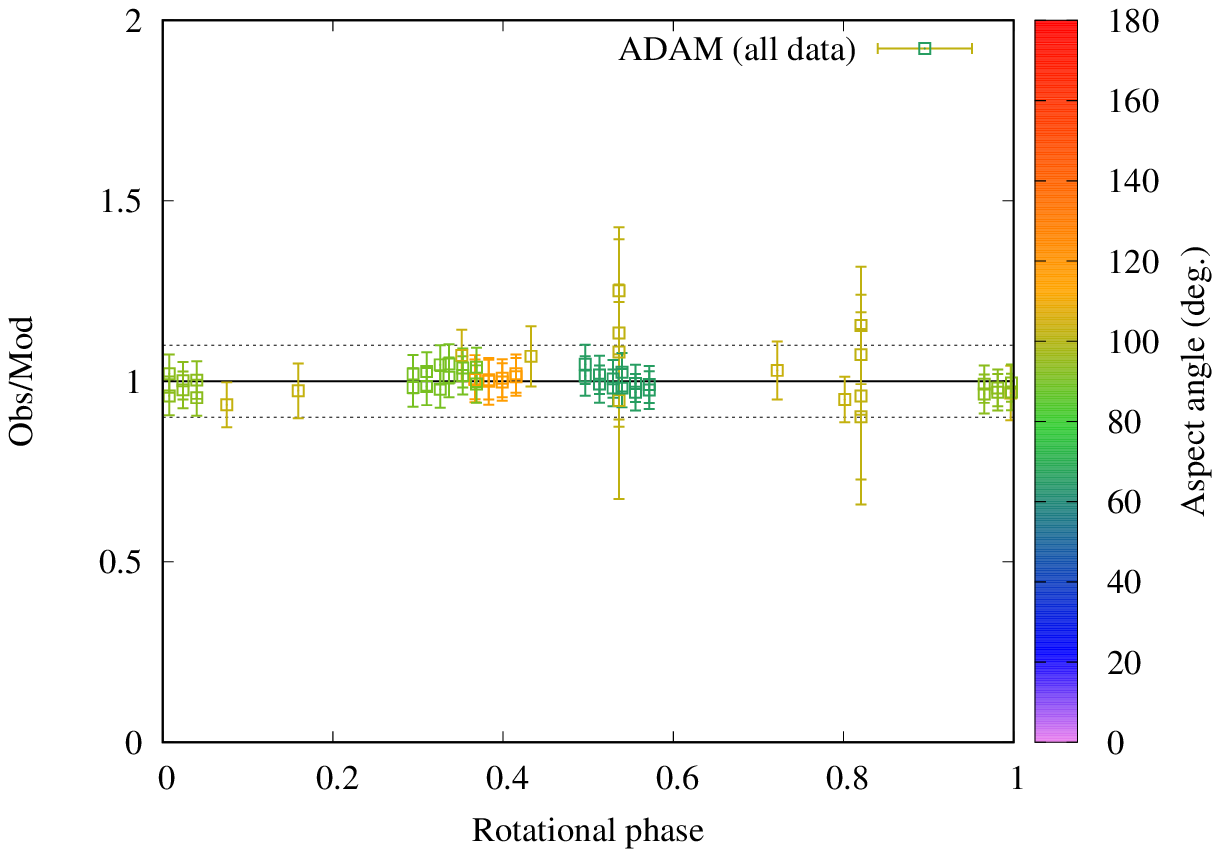}

  \includegraphics[width=0.7\linewidth]{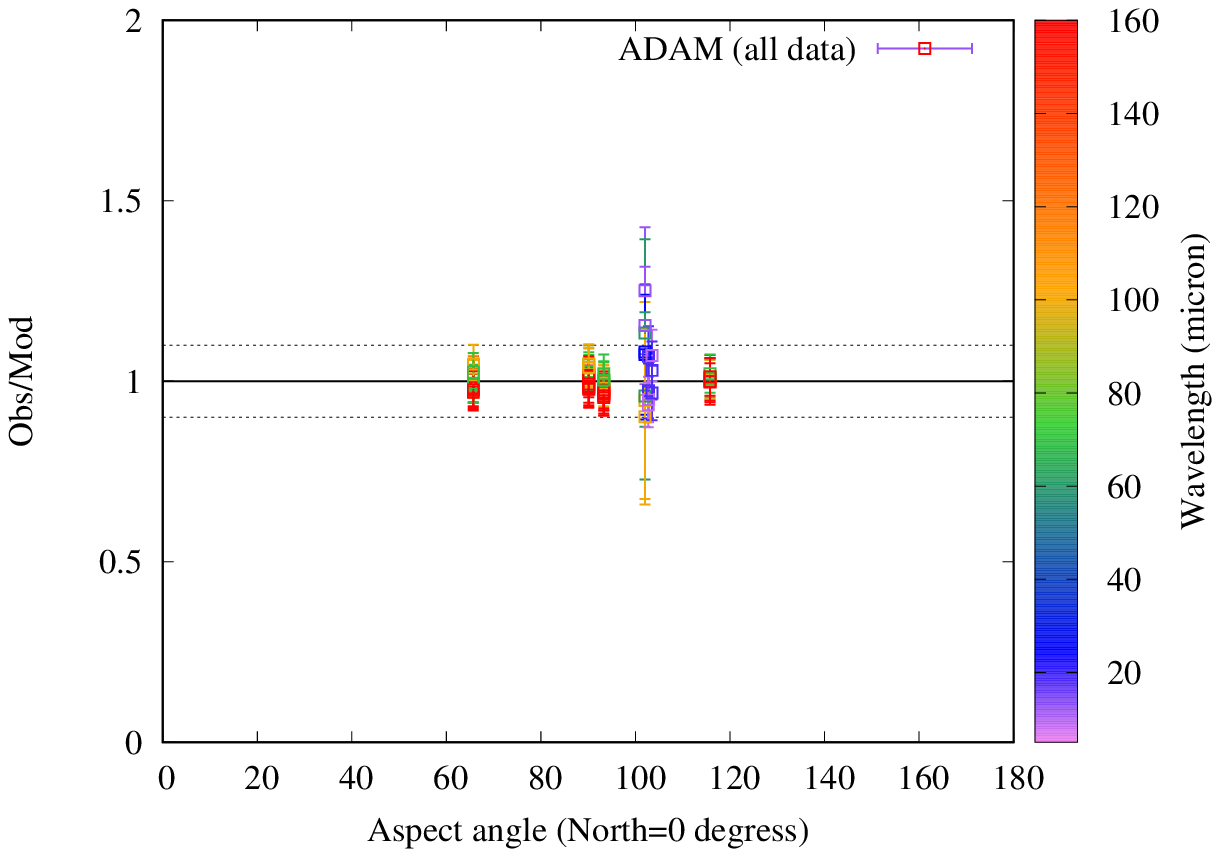}

  \includegraphics[width=0.7\linewidth]{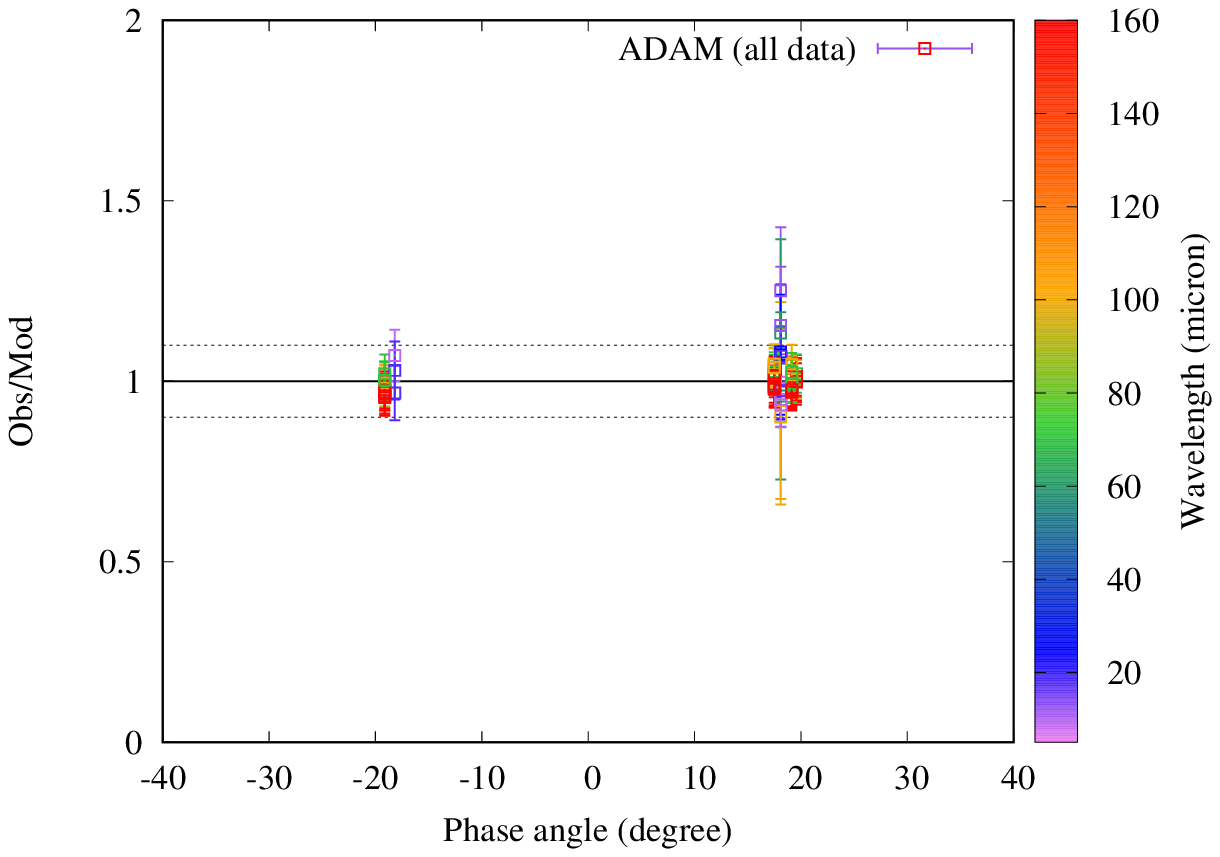}

  \caption{(88) Thisbe. See the caption in Fig.~\ref{fig:00001_OMR}. 
  }\label{fig:00088_OMR}
\end{figure}

\begin{figure}
  \centering
  \includegraphics[width=0.7\linewidth]{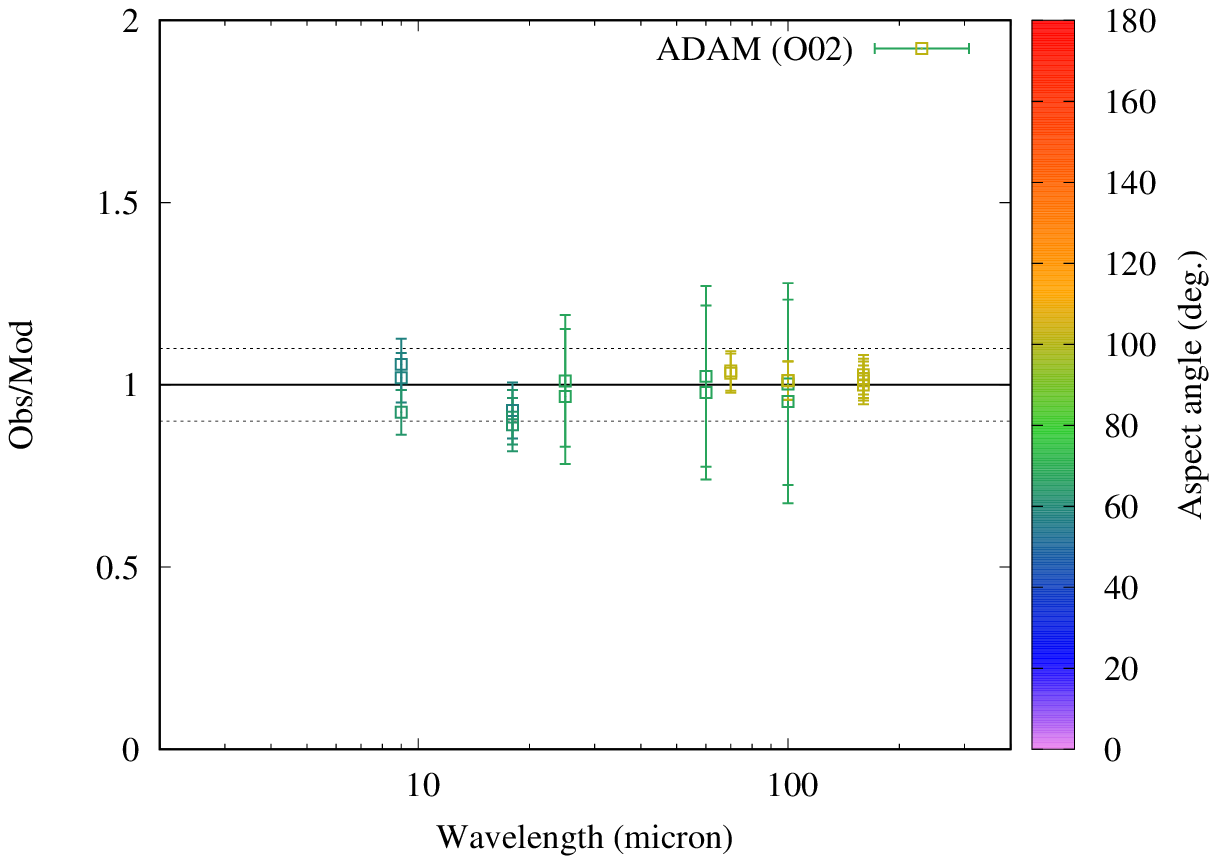}

  \includegraphics[width=0.7\linewidth]{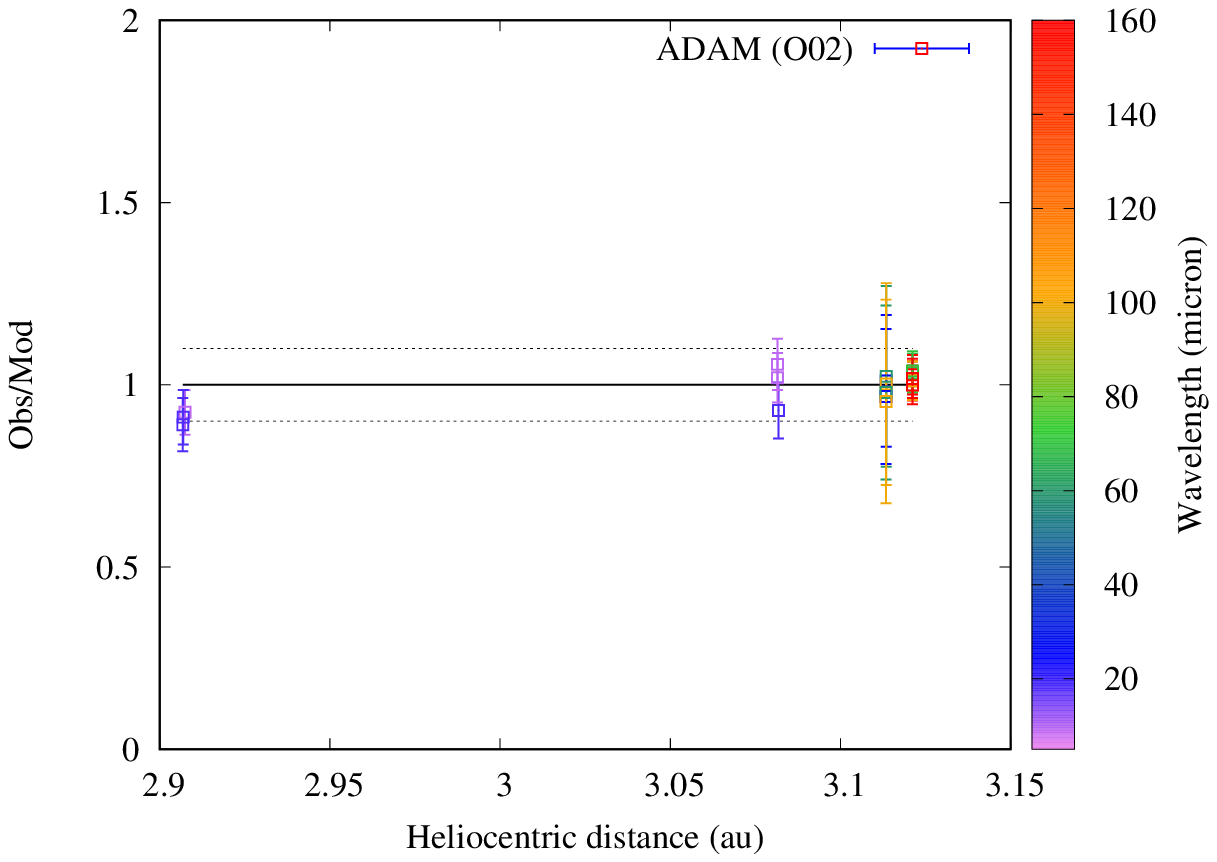}

  \includegraphics[width=0.7\linewidth]{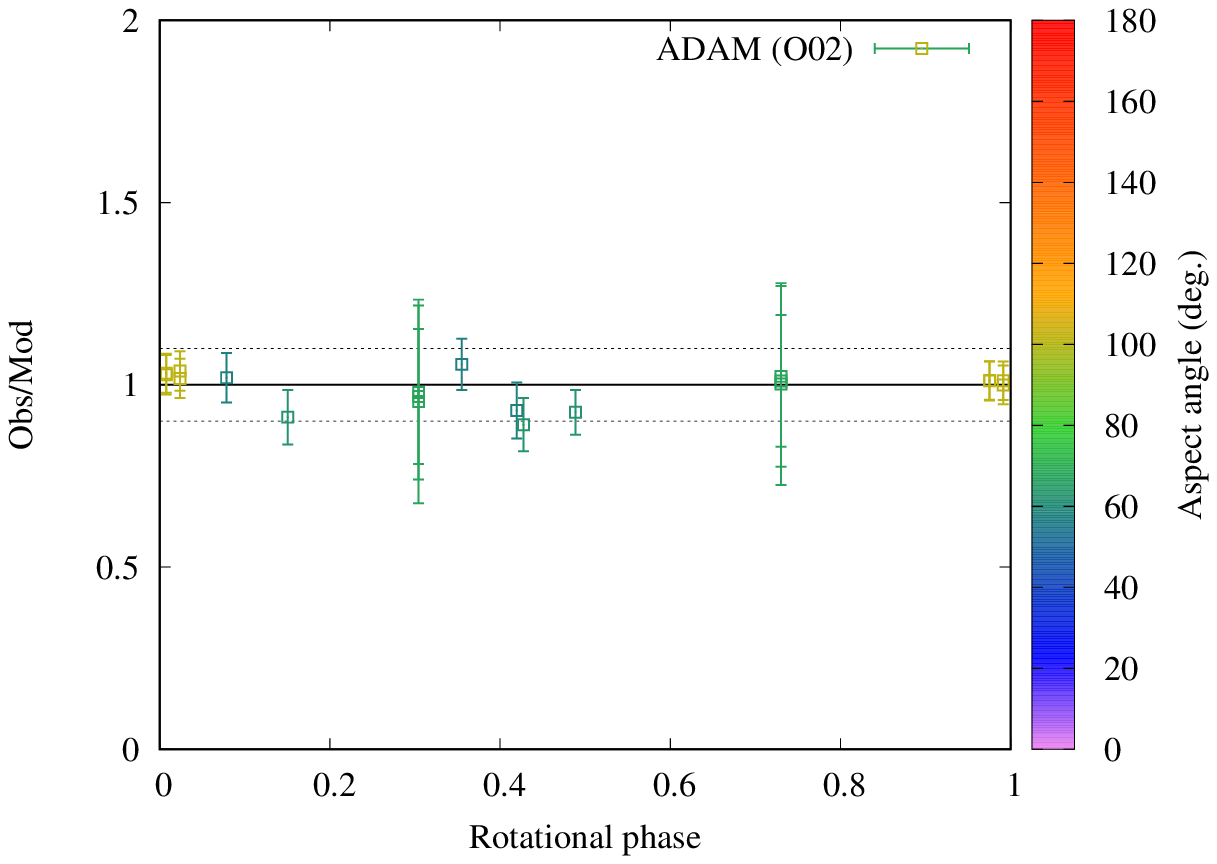}

  \includegraphics[width=0.7\linewidth]{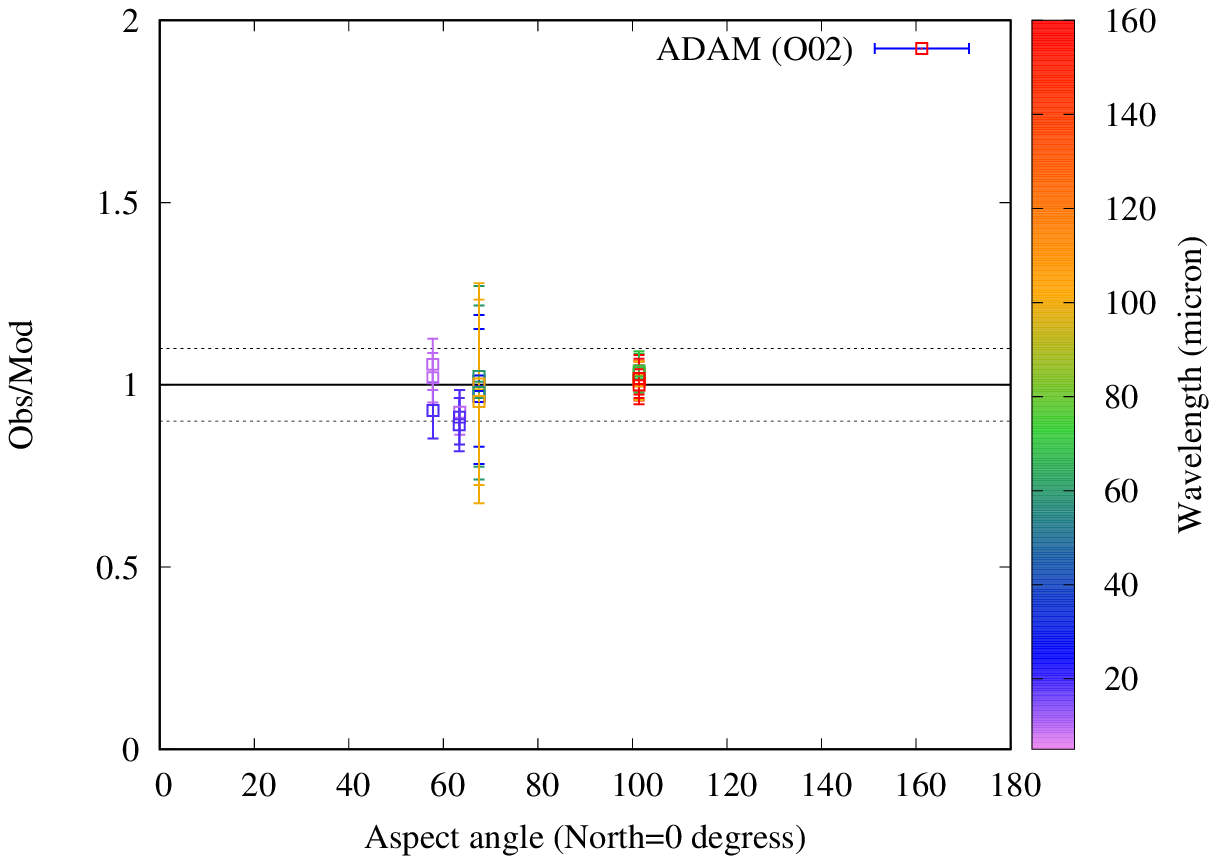}

  \includegraphics[width=0.7\linewidth]{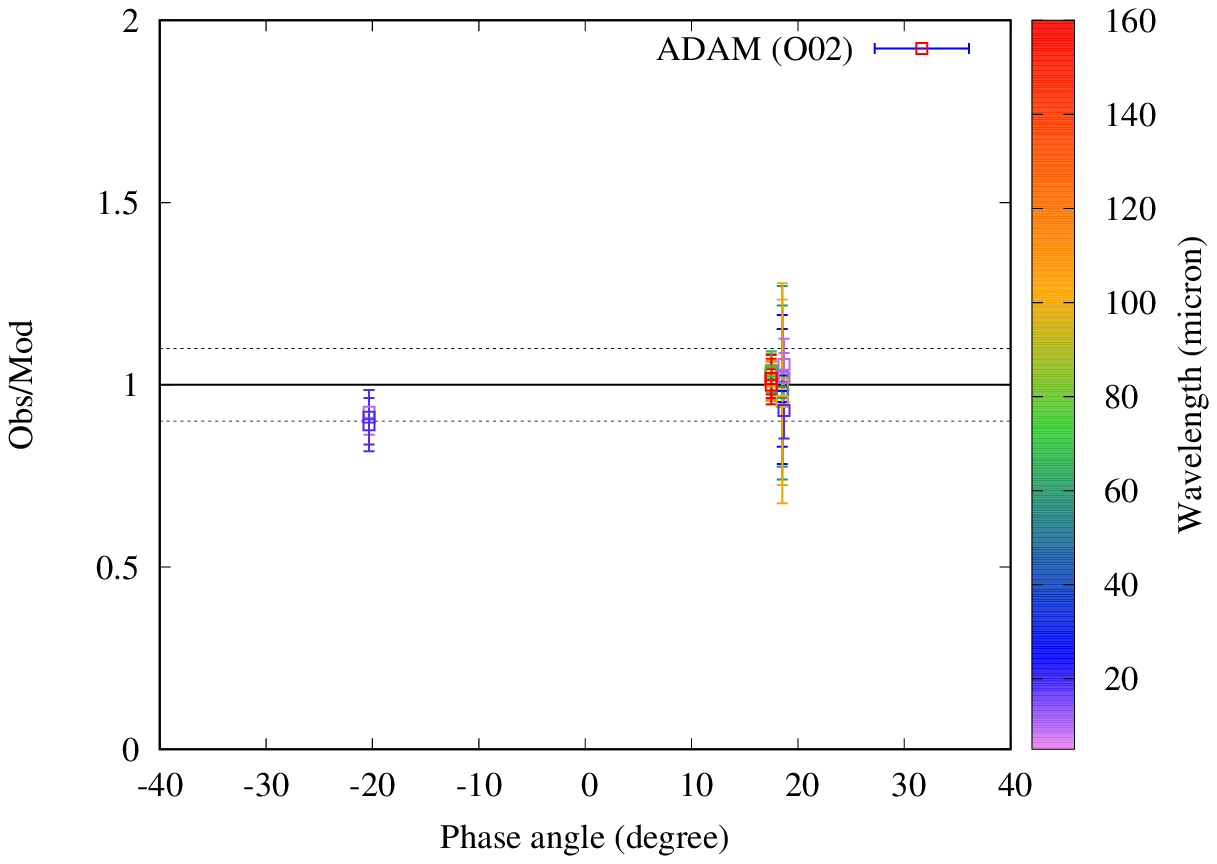}

  \caption{(93) Minerva. See the caption in Fig.~\ref{fig:00001_OMR}. Here, O02 is an internal code referring to the best-fitting model to a subset of data. In this case, all MSX observations and 12-micron IRAS data were not modelled because they could not be approximated by any model and they led to extremely high values of thermal inertia and surface roughness. 
  }\label{fig:00093_OMR}
\end{figure}

\begin{figure}
  \centering
  \includegraphics[width=0.7\linewidth]{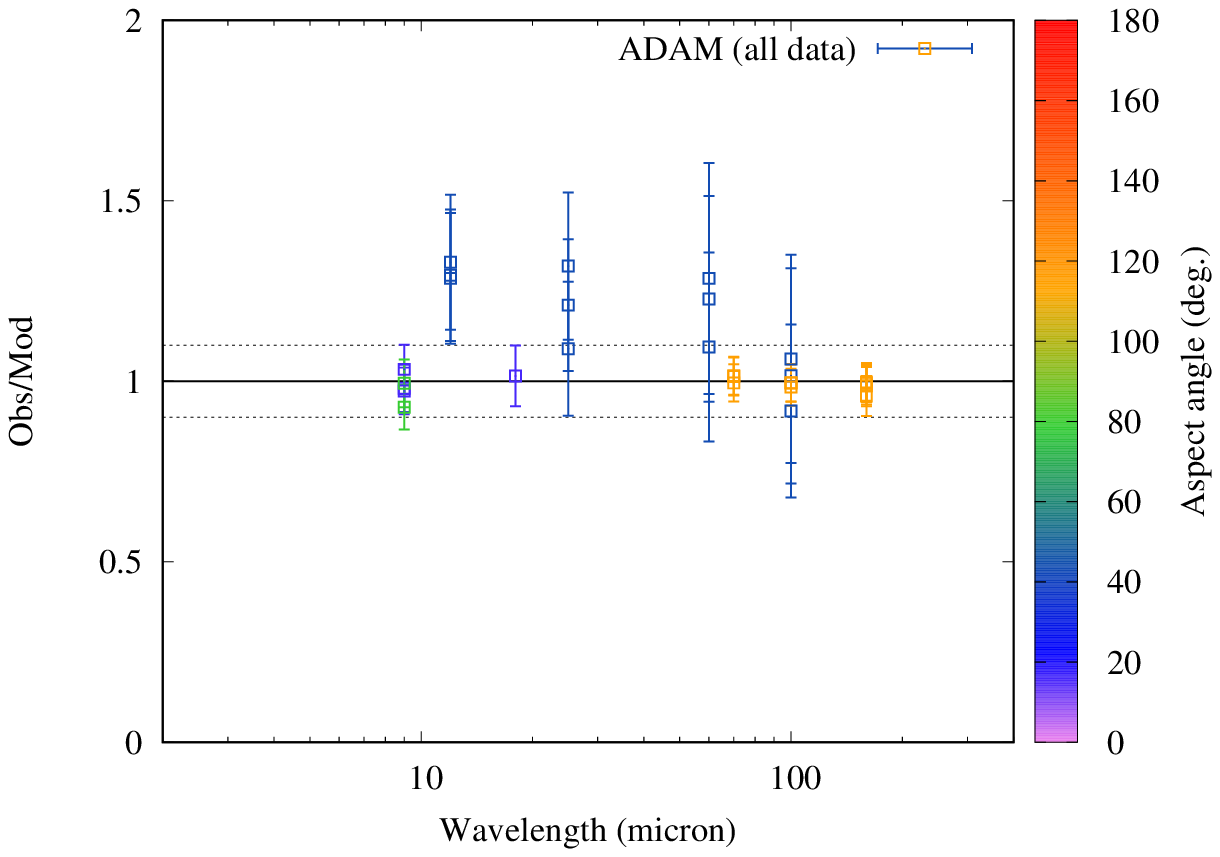}

  \includegraphics[width=0.7\linewidth]{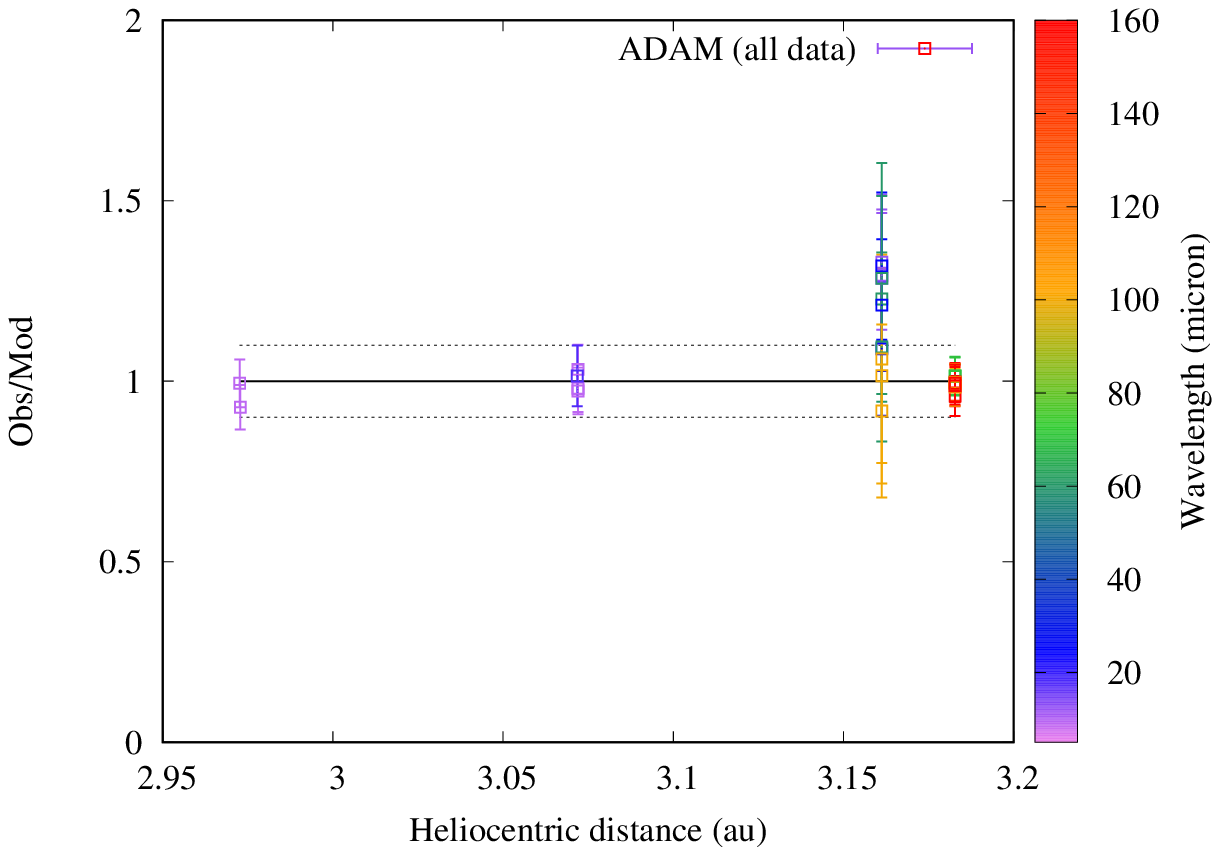}

  \includegraphics[width=0.7\linewidth]{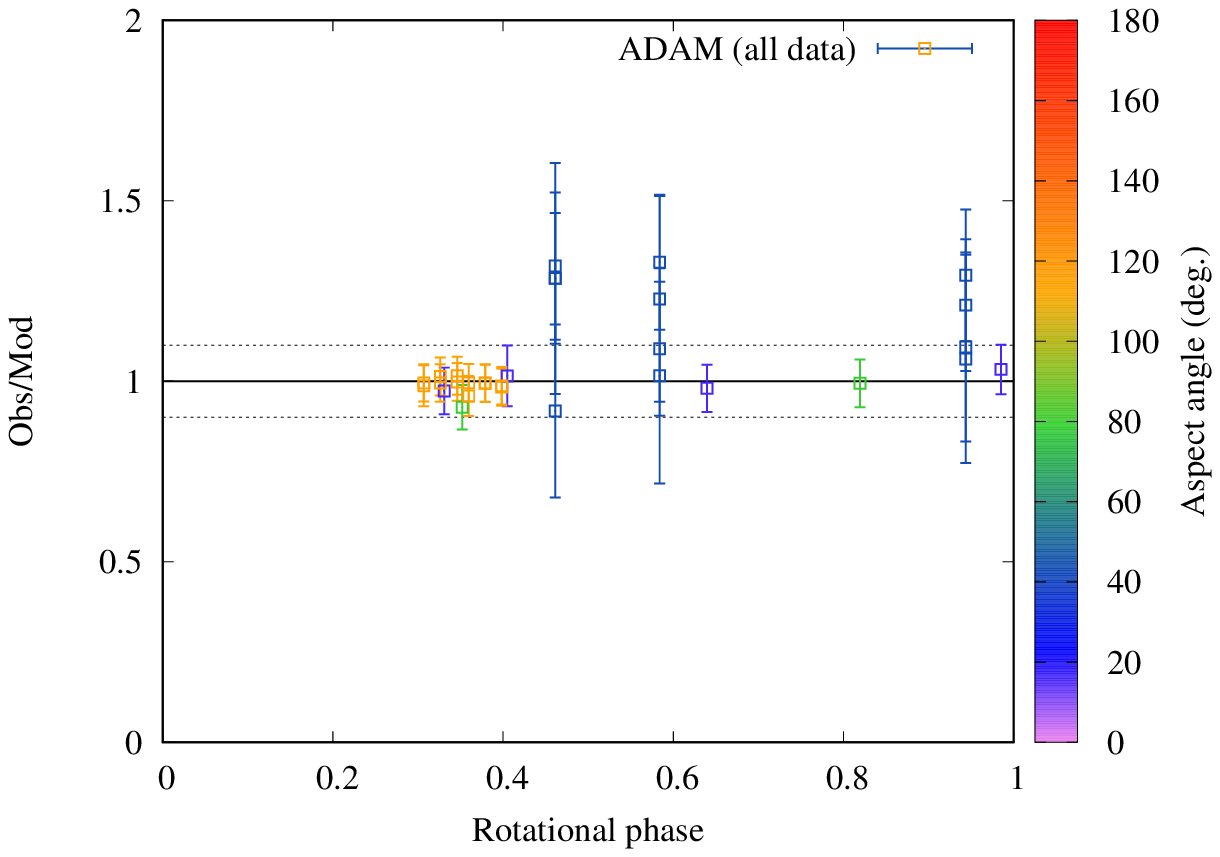}

  \includegraphics[width=0.7\linewidth]{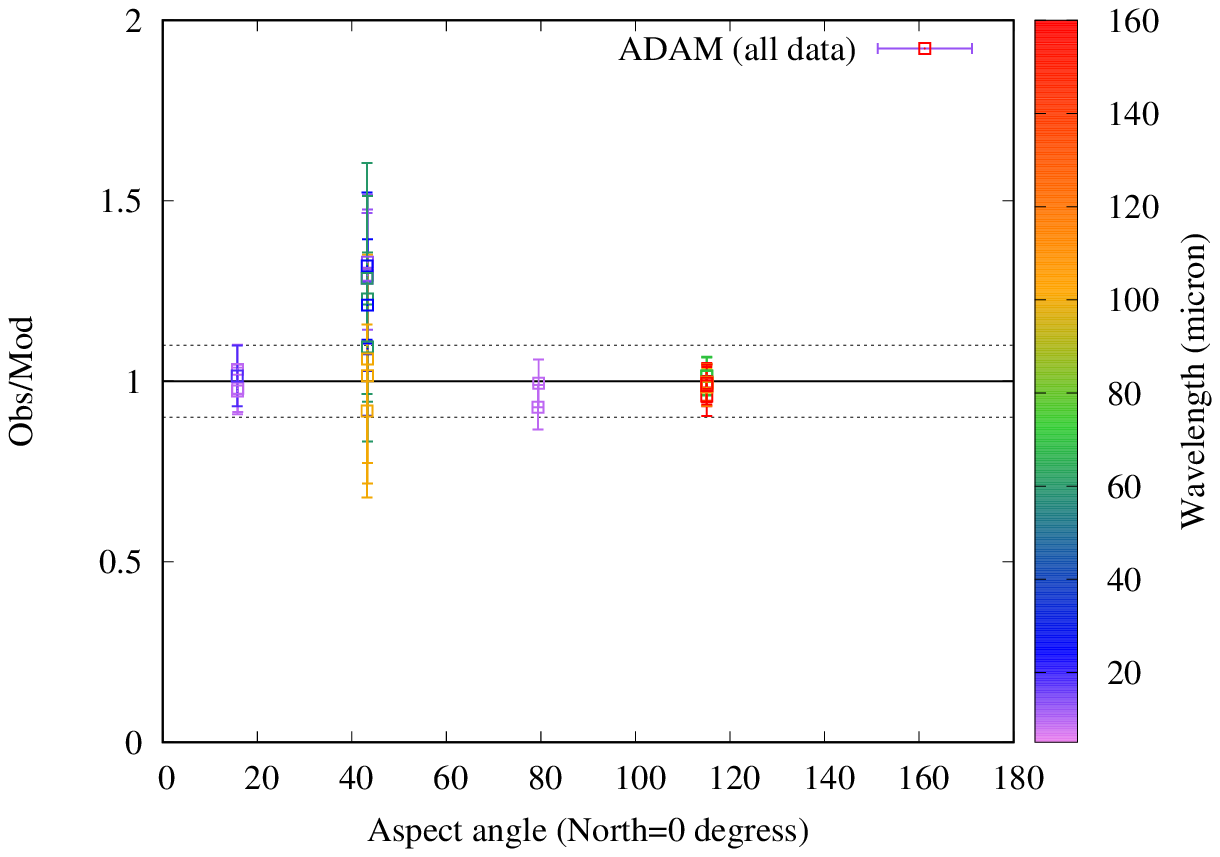}

  \includegraphics[width=0.7\linewidth]{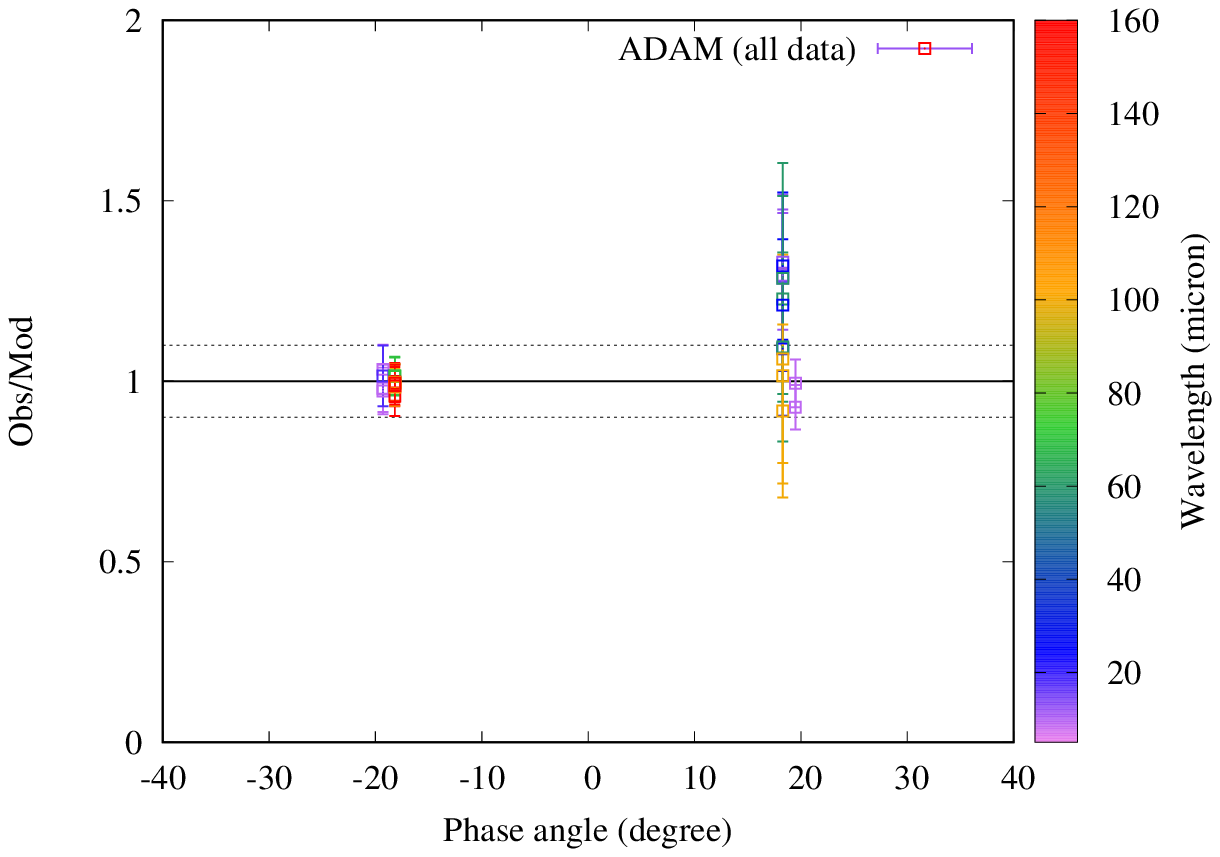}

  \caption{(423) Diotima. See the caption in Fig.~\ref{fig:00001_OMR}. 
  }\label{fig:00423_OMR}
\end{figure}

\begin{figure}
  \centering
  \includegraphics[width=0.7\linewidth]{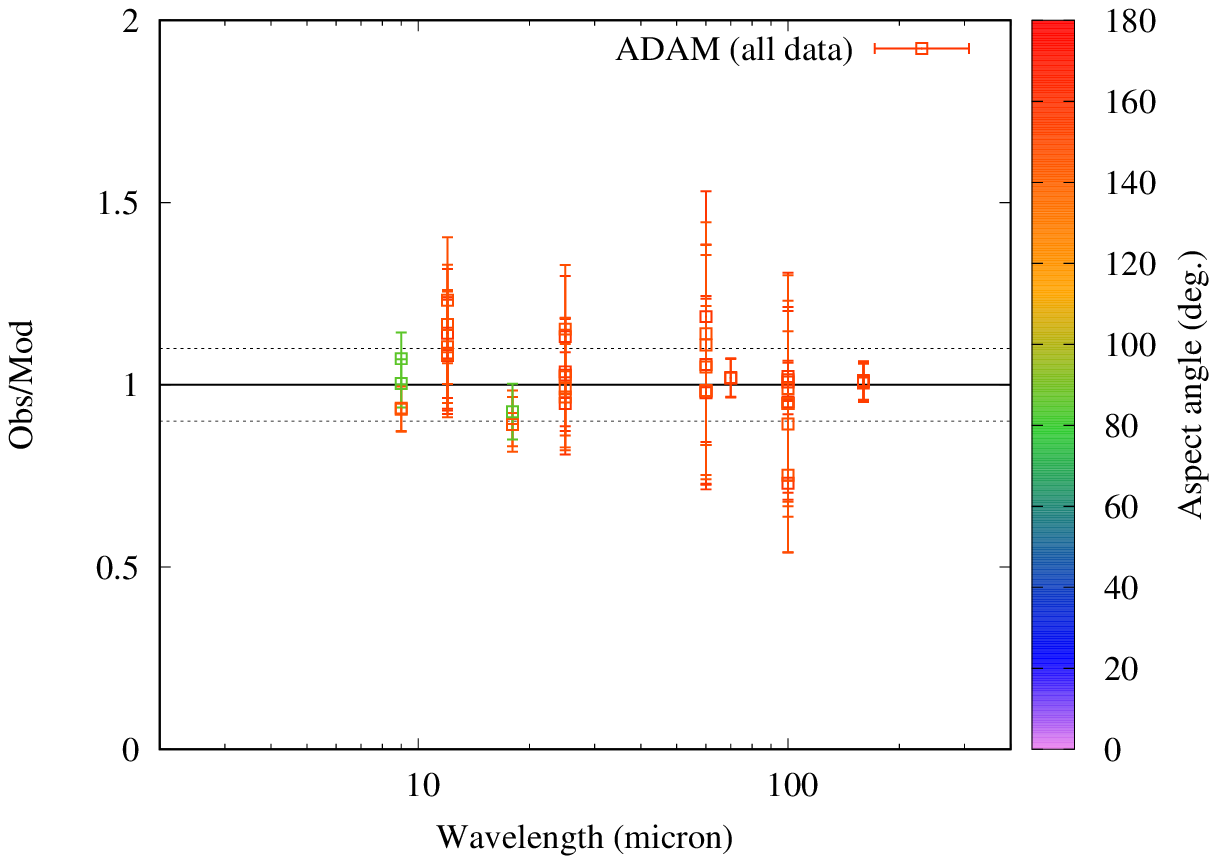}

  \includegraphics[width=0.7\linewidth]{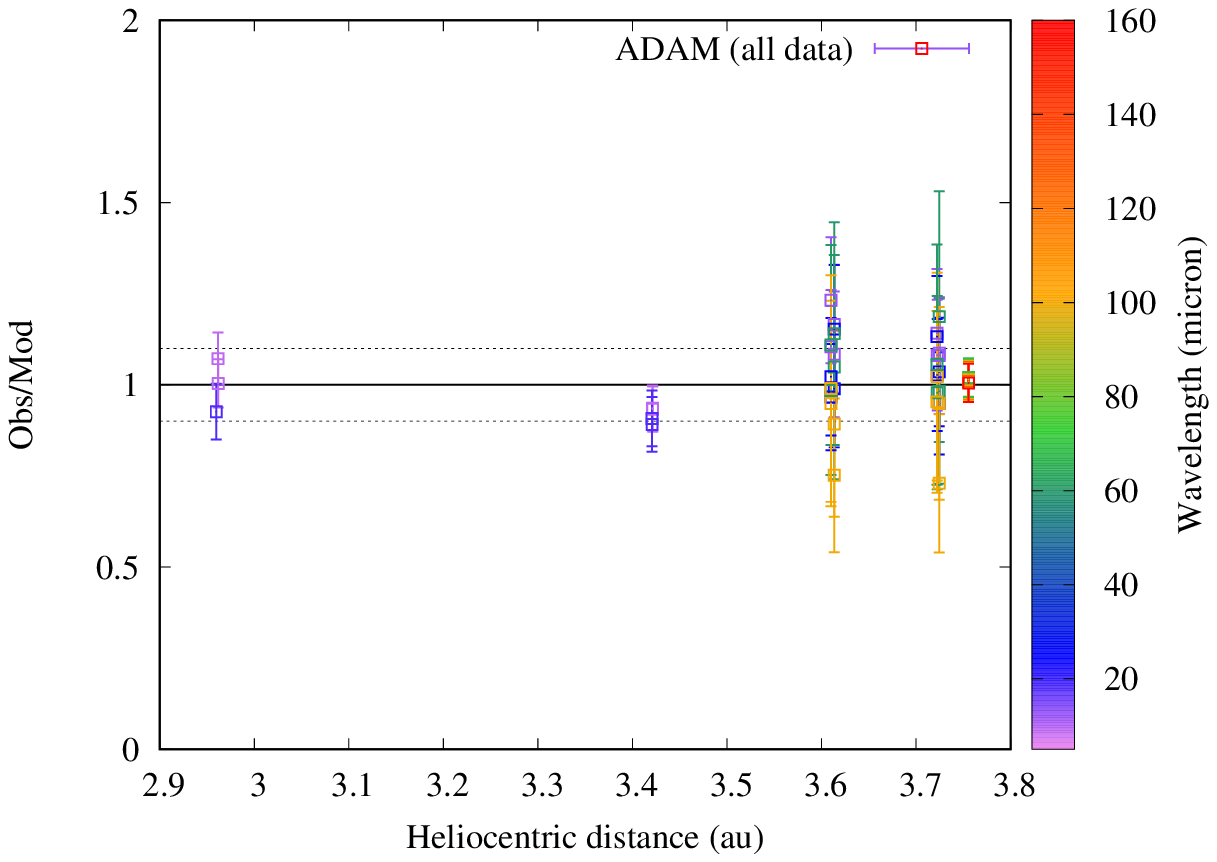}

  \includegraphics[width=0.7\linewidth]{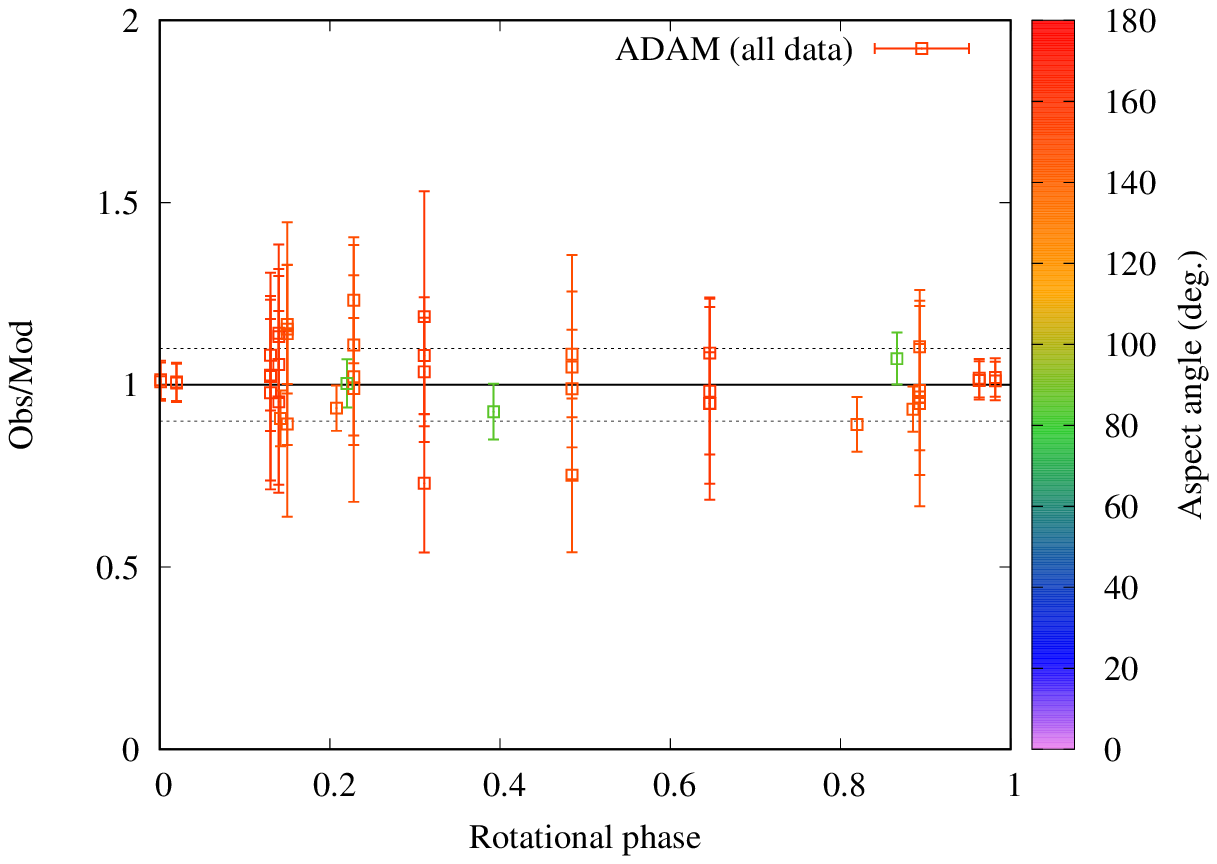}

  \includegraphics[width=0.7\linewidth]{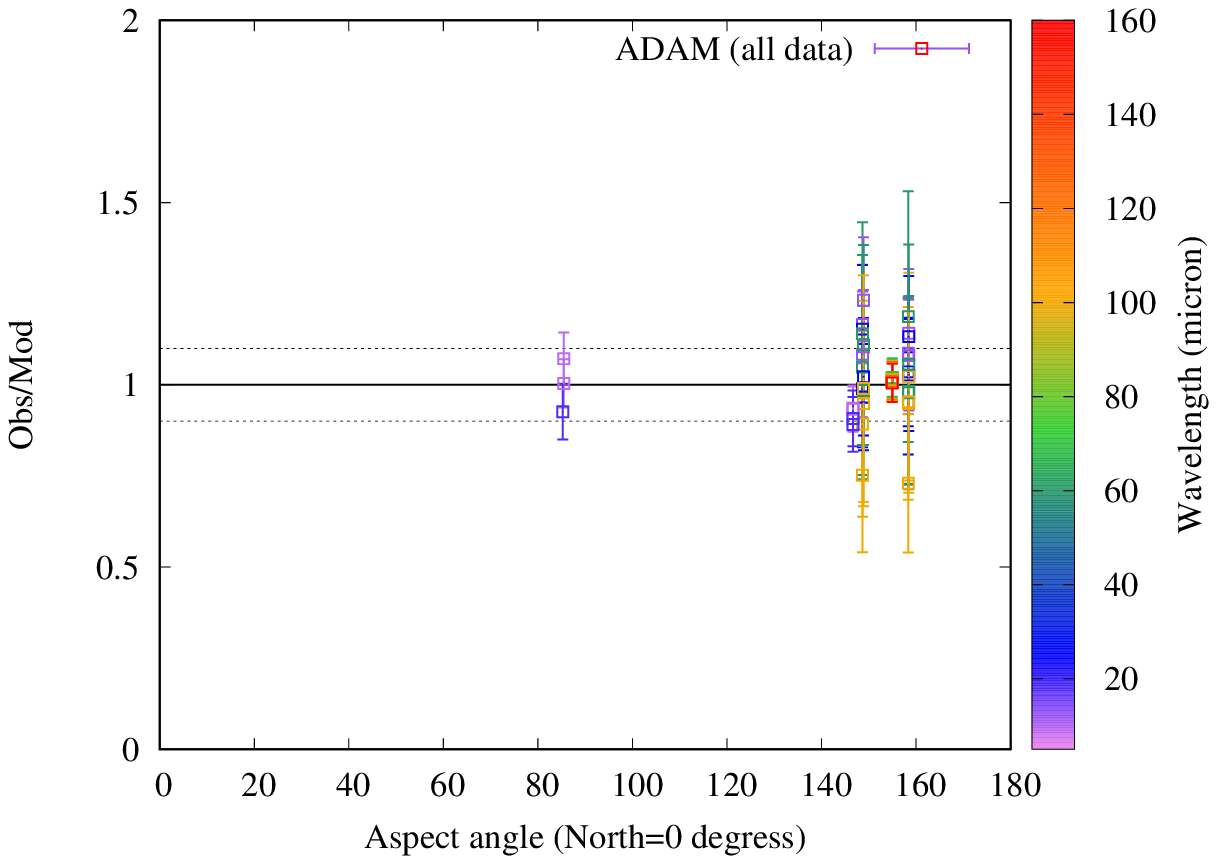}

  \includegraphics[width=0.7\linewidth]{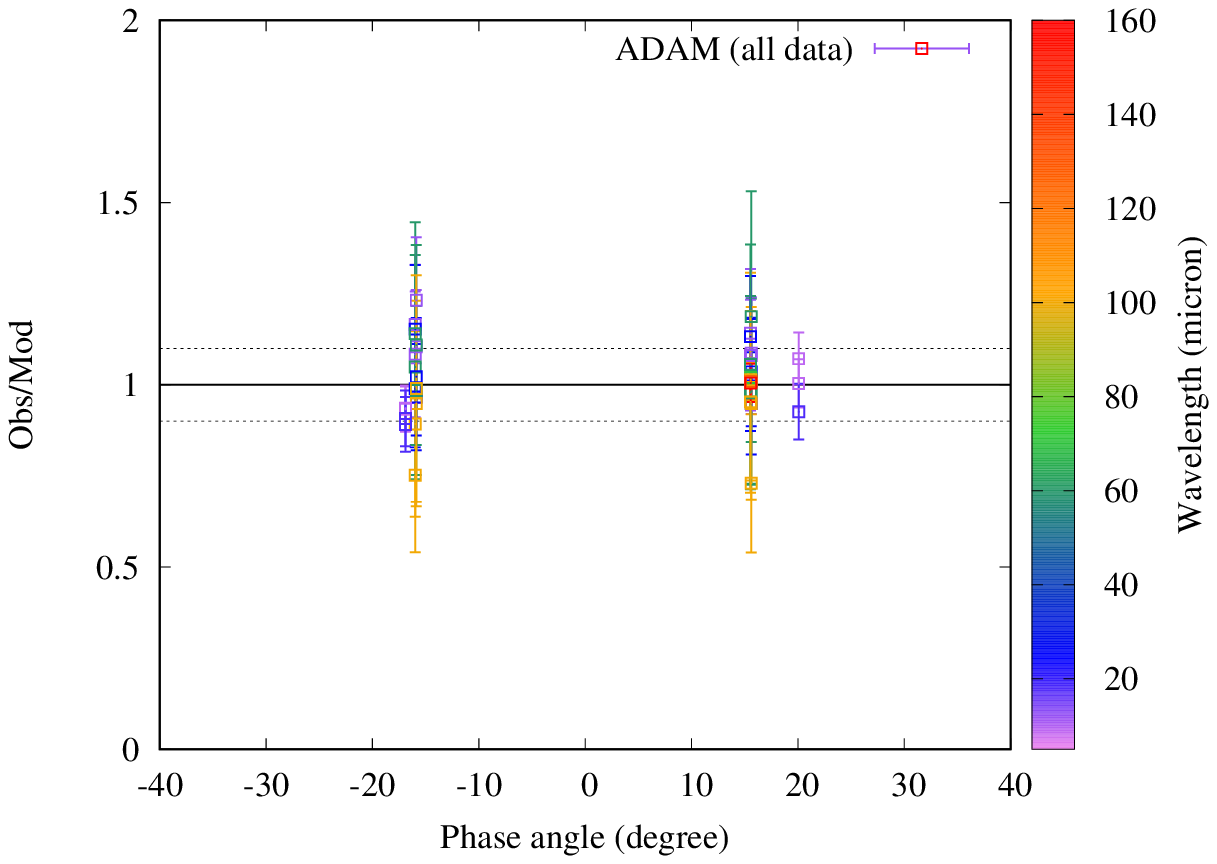}

  \caption{(511) Davida. See the caption in Fig.~\ref{fig:00001_OMR}. 
  }\label{fig:00511_OMR}
\end{figure}

\section{Herschel PACS observations}\label{app:fluxes}

We refer the reader to the caption of Table \ref{tab:fluxes}.
\longtab[1]{
  \footnotesize
  \begin{landscape}
\begin{longtable}{l l l l l l r c r r c c c c }
      \caption{\label{tab:fluxes} Herschel Space Observatory PACS observations used in this work. See \citet{Mueller2005}, \citet{Nielbock2013},\citet{Balog2014}, and \citet{Mueller2014} for details on the data set and observing modes (Obs. mode), and \citet{Kiss2014} on the reduction approach, which required non-standard strategies for a subset of the observations. These fluxes will also be retrievable from the SBNAF database \citep{Szakats2020} along with the other data and all relevant references. The start and end times of the observations are denoted by $t_s$ and $t_e$, $\lambda$ is the reference wavelength of the band, $\langle f\rangle_{\lambda}$ and $\sigma\langle f\rangle_{\lambda}$ the calibrated in-band flux and its error, $f_\lambda$ the colour-corrected flux density, $\sigma f_\lambda$ the absolute calibration error, $r$ the heliocentric distance, $\Delta$ the distance to the observer, and $\alpha$ phase angle.
        The sign of the phase angle is given by that of the vector product between versors from the asteroid to the Sun and the observer. 
      }\\
      \hline\hline
      Target & Obs. ID & Obs. mode & $t_s$ (JD) & $t_e$ (JD) & Band & $\langle f\rangle_{\lambda}$ (Jy) & $\langle f\rangle_{\lambda}$ error (Jy) & $\lambda$ ($\mu$m) & $f_\lambda$ (Jy) & $\sigma f_\lambda$ (Jy) & $r$ (au) & $\Delta$ (au) & $\alpha$ (deg.) \\
      \hline
      \endfirsthead
      \caption{continued.}\\
      \hline\hline
      Target & Obs. ID & Obs. mode & $t_s$ (JD) & $t_e$ (JD) & Band & $\langle f\rangle_{\lambda}$ (Jy) & $\sigma\langle f\rangle_{\lambda}$ (Jy) & $\lambda$ ($\mu$m) & $f_\lambda$ (Jy) & $\sigma f_\lambda$ (Jy) & $r$ (au) & $\Delta$ (au) & $\alpha$ (deg.) \\
      \hline
      \endhead
      \hline
      \endfoot
(3) Juno & 1342188358 & ChopNod & 2455186.74250 & 2455186.74297 & blue & 51.7010 &  0.5174 & 70 &  51.444 &   2.673 &  2.037 &  1.765 & -29.2 \\
(3) Juno & 1342188358 & ChopNod & 2455186.74250 & 2455186.74297 & red & 13.3162 &  0.1430 & 160 &  12.562 &   0.655 &  2.037 &  1.765 & -29.2 \\
(3) Juno & 1342188359 & ChopNod & 2455186.74510 & 2455186.74557 & green & 29.5331 &  0.2960 & 100 &  28.813 &   1.497 &  2.037 &  1.765 & -29.2 \\
(3) Juno & 1342188359 & ChopNod & 2455186.74510 & 2455186.74557 & red & 13.1645 &  0.1415 & 160 &  12.419 &   0.647 &  2.037 &  1.765 & -29.2 \\
(3) Juno & 1342188360 & Scanmap & 2455186.74776 & 2455186.75040 & blue & 52.2075 &  0.5221 & 70 &  51.948 &   2.699 &  2.037 &  1.765 & -29.2 \\
(3) Juno & 1342188360 & Scanmap & 2455186.74776 & 2455186.75040 & red & 13.3157 &  0.1338 & 160 &  12.562 &   0.653 &  2.037 &  1.765 & -29.2 \\
(3) Juno & 1342188361 & Scanmap & 2455186.75143 & 2455186.75407 & blue & 51.9514 &  0.5195 & 70 &  51.693 &   2.686 &  2.037 &  1.765 & -29.2 \\
(3) Juno & 1342188361 & Scanmap & 2455186.75143 & 2455186.75407 & red & 13.2663 &  0.1333 & 160 &  12.515 &   0.650 &  2.037 &  1.765 & -29.2 \\
\end{longtable}
  \end{landscape}      
}

\end{appendix}

\end{document}